\documentclass[twoside]{article}

\usepackage{arxiv}
\renewcommand{\undertitle}{}
\renewcommand{\headeright}{}

\usepackage{natbib}

\usepackage{amsmath, amssymb, graphicx, bm}
\usepackage{amsthm}

\usepackage{bbm}
\usepackage{hyperref}
\usepackage{xcolor}
\usepackage{float}
\usepackage{booktabs}
\usepackage{bbding}
\usepackage{makecell}
\usepackage{comment}

\usepackage{graphicx}
\usepackage[caption=false]{subfig}

\allowdisplaybreaks

\usepackage{macros}

\newtheorem*{lem}{Lemma}

\newcommand{\rp}[1]{{\color{black}#1}}
\newcommand{\na}[1]{{\color{black}#1}}

\newcounter{savedfigure}
\newcounter{algorithm}
\renewcommand{\thealgorithm}{A\arabic{algorithm}}

\usepackage{bibunits}
\defaultbibliography{references}
\defaultbibliographystyle{apalike}

\usepackage{authblk}
\title{Order-Restricted Bayesian Ordinal Regression \\
for the Modeling of Neuron Degeneration in \textit{Caenorhabditis elegans}}
\author[1]{Rick Presman}
\author[1]{Niccol\`o Anceschi$^*$}
\author[2]{Javier Huayta}
\author[2]{Joel N.\ Meyer}
\author[1]{Amy H.\ Herring}
\affil[1]{Department of Statistical Science, Duke University, Durham, NC}
\affil[2]{Nicholas School of the Environment, Duke University, Durham, NC}
\date{}
\renewcommand{\shorttitle}{Neuron Degeneration in {\it C. elegans} Assays: Direct vs Parental Exposure}
\begin{document}

\maketitle
\renewcommand{\thefootnote}{$~$}
\footnotetext{{\normalsize $^*$} Corresponding author: \texttt{niccolo.anceschi@duke.edu}}
\renewcommand{\thefootnote}{\arabic{footnote}}

\begin{bibunit}

\vspace{-20pt}

\begin{abstract}
    Neuron degeneration is the underlying mechanism for the development of many diseases. Quantifying the association between increasing levels of toxic exposure and progressive neuronal damage is a critical component of understanding this development. We investigate this association by analyzing a novel dataset of ordinal neuronal damage scores derived from a series of toxicological assays of {\it C. elegans}, including variables such as 
    toxicant concentration, maternal treatment, and direct chemical exposure.
    \rp{We propose a computationally efficient parameter-constrained Bayesian ordinal regression that captures the monotonic association between neuron damage scores and corresponding treatments.} 
    \rp{Power analysis via simulation studies reinforces the advantages of our model over standard alternatives used in existing work by practitioners.}
    \na{Analysis of the novel {\it C. elegans} assays indicates that maternal toxicity increases susceptibility in progeny, with the offspring generation exhibiting amplified neuronal damage upon later-life rotenone exposure even under mild parental developmental treatment.}
\end{abstract}


\section{Introduction}

Neuron degeneration is a key factor in the progression of many diseases. Dopaminergic (DA) neurons, in particular, are a critical component of the central nervous system (CNS) \citep{iversen2007dopamine}. As neurons are exposed to toxic environmental factors at increasing levels of severity, their structure and function atrophy, resulting in the development of neurologic disease, such as Parkinson's disease \citep{xi2011modeling}. Thus, it is critical to understand the association between toxic exposure and the ensuing neuronal damage.

\begin{figure}[H]
    \centering
    \includegraphics[width = 0.8\linewidth]{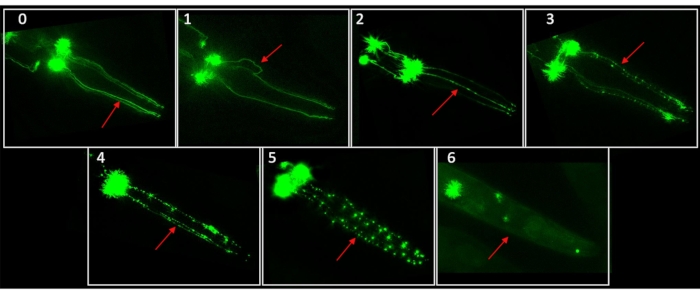}
    \vspace{-5pt}
    \caption{
    Fluorescence microscopy images of {\it C. elegans} neurons, illustrating the ordinal scale developed in \cite{bijwadia2021quantifying} to quantify dopaminergic degeneration. Used with permission.}
    \label{fig:neuron_pic}
\end{figure}

We investigate this association by developing a novel statistical model to analyze experimental data from \citet{bergemann2026progeny},
\na{where {\it C. elegans} nematodes were developmentally exposed to a low or high dose of rotenone, or otherwise assigned to a control group.
The experiment spans two generations: a parental generation ($P0$) directly exposed to rotenone during larval development, and an offspring generation ($F1$) without development-stage exposure to rotenone and born after a chemical washout period.
This protocol was specifically designed to minimize the potential for maternal loading of pollutants to gametes or toxic effects to parents, which could otherwise confound offspring outcomes through indirect effects on germ cell or gamete health.
Both generations were independently examined with and without a late-life rechallenge dose of substantially higher rotenone concentration, allowing joint assessment of direct dose-response effects and the influence of parental exposure on neuronal susceptibility in progeny.}
Damage in each neuron of the {\it C. elegans} nematodes is observed through a fluorescence microscopy-based morphological analysis, and then converted to an ordinal damage scale ranging from 0 (no observable damage) to 6 (a neuron with totally disrupted morphological integrity). 
This scoring system, developed in \cite{bijwadia2021quantifying}, combines qualitative and quantitative information, such as the intensity of fluorescence and the number of breaks observed along a neuron.
\na{
Figure~\ref{fig:neuron_pic} illustrates representative neurons at each level of such damage scale.}

Our aim is to assess the association between rotenone and damage score while enforcing monotonicity in the dose-response ``curve'' and accounting for the dependence
among neurons belonging to the same nematode.
\na{This reflects the biological expectation that higher concentrations of a toxic compound can only sustain or exacerbate cellular damage, as biological repair mechanisms become progressively overwhelmed.}
\na{From a statistical methods perspective, foundational approaches to modeling associations between ordinal variables under monotonicity constraints date back to \cite{mccullagh1980regression}, who introduced cumulative ordinal regression as the natural framework for ordinal responses, and \cite{agresti2002analysis}, who discussed order-restricted parameterizations of log-linear and categorical models for contingency tables under inequality constraints. Complementary approaches drop the constraints entirely and instead rely on targeted trend tests \citep{williams1971test, marcus1976powers, farrar2022comparison}. However, inference under order restrictions is complicated by the likelihood-ratio statistic lacking a standard closed-form asymptotic distribution \citep{davidov2011constrained, davis2011constrained}.
Additionally, existing software jointly accommodates at most two of the key features relevant for our experimental setup -- i.e. ordinal responses, parameter constraints, and random effects for within-subject dependence -- but never all three simultaneously.
}

\na{
Bayesian methods offer a more natural route to constraint incorporation \citep{klugkist2010bayesian, sen2018constrained, presman2023distance}, while bypassing the asymptotic distribution challenges in quantifying uncertainty. However, general-purpose tools such as \texttt{stan} \citep{carpenter2017stan} and the \texttt{R} package \texttt{brms} \citep{burkner2017brms} can be computationally demanding and sensitive to prior specification and initialization, making them cumbersome to deploy for domain scientists.
Faced with these challenges, toxicologists often rely on simpler alternatives, typically 
t-tests and ANOVA \citep{berkowitz2008application, luo2019age, bijwadia2021quantifying}, which are straightforward to implement but assume Gaussianity, ignore the ordinal structure of both dose and response, and neglect within-subject dependence. 
More powerful alternatives that exploit the monotonicity of the dose-response relationship include the Cochran-Armitage test \citep{cochrane1954some, armitage1955tests} and its nonparametric extension, the Jonckheere-Terpstra test \citep{terpstra1952asymptotic, jonckheere1954distribution}, though these remain hypothesis tests rather than models and similarly fail to account for the full structure of the data.
}

\na{
To overcome these limitations, we propose a Bayesian ordinal regression model that enforces a monotone dose-response relationship via non-negative treatment increments, while allowing for flat regions where additional exposure elicits no further damage. Within-worm dependence is captured through worm-level random effects, and inference is carried out via the rank likelihood framework \citep{hoff2009first, Murray2013}, which relaxes the parametric assumptions of standard cumulative ordinal models. The resulting posterior admits an efficient Gibbs sampler that is substantially faster than \texttt{stan}-based alternatives and, as we demonstrate in our simulation study, achieves higher statistical power than both Bayesian and frequentist competitors.
Applied to the {\it C. elegans} assays, our model reveals that parental rotenone exposure sensitizes offspring neurons, with the $F1$ generation exhibiting amplified neuronal damage upon adult-life rotenone exposure even for low parents' developmental treatment.
}

\na{The rest of this paper is organized as follows. Section~\ref{sec:data} describes the {\it C. elegans} assay data in detail.
Section~\ref{sec:method} introduces the proposed Bayesian isotonic rank likelihood model.
Section~\ref{sec:simulations} benchmarks the statistical power of our approach against both Bayesian and frequentist competitors in simulation studies.
Section~\ref{sec:analysis} applies the model to the {\it C. elegans} data, with results at the neuron and worm level. Section~\ref{sec:discussion} concludes with directions for future work.
}

\section{The {\it C. elegans} Neuron Data}\label{sec:data}

The data consists of 559 worms, each with up to 4 neurons, with 2,099 total neurons observed. \na{Damage scores are recorded on each neuron based on the system developed in \cite{bijwadia2021quantifying}, with scores ranging from 0 (no damage) to 6 (full loss of neuron morphological integrity). Worms belong to either the parental generation ($P0$, 46.7\%), developmentally exposed to rotenone at 0 $\mu M$ (control), 0.03 $\mu M$, or 0.5 $\mu M$, or the offspring generation ($F1$, 53.3\%), born after a chemical washout and never directly exposed to rotenone during development. Specific parent-offspring pairs are not tracked, but are aggregated into generation cohorts. Thus, $P0$ and $F1$ groups reflect known exposure histories at the population-level ancestry -- rather than individual lineages -- enabling generation-level inference. Approximately half the worms in each generation (42.9\% overall) additionally received a late-life rechallenge dose of 25 $\mu M$ rotenone at day 4 of adulthood, to unmask latent neuronal susceptibility. Experiments were conducted across three independent replicates (33.3\%, 35.2\%, and 31.5\% of worms, respectively), which may introduce batch effects due to exogenous experimental factors.}

\begin{figure}[t!]
\centering
\includegraphics[width=\textwidth]{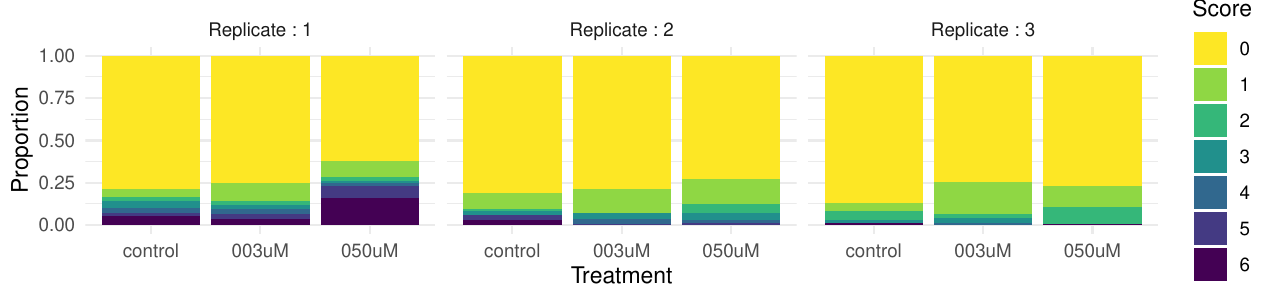}
\put(-468,98){\makebox(0,0){\textbf{(A)}}}
\vspace{1pt}
\noindent
\begin{minipage}[t]{0.63\textwidth}
\vspace{-200pt}
    \includegraphics[width=\textwidth, trim={0 0 50 0}, clip]{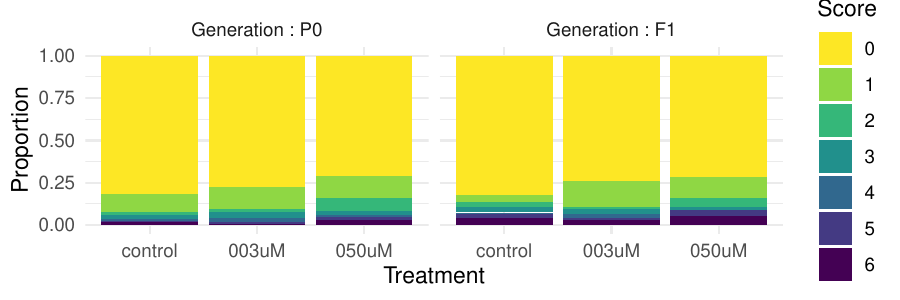}
    \put(-295,100){\makebox(0,0){\textbf{(B)}}}
    \vspace{2pt}
    \includegraphics[width=\textwidth, trim={0 0 50 0}, clip]{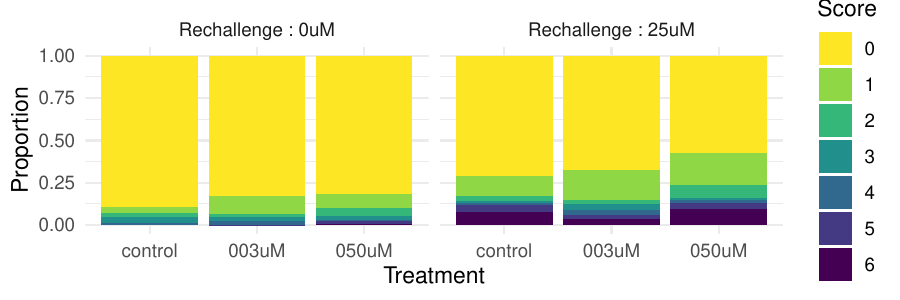}
    \put(-295,100){\makebox(0,0){\textbf{(C)}}}
\end{minipage}%
\hfill
\begin{minipage}[t]{0.32\textwidth}
    \includegraphics[width=\textwidth, height=7cm]{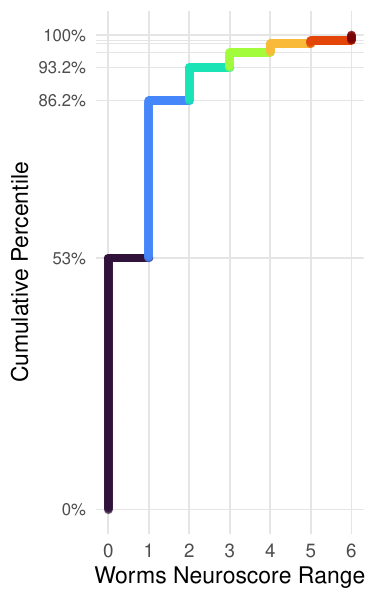}
    \put(-153,186){\makebox(0,0){\textbf{(D)}}}
\end{minipage}
\vspace{-5pt}
\caption{(A,B,C) Proportions of damage scores against treatment levels, grouped by different experimental conditions. (A) Distributions across replicate-level suggest non-negligible batch effects. (B,C) Stratification by generation (parental $P0$ vs.\ offspring $F1$) and retreatment status (control vs.\ 25\,$\mu$M rotenone), respectively, underscores their biological relevance to the observed variation. (D) Empirical distribution of within-worm damage score range, computed as the difference between the maximum and minimum observed damage scores in each worm. The y-axis reports the cumulative percentile of worms attaining a given range width. More than 85\% of worms exhibit scores spanning at most two consecutive ordinal categories.}
\label{fig:EDA_all}
\end{figure} 

\na{Figure~\ref{fig:EDA_all} displays observed damage scores at the neuron level, stratified by treatment, generation, retreatment status, and experimental replicate.
Across all groups, damage scores are heavily concentrated in the lowest categories, with the majority of neurons showing no observable damage. 
As expected, higher rotenone doses shift the distribution toward greater damage, with the effect most pronounced at the 0.5 $\mu M$ level. 
Stratifying by generation alone, the distributions appear broadly similar across parental ($P0$) and offspring ($F1$) generations (Figure 2B), which seems to suggest no strong marginal difference in damage levels.
By contrast, retreated worms show markedly higher damage levels across all dose groups (Figure 2C), consistent with the hypothesis that prior exposure sensitizes neurons to subsequent insult. 
Replicate 1 exhibits systematically higher damage than Replicates 2 and 3 (Figure 2A), suggesting non-negligible batch effects attributable to exogenous experimental factors such as variability in laboratory conditions or operator handling; controlling for replicate in the model helps isolate the treatment effect of interest. 
}

\na{Despite being recorded as numerical values, damage scores form an ordinal scale: a higher score indicates greater neuronal damage, but equal spacing between categories cannot be assumed, as each score reflects a qualitative description of damage severity rather than a continuous measured quantity. 
Treating them as continuous risks distorting inference by imposing an unwarranted metric structure.
Similarly, while rotenone concentrations originate from a continuous scale, the availability of only a few discrete dose levels makes it more natural to treat the exposure covariate as ordinal rather than continuous. 
Beyond ordinality, the dose-response relationship is conventionally assumed to be monotonically non-decreasing \citep{hill2018nonmonotonic}, an assumption well-supported by the empirical patterns visible in Figure~\ref{fig:EDA_all}.
Finally, neurons from the same worm may share similar susceptibility to damage -- a biologically plausible feature, as weaker worms may tend to have uniformly more vulnerable neurons and more resilient worms uniformly more robust ones. Despite its relevance, this dependence has been largely overlooked in existing analyses, and its explicit modeling can improve both inference on treatment effects and uncertainty quantification.}

\section{Bayesian Rank-based Isotonic Regression}\label{sec:method}

\na{
Our primary goal is to model the monotone relationship between a set of ordinal random variables $Y_i \in \{0,\dots,L\}$ and $p$-dimensional predictors $\bx_i$, for $i \in \{1,\dots,n \}$.
This can be addressed via a cumulative regression model \citep{mccullagh1980regression} of the form
\begin{equation}\label{eq_ordinal_reg}
\bbP[Y_i\leq \ell] = G\big( \delta_\ell -\bx_i^\top\bbeta \big).
\end{equation}
Here $G:[0,1]\rightarrow \Re$ is a monotone non-decreasing link function, and the cutpoints $-\infty=\delta_{-1}<\delta_0 < \dots < \delta_{L-1} < \delta_L = \infty$ partition the real line into ordered categories. 
We further assumed proportional effects of the predictors, by having the same coefficient vector $\boldsymbol{\beta} \in \Re^p$ governing all category boundaries.
Common model choices for the link function include the ordinal logit $G(r)=e^{r}/(1+e^{r})$ and the ordinal probit $G(r)=\Phi(r)$.
Equivalently, these models can be expressed via a data augmentation strategy \citep{albert1993bayesian} which introduces latent continuous scores $Z_i \in \Re$ such that $Y_i = \ell$ if and only if $\delta_{\ell-1} < Z_i < \delta_\ell$.
Under this formulation, the probit model corresponds to a Gaussian latent structure $Z_i \sim \mathcal{N}(\bx_i^\top \bbeta ,\sigma^2)$, with $\sigma^2$ typically fixed to $1$. This representation is particularly attractive in Bayesian inference, due to the resulting conditional conjugacy and the availability of efficient Gibbs updates.
In contrast, the logit model does not admit such direct conjugacy and typically relies on scale-mixture representations of the logistic distribution via Pólya–Gamma augmentation \citep{polson2013bayesian}.
}

\na{
Jointly estimating cutpoints $\delta_\ell$ and the latent scores $Z_i$ leads to redundancies and nonidentifiability, which can hinder posterior exploration and result in poor mixing.
Instead, we adopt the rank likelihood framework \citep{hoff2009first}, which removes the need to estimate cutpoints and avoids the specification of a parametric link function.
The core idea is to replace the full likelihood with a probability statement on the ranks of the latent variables.
The observed responses $Y_i$ induce a partial ordering on the latent scores $Z_i$, so that any admissible latent configurations must be consistent with the observed ranks.
More formally, the full likelihood resulting from \eqref{eq_ordinal_reg} can be rewritten as 
\begin{equation}\label{eq_rank_likelihood}
    p( \bY \! \mid \bbeta ) = 
    p\big( \bY, \bZ \in \mathcal{R}(\bY) \mid \bbeta\big) =
    P\big[ \bZ \in \mathcal{R}(\bY) \mid \bbeta \big] \,
        p\big( \bY \! \mid \bZ \in \mathcal{R}(\bY), \bbeta \big),
\end{equation}
where $\bZ=(Z_1,\dots,Z_n)^\top$ and $\bY=(Y_1,\dots,Y_n)^\top$.
Here, we defined the feasible region
$ \mathcal{R}(\bY) := \{\bZ\in \Re^n \mid Y_{i} < Y_{i'} \Rightarrow Z_{i} < Z_{i'} \}$ and we leveraged $P\big[ \bZ \in \mathcal{R}(\bY) \!\mid\! \bY \big] =1 $.
\citet{d2007extending} proposed dropping the last term in \eqref{eq_rank_likelihood}, expected to be less informative, and using only $P\big[ \bZ \!\in\! \mathcal{R}(\bY) \!\mid\! \bbeta \big]$ as effective likelihood.
The rank likelihood thus enforces on $\bZ$ the same ranks observed in the ordinal data, by restricting the support of a tractable Gaussian latent variable model $\bZ \sim \mathcal{N}_n(\bX\bbeta, \sigma^2 \mathbb{I}_n)$ to $\mathbf{Z} \in \mathcal{R}(\bY)$, with $\bX=(x_1^\top,\dots,x_n^\top)^\top$. \\
}

\na{
Recall that the exposure covariate is itself ordinal in our setting, and we seek to enforce a monotone dose–response relationship. 
Regression with such a monotonicity constraint is a well-studied problem known as isotonic regression \citep{barlow1972isotonic}.
Consider a single predictor $x_i$ for ease of exposition.
If $x_i$ were to be continuous, monotonicity would readily be reflected in a nonnegative regression coefficient $\beta \geq 0$.
Conversely, we follow \cite{neelon2004bayesian} in modeling monotonicity in an ordinal predictor $x_{i} \in \{0,\dots,T\}$ as
$\textstyle{\sum_{t=1}^T} \alpha_t \, \mathbbm{1}_{(x_i \geq t)}$ with incremental effects $\alpha_t \! \geq 0$, $ \forall \, t \geq 1
$.
Here $\mathbbm{1}_{(A)}$ is the indicator variable, taking values $1$ if the event $A$ is true and $0$ otherwise, and we set to zero the contribution of the control level $x_i=0$.
The equivalent parametrization in terms of cumulative effects $\beta_t = \textstyle{\sum_{1 \leq s \leq t}} \alpha_s$ -- were now $ 0 \leq \beta_t \leq \beta_{t+1} \; \forall \, t \geq 1$ -- is then readily retrieved by
$\textstyle{\sum_{t=1}^T} \alpha_t \, \mathbbm{1}_{(x_i \geq t)} \,=\, \textstyle{\sum_{t=1}^T} \beta_t \, \mathbbm{1}_{(x_i=t)} $.
}

\subsection{\na{Model Specification for the {\it C. elegans} Assay}}

\na{
The isotonic rank likelihood framework can be further tailored to capture the multi-level structure of our experimental design through the linear predictor.
Suppose the data comprise $n$ worms, with worm $i=1,\ldots,n$ contributing $n_i$ neurons.
Let us denote by $Y_{ij}\in\{0,\ldots,L\}$ the damage score of neuron $j=1,\ldots,n_i$ from the $i$-th worm.
Each worm belongs to one experimental replicate $r \in \{1,\ldots,R\}$, is associated with a specific generation $g \in \{0,\ldots,G\}$ and a rechallenge group $c \in \{0,\ldots,C\}$, and belongs to a lineage where parent worms developmentally exposed to a dose level $t \in \{0,\ldots,T\}$.
Following the preliminary analysis in Section~\ref{sec:data}, we incorporate batch effects $\mu_r \in \Re$ for replicate group $r$ and random effects $\eta_i \overset{iid}{\sim} \mathcal{N}(0, \rho^2)$ to account for correlation within neurons of worm $i$.}

\na{
Our scientific goal is to compare treatment effects both within and across groups. To this end, we adopt a cell-mean parameterization with three-way interaction effects $\alpha_{tgc} \geq 0$ indexed by treatment $t$, generation $g$, and rechallenge group $c$, giving
\begin{equation}\label{Z_model_v0}
\begin{aligned}
Z_{ij} &\sim \mathcal{N} \Big( \, \psi_i + \eta_i \, , \, \sigma^2 \, \Big) \\
\psi_i &= \sum_{r=1}^R \mu_r \, \mathbbm{1}_{\big(\text{rep}(i)= r \big)} + \sum_{t=1}^T \sum_{g=0}^G \sum_{c=0}^C \alpha_{tgc} \,
    \mathbbm{1}_{\big(\text{treat}(i) \geq t \big)}
    \mathbbm{1}_{\big(\text{gen}(i) = g \big)}
    \mathbbm{1}_{\big(\text{rech}(i) = c \big)} \;.
\end{aligned}
\end{equation}
Notice that, rather than imposing a single global null category, we facilitate interpretation of treatment effects across and within strata by setting to zero the contribution of the dose-control level within each stratum.
This amounts to dropping the three-way terms for any zero-dose instance, having the treatment-covariate summation start from $t=1$ rather than $t=0$. 
}

\na{
Consistent with the rank likelihood formulation, the latent score vector $\mathbf{Z}$ is then restricted to the feasible region $\mathcal{R}(\bY) := \{\bZ\in \Re^N \mid Y_{ij} < Y_{i'j'} \Rightarrow Z_{ij} < Z_{i'j'} \}$,
where now $N=\sum_{i=1}^n n_i$, $\bY=\operatorname{vec}\big(\{ \{ Y_{ij} \}_{j=1}^{n_i} \}_{i=1}^n\big)$ and $\bZ=\operatorname{vec}\big(\{ \{ Z_{ij} \}_{j=1}^{n_i} \}_{i=1}^n\big)$.
By leveraging only information on the ranks, the latent scores $Z_{ij}$ are identifiable only up to monotone transformations, such as linear shifts and rescaling.
To bypass this indeterminacy, we fix $\sigma^2=1$ and enforce $\sum_{r=1}^R \mu_r=0$, both standard in such settings.
}

\subsection{\na{Bayesian Inference via Conjugate Priors}}

\na{
Bayesian inference for the proposed model can be carried out efficiently via Gibbs sampling. To this end, we specify conjugate priors that facilitate straightforward implementation.
We assume independent priors across the model parameters $\rho^2$, $\{\mu_r\}_{r=1}^R$ and $\{\{\{ \alpha_{tgc} \}_{t=1}^T \}_{g=0}^G \}_{c=0}^C $,
and set 
\begin{equation}\label{eq_priors}
\begin{aligned}
    \rho^{2} \sim \mathcal{I}nv\mathcal{G}a(a,b) \qquad 
    \mu_r \sim \mathcal{N}(0, \varphi^2) \qquad
    \alpha_{tgc} \sim \pi_0 \, \delta_0 + (1-\pi_0) \, \mathcal{TN}\big(\lambda,\nu^2, (0,\infty)\big) \; ,
\end{aligned}
\end{equation}
where $\mathcal{I}nv\mathcal{G}a$ and $\mathcal{TN}$ denotes inverse gamma and truncated normal distributions, respectively.
A priori independent and identically distributed replicate-specific intercepts $\mu_r$ reflect the absence of strong prior knowledge on batch effects, while the inverse gamma prior on the within-worm variance $\rho^2$ is a common convenient choice. 
The prior on each treatment coefficient $\alpha_{tgc}$ corresponds to the zero-inflated truncated normal prior proposed by \cite{neelon2004bayesian}, whose structure deserves a closer look.}

\na{
Ensuring monotonicity of the treatment effects amounts to requiring $\mathbb{P}(\alpha_{tgc} < 0)=0$, which is readily satisfied.
Beyond this, we want to allow for flat regions in the dose–response relationship, rather than enforcing strictly increasing effects.
Such plateaus may arise when additional exposure does not lead to further damage, for instance, due to biological saturation or resilience mechanisms.
Mathematically, this coincides with allowing exact zeros $\mathbb{P}(\alpha_{tgc} = 0)>0$, which can also improve robustness and power by avoiding spurious detection of small incremental effects.
The zero-inflated truncated normal prior accommodates all these features while preserving conditional conjugacy.
It can be viewed as a spike-and-slab-type prior \citep{mitchell1988bayesian} with a point mass at zero as the spike and a positive truncated normal distribution as the slab.
}

\na{
We complete the prior specification with calibrated but weakly informative hyperparameter choices.
For the within-worm precision, we set $a = b = 2$, yielding a prior intra-class correlation of $\mathbb{E}[\rho^{2}] / \big(\mathbb{E}[\rho^{2}] + \sigma^2\big) = 2/3$, in line with the strong within-neuron dependence observed in the data.
The replicate intercepts are assigned variance $\varphi^2 = 1$, on the same scale as the residual variance.
For each treatment coefficient $\alpha_{tgc}$, we set $\pi_0 = 0.5$, reflecting equal prior weight on null and non-null effects, and center the truncated normal slab at $\lambda = 0$ with variance $\nu^2 = 1$.
}
\na{Samples of the posterior are then generated via a Gibbs sampler scheme, with full details provided in \ref{sec_gibbs}.
Remarkable computational efficiency is achieved through updates' vectorization, while random effect marginalization and parameter expansion techniques \citep{Murray2013} are employed to improve mixing.}

\section{Power Analysis in Simulation Studies}\label{sec:simulations}

\na{Before analyzing the {\it C. elegans} data, we conduct a simulation study to empirically assess how our model's power compares to different alternatives.}
Recall that, in the frequentist interpretation, power represents the probability of a statistical procedure to reject a null hypothesis when the null is in fact false. 
However, our model is Bayesian, so our analysis aims to define an analogous rejection rule that we hope provides similar behavior to the frequentist decision rule of rejecting the null when the $p$-value falls below a specified significance level.

\na{
For a given rechallenge--generation stratum, our inferential goal of detecting a non-zero monotone dose–response can be formalized as a hypothesis test.
Specifically, we compare the null $H_0: \bigcap_{t=1}^T \big\{ {\beta}_{tgc}=0 \big\}$ to the alternative $H_1: \bigcup_{t=1}^T \big\{ {\beta}_{tgc}>0 \big\} $, where we recall that $\beta_{tgc} = \sum_{s=1}^t \alpha_{sgc}$.
The monotonicity constraints $\alpha_{sgc} \geq 0$ induce the ordering $\beta_{1gc} \leq \beta_{2gc} \leq \cdots \leq \beta_{Tgc}$, so that $\beta_{Tgc} = 0$ is both necessary and sufficient for the event $\bigcap_{t=1}^T \big\{ {\beta}_{tgc}=0 \big\}$.
The hypothesis test thus reduces to
$H_0: \beta_{Tgc} = 0$ versus $H_1: \beta_{Tgc} > 0$,
allowing us to adopt the posterior probability $
\mathbb{P}\big[ \, \textstyle{\sum_{t=1}^T} \alpha_{tgc} = 0 \mid \bY \,\big]$
as test statistic.
}
\na{For ease of exposition, we introduce the acronym \textsc{bayes-rank} to denote our modeling choice.}

\subsection{\na{Alternative Modeling Approaches}}

\na{
We consider a range of alternative methods from both frequentist and Bayesian perspectives to benchmark our approach.
The most natural alternative is an isotonic cumulative regression model as in equation~\eqref{eq_ordinal_reg}, where the linear predictor $\bx_i^\top\bbeta$ is replaced with $\psi_i + \eta_i$ as in equation~\eqref{Z_model_v0}.
Below, we present three alternative strategies for fitting such a model under a logistic link $G(r)$ and performing the corresponding hypothesis test.
}

\na{
\begin{itemize}
    \item[--] (\textsc{bayes-cum-logit})
    A Bayesian isotonic cumulative-logit mixed model fit via the \texttt{R} package \texttt{brms} \citep{burkner2017brms}. Monotonicity in the ordinal predictor is enforced via the \texttt{mo($\,$)} wrapper of \citet{burkner2020mo}, which reparametrizes the cumulative effects as $\beta_{tgc} = T \, \widetilde{\beta}_{gc} \, \textstyle{\sum_{s=1}^t} \zeta_{sgc}$, with overall magnitude $\widetilde{\beta}_{gc} \geq 0$ and normalized increments on the simplex $\sum_{t=1}^T \zeta_{tgc} = 1$ -- separately for each rechallenge--generation stratum. Hypothesis testing then compares $H_0: \widetilde{\beta}_{gc} = 0$ against $H_1: \widetilde{\beta}_{gc} > 0$ using the posterior mean $\mathbb{E}[\widetilde{\beta}_{gc} \mid \bY]$ as test statistic.
    \item[--] (\textsc{iso-cum-logit})
    A frequentist isotonic cumulative-logit mixed model that explicitly enforces $\alpha_{tgc}\geq 0$, with formal hypothesis testing of $H_0: \boldsymbol{\alpha} = \boldsymbol{0}$ against $H_1: \boldsymbol{\alpha} \geq \boldsymbol{0}$ via the likelihood-ratio statistic \citep{agresti2002analysis,davidov2011constrained,davis2011constrained}. Its asymptotic null distribution is a chi-bar-squared $\bar{\chi}^2 = \sum_{d=0}^{T \cdot G \cdot C} \rho_d \, \chi^2_d$, with mixing weights $\{\rho_d\}_d$ entailing orthant probabilities of a multivariate Gaussian and computable via the \texttt{R} package \texttt{ic.infer} from a plug-in estimate of the Fisher information.
    \item[--] (\textsc{trend-cum-logit})
    A frequentist unconstrained cumulative-logit mixed model with $\beta_{tgc}\in\Re$, deferring the positivity analysis to a trend-based test on the contrast $\psi_{gc} = \sum_{t=1}^T \varpi_t \, \beta_{tgc}$ -- for some prespecified weights $\boldsymbol{\varpi}$ \citep{farrar2022comparison}. For a given rechallenge--generation stratum, we test $H_0: \psi_{gc} \leq 0$ against $H_1: \psi_{gc} > 0$.
    The test exploits the asymptotic normality of the estimator of $\psi_{gc}$, being a linear combination of unconstrained maximum-likelihood estimates. Since the choice of $\boldsymbol{\varpi}$ can substantially affect power, we consider different popular alternatives (uniform weights and aggregating low-doses to control) and adjust for multiple testing via a Bonferroni correction.
\end{itemize}
}

\na{
More details on all considered competitors are available in \ref{app:sim_study_details}.
\texttt{brms} provides an interface to Hamiltonian Monte Carlo sampling through \texttt{stan}, where the constraint $\widetilde{\beta}_{gc} \geq 0$ can be enforced via a prior with positive support.
However, such a prior cannot accommodate an exact point mass at zero, since \texttt{stan} does not handle discrete distributions.
This is in contrast with our zero-inflated truncated normal prior, whose spike component can yield improved power. 
Additionally, \texttt{stan} is substantially slower in absolute terms, with runtimes almost 10-times larger than the Gibbs sampler for our model (see Table~\ref{tab:test_desc} in \ref{app:sim_study_details}). 
We also found mixing of \texttt{brms} to be sensitive to prior choices and initialization, with tuning rendered burdensome by the computational cost. 
}

\na{
The two frequentist competitors also pose non-trivial implementation challenges: no off-the-shelf software jointly accommodates ordinal responses, random effects, and order restrictions on regression coefficients, with available packages covering at most two of these three components (see Table~\ref{tab:packages_coverage} in \ref{app:sim_study_details}).
To circumvent this, we accessed internal routines of \texttt{ordinal::clmm} to extract the marginal likelihood of an unconstrained cumulative-logit mixed model -- with random effects integrated out
-- and used \texttt{nlminb} to optimize it under the desired constraints and to obtain its Hessian at specified parameter values.
}

\subsubsection{Standard Frequentist Competitors}

\na{
Given the non-trivial practical challenges entailed by all competitor strategies presented above, we additionally consider five simpler competitors that, while less faithful to the modeling desiderata, are commonly adopted by practitioners in routine analyses of these types of assays.
}

\begin{itemize}
    \item[--] (\textit{Multinomial}) Three nested unconstrained cumulative-logit mixed models on the ordinal damage scores are fit via \texttt{clmm2}: a baseline with no treatment effects in the target stratum, one adding the low-dose effect, and one adding both low- and high-dose effects. Each augmented model is compared to the baseline via a likelihood-ratio $\chi^2$ test using \texttt{anova}, and the resulting $p$-values are combined under a Bonferroni correction.
    \item[--] (\textit{ANOVA}) Nested linear models on the log-damage score, including versus excluding selected treatment interaction terms, are compared via a likelihood-ratio $F$-test using \texttt{anova} and controlling for other covariates.
    Together with $t$-\textit{test}, \textit{ANOVA} is arguably the most common method applied in current practice.
    \item[--] (\textit{Lin.\ Reg.}) A linear regression of the log-damage score on treatment exposure -- treating both response and predictor as continuous variables and controlling for replicate effects -- is fit via \texttt{lm}, with the $p$-value for the treatment coefficient of interest extracted directly from its output.
    \item[--] ($t$-\textit{test}) Within a target rechallenge–generation stratum, log-damage scores are compared pairwise across treatments via \texttt{pairwise.t.test}, with $p$-values adjusted by Bonferroni correction.
    \item[--] ($\chi^2$-\textit{test}) Within the same stratum, a two-way contingency table test of association between log-damage scores and treatment is fit via \texttt{chisq.test}.
\end{itemize}

\begin{figure}[hp!]
\centering
    \includegraphics[width=0.49\linewidth,trim={0 40pt 0 0}, clip]{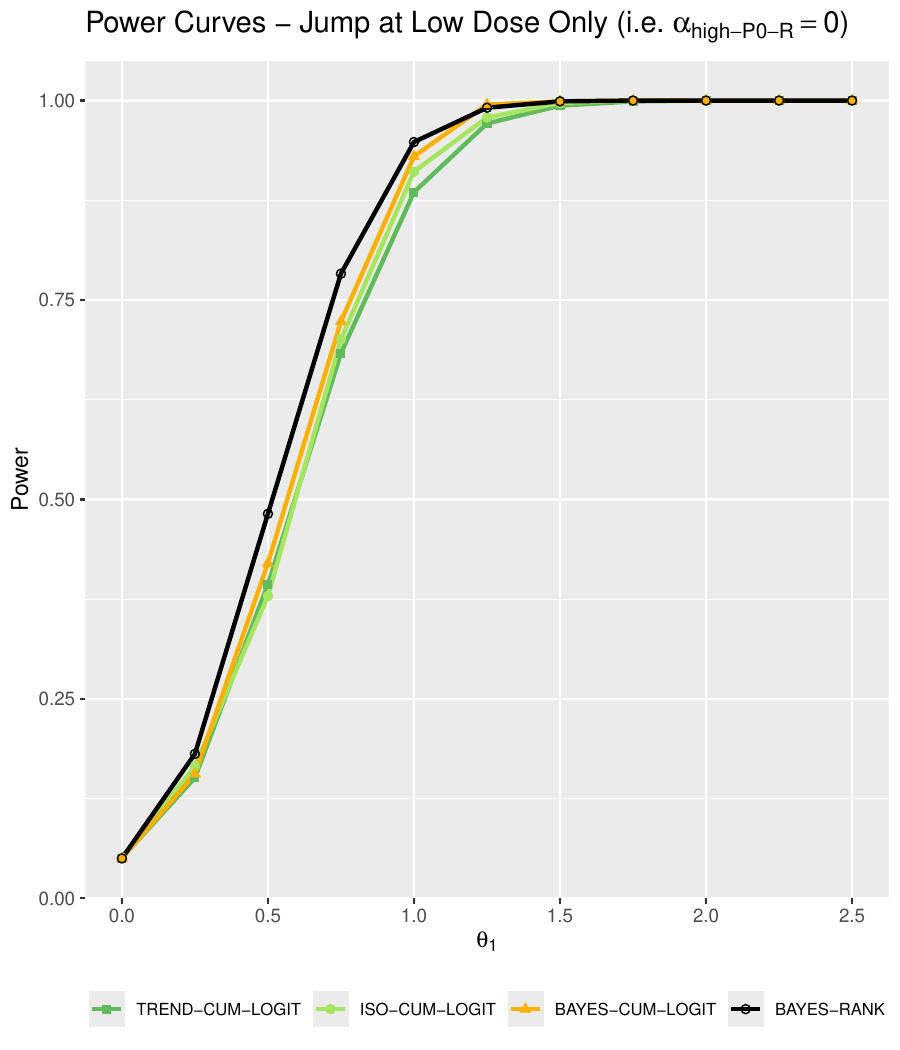}
    \includegraphics[width=0.49\linewidth,trim={0 40pt 0 0}, clip]{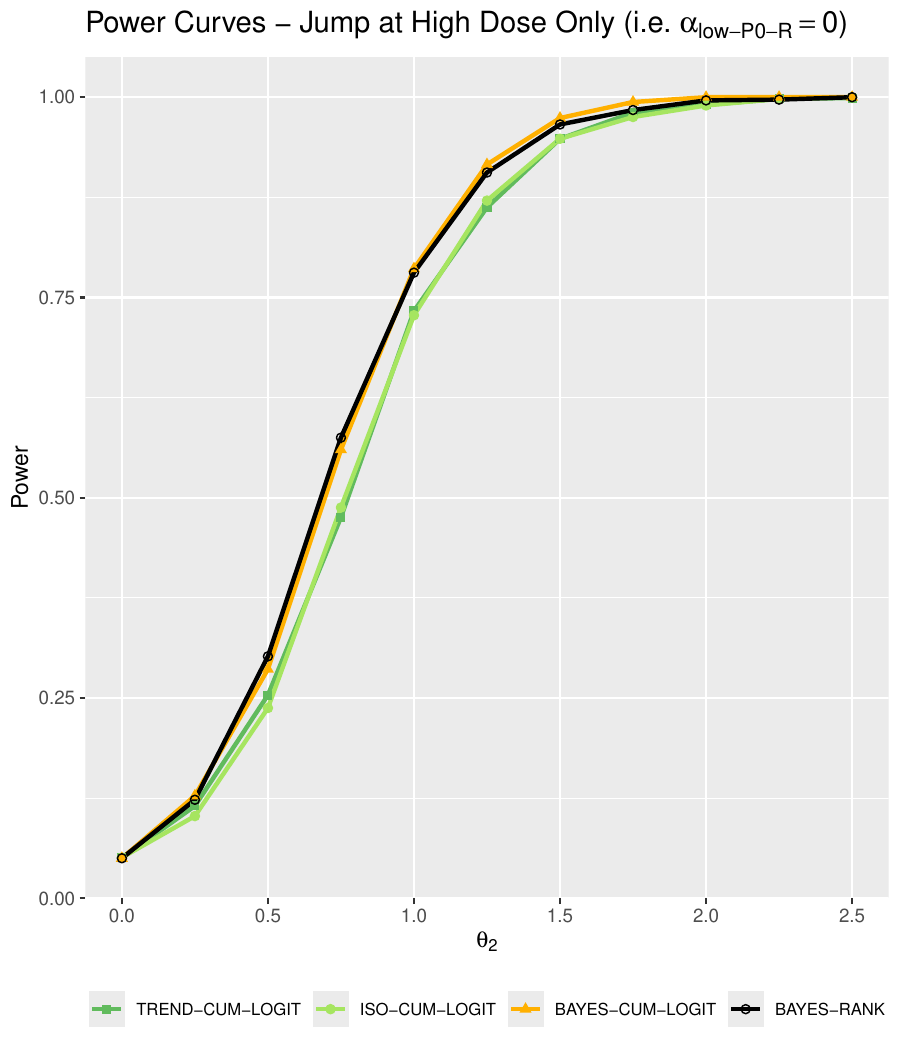}\\
    \includegraphics[width=0.95\linewidth,trim={0 0 0 470pt}, clip]{New_Images/Sim_Power/Power_vs_highDose_New_Competitors.pdf}\\
    \includegraphics[width=0.49\linewidth,trim={0 40pt 0 0}, clip]{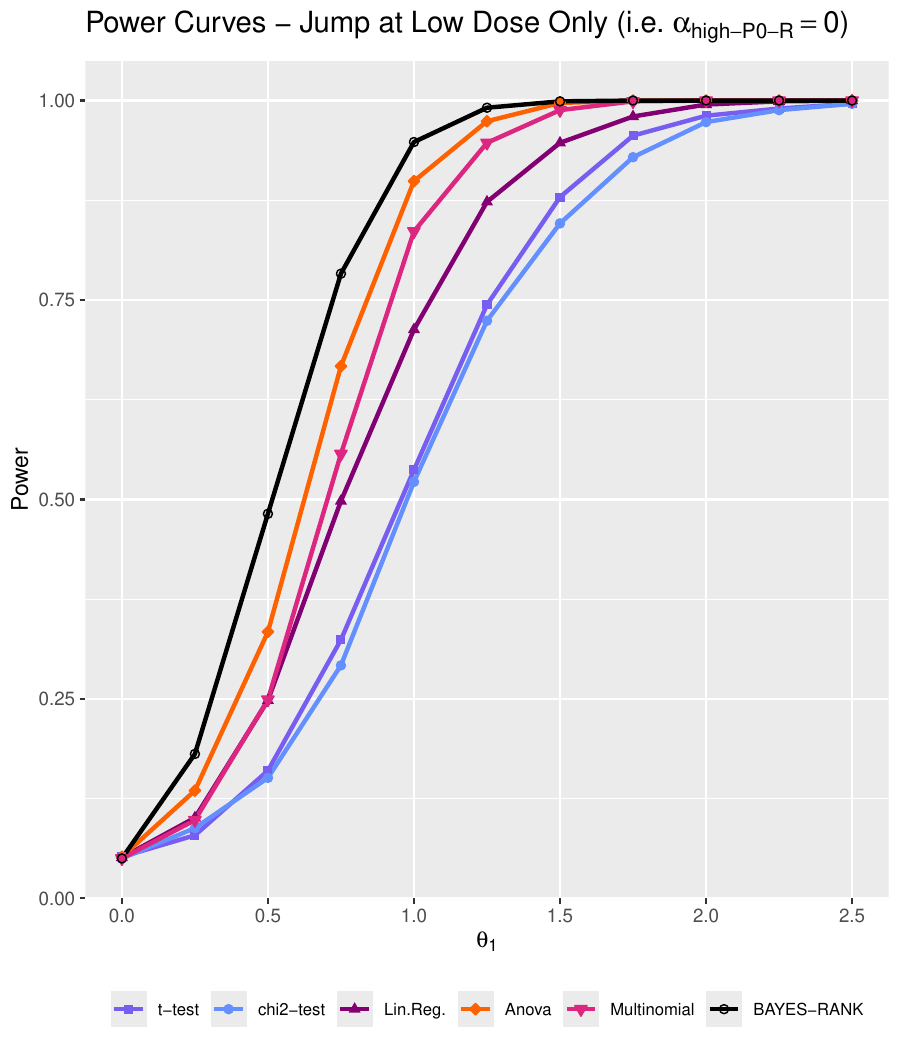}
    \includegraphics[width=0.49\linewidth,trim={0 40pt 0 0}, clip]{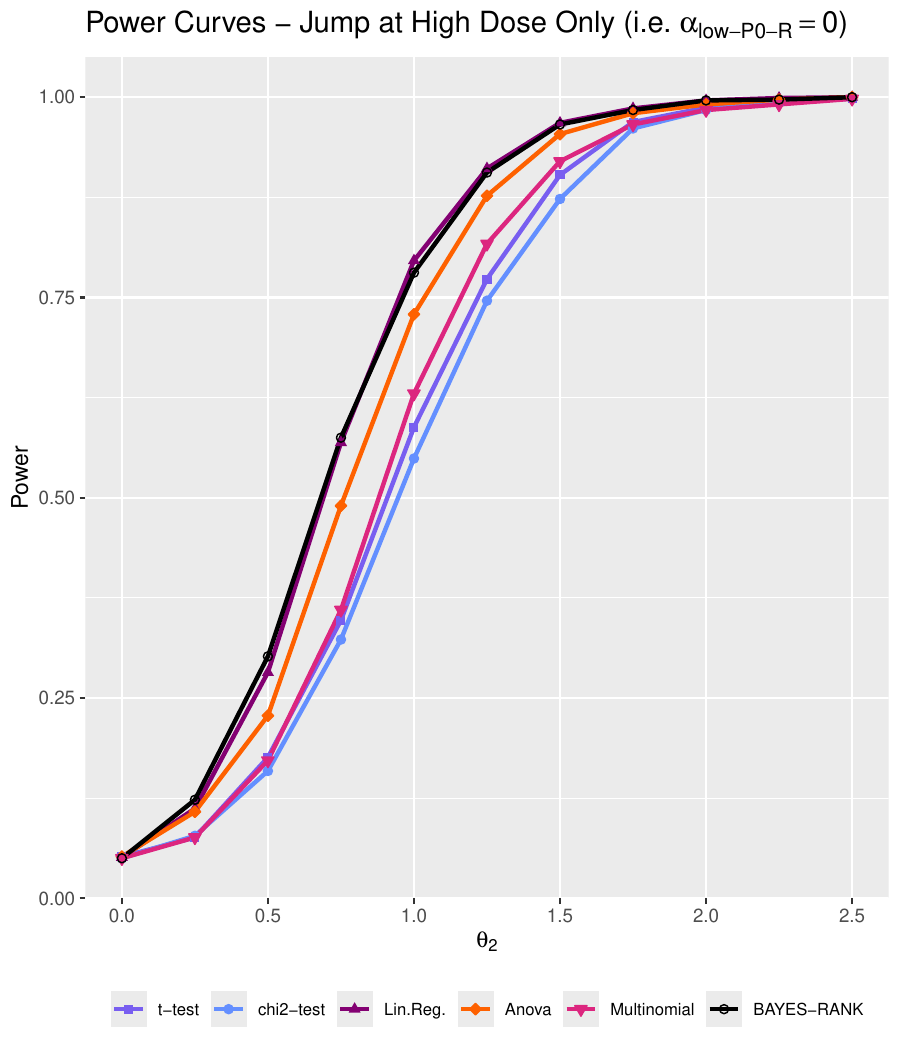} \\
    \includegraphics[width=0.98\linewidth,trim={0 0 0 470pt}, clip]{New_Images/Sim_Power/Power_vs_highDose_Old_Competitors.pdf}
    \vspace{-10pt}
    \caption{Power curves in simulation studies against structured competitors (top) and standard frequentist alternatives (bottom), assuming a step-wise dose-response curve. 
    The left column considers a single jump at the low dose ($\alpha_{\text{low}\text{-}P0\text{-}R} > 0$ and $\alpha_{\text{high}\text{-}P0\text{-}R} = 0$).
    The right column considers a single jump at the high dose ($\alpha_{\text{low}\text{-}P0\text{-}R} = 0$ and $\alpha_{\text{high}\text{-}P0\text{-}R} > 0$).
    All rejection thresholds are calibrated to achieve a nominal Type-I error of 5\%.    }
    \label{fig:power}
\end{figure}

\subsection{Simulation Study Setup}

\na{
Recall that, in the {\it C.\ elegans} assay, we assess rotenone effect across $G=2$ generation groups (parents ``$P0$'' and offspring ``$F1$''), stratified by parental developmental exposure to $T=2$ non-zero drug doses (``low'' dose $0.03\,\mu\mathrm{M}$ and ``high'' dose $0.5\,\mu\mathrm{M}$), and $C=2$ rechallenge groups (un-rechallenged ``$U$'' and rechallenged ``$R$'').
To ease readability, we will henceforth use these mnemonic labels in place of the abstract indices, writing $\alpha_{[\text{low}/\text{high}]\text{-}[P0/F1]\text{-}[U/R]}$ for the treatment coefficients $\alpha_{tgc}$.
To make the power analysis more interpretable, we restrict attention to a single generation--rechallenge stratum, namely the rechallenged parent generation.
The choice of this specific stratum is arbitrary, as the analysis could proceed analogously for any other.
Accordingly, we set to zero all treatment coefficients except $\alpha_{\text{low}\text{-}P0\text{-}R}$ and $\alpha_{\text{high}\text{-}P0\text{-}R}$, and impose the monotonicity constraints only on these in the model fitting.
}
Since the signal is entirely encoded in these two coefficients, we condition each testing procedure on the knowledge that the relevant effects lie exclusively within the rechallenged parent stratum.

\na{
We generated 1,000 synthetic datasets from a cumulative probit model as in equation~\eqref{eq_ordinal_reg}, with hyperparameters $\boldsymbol{\mu}$,
$\boldsymbol{\delta}$,
$\sigma^2$,
and $\rho^2$ calibrated to the original {\it C.\ elegans} assay, and worm counts per stratum drawn from its empirical distribution. All treatment coefficients are set to zero except $\alpha_{\text{low}\text{-}P0\text{-}R}$ and $\alpha_{\text{high}\text{-}P0\text{-}R}$, which are varied one at a time over a grid of eleven equally spaced values in $[0,2.5]$, with the other one kept at zero. Full details are provided in \ref{app:sim_study_details}. 
We run 15,000 Gibbs sampling iterations of \textsc{bayes-rank}, discarding the first 5,000 as burn-in.
Due to its high computational burden, we only generate 6,000 posterior samples from \textsc{bayes-cum-logit}, discarding the first 1,000 as burn-in.
Typical runtimes for all models are reported in Table~\ref{tab:test_desc} in \ref{app:sim_study_details}.
}

\na{Most considered methods are misspecified to varying degrees, making unadjusted power comparisons unreliable.
We therefore calibrate the rejection threshold of each method via simulation under the null, targeting an empirical Type-I error of approximately 5\%.}
In practice, this amounts to computing the rejection rate of each method under \na{$\alpha_{\text{low}\text{-}P0\text{-}R} = \alpha_{\text{high}\text{-}P0\text{-}R} = 0$} and using it to determine the calibrated thresholds (reported in \ref{app:sim_study_details}).
By varying the levels of the treatment effect in the simulations, we can tabulate the proportion of true rejections for the model.

\subsection{Power Assessment}

\na{
The resulting power curves are shown in Figure~\ref{fig:power}, separately for sweeps along $\alpha_{\text{low}\text{-}P0\text{-}R}$ (left column, with $\alpha_{\text{high}\text{-}P0\text{-}R} = 0$) and $\alpha_{\text{high}\text{-}P0\text{-}R}$ (right column, with $\alpha_{\text{low}\text{-}P0\text{-}R} = 0$).
At the origin, where both coefficients are zero, all methods exhibit a rejection rate of approximately $5\%$, confirming that the calibration successfully equalizes Type-I errors across competitors.
As the effect size grows, \textsc{bayes-rank} dominates all alternatives, with the gap widening most clearly in the small-to-moderate effect regime ($\alpha \in [0.5, 1.25]$) where power differences are most informative.
Among the structured competitors (top row), the advantage of \textsc{bayes-rank} over \textsc{bayes-cum-logit} is modest but consistent, while the gap over the two frequentist alternatives (\textsc{iso-cum-logit} and \textsc{trend-cum-logit}) is more pronounced.
}

\na{
Against the standard competitors (bottom row), the power gap is substantially larger, with \textsc{bayes-rank} achieving near-unit power at effect sizes where the simplest baselines ($t$-test and $\chi^2$-test) recover only $30$--$50\%$ of true effects.
This is consistent with the expectation that methods incorporating ordinal scale, random-effects, and monotonicity constraint should outperform those that ignore one or more of these features.
Among the standard baselines, \textit{Multinomial} performs best -- the only competitor that retains both the ordinal nature of the response and the within-worm dependence -- followed by \textit{ANOVA} and \textit{Lin.\ Reg.} -- both collapsing the ordinal scale to a continuous one.
}

\section{Analysis of {\it C. elegans} Neuron Data}\label{sec:analysis}

We now proceed to apply the proposed Bayesian ordinal-isotonic regression model to the {\it C. elegans} assay data.
We run our analysis on \na{20,000} iterations of the Gibbs sampler and then remove the first \na{10,000} samples as burn-in. 
Below, we summarize the findings and insights that our model produces for this data set. 
\na{Further quantitative analysis and model diagnostics can be found in \ref{app:performance}, including goodness-of-fit assessment, posterior traceplots, effective sample size, and autocorrelation functions -- consistently supporting the good performance of the proposed approach.
Reproducibility of our analysis is supported by the \texttt{R} package \texttt{BayesRank} -- openly available at \url{https://github.com/niccoloanceschi/BayesRank} -- including model fitting and visualization of the main results.}

\subsection{Treatment Effects}

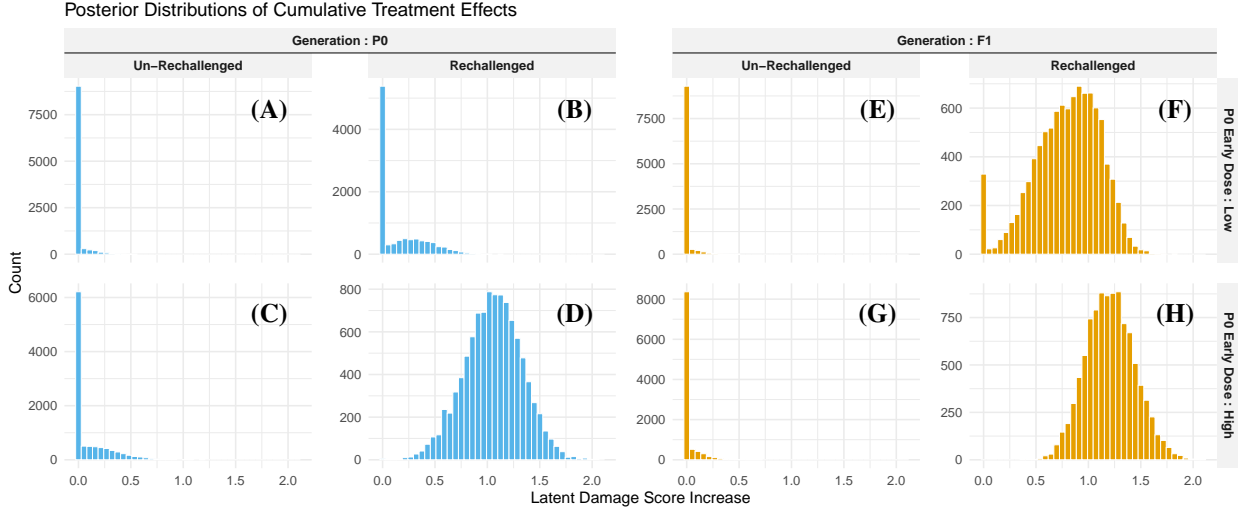
\begin{figure}[ht!]
        \centering
        \includegraphics[width = \linewidth]{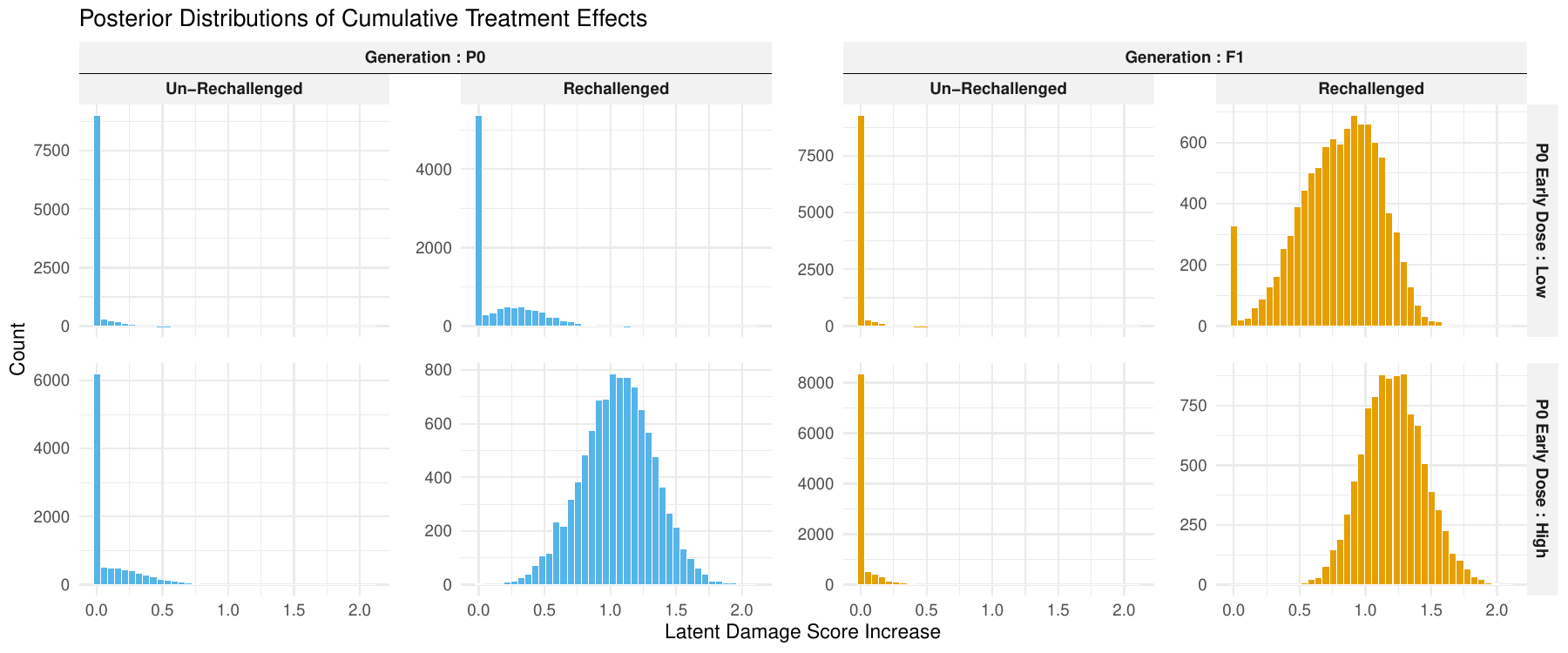}
        \put(-369,152){\makebox(0,0){\textbf{(A)}}}
        \put(-254,152){\makebox(0,0){\textbf{(B)}}}
        \put(-369,75){\makebox(0,0){\textbf{(C)}}}
        \put(-254,75){\makebox(0,0){\textbf{(D)}}}
        \put(-140,152){\makebox(0,0){\textbf{(E)}}}
        \put(-28,152){\makebox(0,0){\textbf{(F)}}}
        \put(-140,75){\makebox(0,0){\textbf{(G)}}}
        \put(-28,75){\makebox(0,0){\textbf{(H)}}}
        \vspace{-20pt}
        \caption{Marginal posterior distributions of cumulative treatment effects (${\beta_{tgc} = \textstyle{\sum_{1 \leq s \leq t}} \alpha_{sgc}}$) for each sub-group of worms, representing the overall shift on the latent damage scale induced by exposure to a particular treatment regimen.
        }
        \label{ref:treatment}
\end{figure}

Figure~\ref{ref:treatment} summarizes the marginal posterior distributions for the cumulative treatment effects in the model, which are our main object of interest.
\na{
As expected, non-rechallenged worms across all subgroups (panels A, C, E, G) show no meaningful damage increase over the baseline, with posterior mass overwhelmingly concentrated at zero -- the corresponding spike probability exceeds $0.5$ for all coefficients, as reported in Table~\ref{tab:rep_icc} in the \ref{app:performance}.
This is consistent with early developmental rotenone exposure alone being insufficient to induce detectable neurodegeneration at the level considered. 
Notably, this holds regardless of early dose level (panels $A\,\text{vs}\,C$, and $E\,\text{vs}\,G$) and regardless of generation (panels $A\,\text{vs}\,E$, and $C\,\text{vs}\,G$), suggesting that parental developmental exposure does not by itself alter the offspring damage profile.
}

\na{
More informative patterns emerge in the rechallenged subgroups (panels $B, D, F, H$). At the high early dose, both parents and offspring show a pronounced response to rechallenge (panels $D$ and $H$), with broadly comparable posterior distributions -- though with a slight upward shift in offspring relative to parents -- indicating that severe early parental exposure predisposes both generations to substantial neurodegeneration upon rechallenge, with only modest additional sensitization effect in the $F1$ generation.
The most striking contrast, however, appears under parental early low exposure.
While rechallenged parents show only a modest response (panel $B$), rechallenged offspring display a substantially more severe damage profile (panel $F$). 
This asymmetry suggests that even mild parental developmental exposure -- too subtle to induce sizeable neurodegeneration in parents themselves -- is sufficient to sensitize offspring neurons, substantially amplifying the neuronal damage elicited by the same later-life exposure level.
}

\begin{figure}[ht!]
\centering
~\vspace{0pt}
\includegraphics[width=\textwidth]{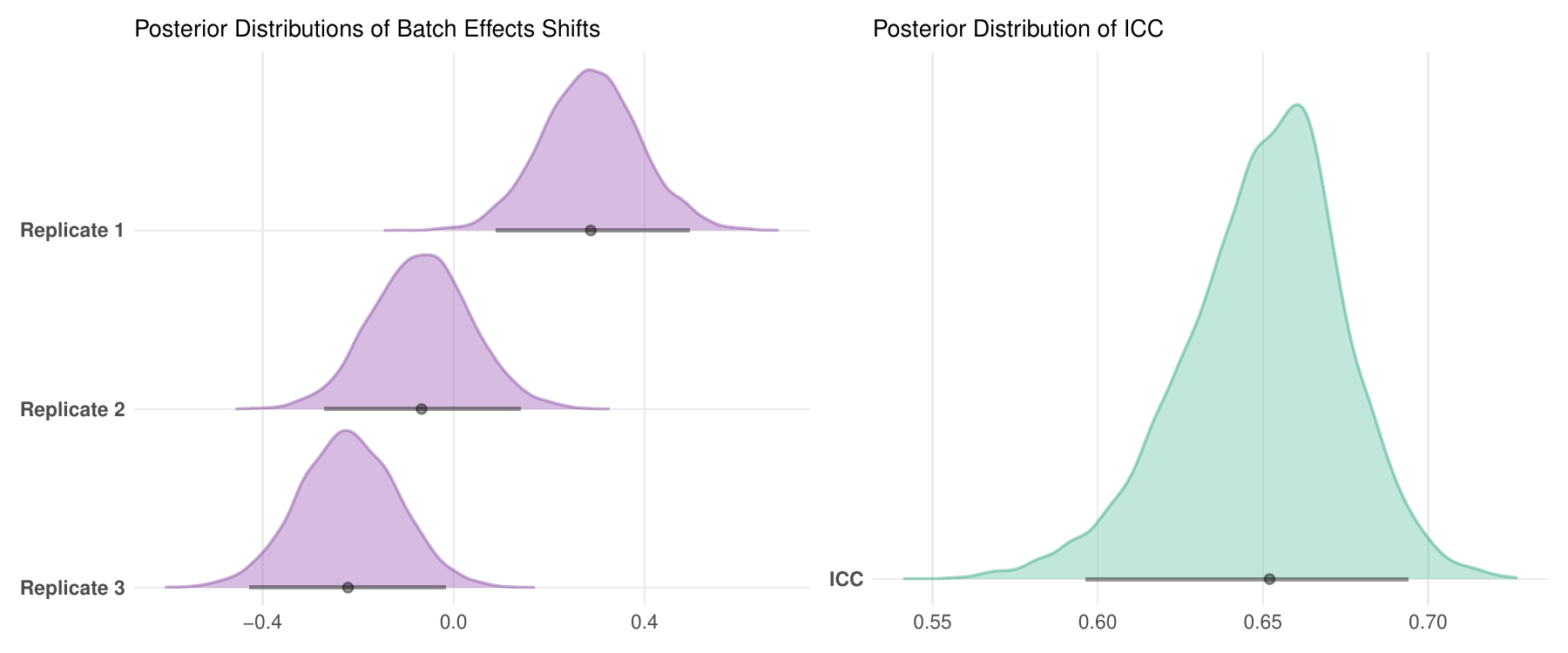}
\put(-455,186){\makebox(0,0){\textbf{(A)}}}
\put(-230,186){\makebox(0,0){\textbf{(B)}}}
\vspace{-20pt}
\caption{(A) Posterior distributions of the intercept $\bmu$, \na{capturing the batch effects for experimental replicates}.
\na{We fit the model enforcing the zero sum constrint $\sum_{r=1}^R \mu_r=0$.}
(B) Posterior distributions of the intra-class correlation coefficient $\rho^2/(\sigma^2 + \rho^2)$, where $\sigma^2=1$, quantifying the proportion of variance explained by the worm effect relative to total variance.}
\label{fig:mcmc_repl_icc}
\end{figure}

\subsection{Replicate Effects and Intra-Class Correlation Coefficient}

The batch effects due to experimental variations between experimental groups are summarized in Figure~\ref{fig:mcmc_repl_icc}. 
Recall that each worm belongs to only one replicate.
\na{
Consistent with the empirical data analysis of Section~\ref{sec:data}, the first replicate shows a shift toward higher damage scores.
This is reflected in a $95\%$ credible interval that does not overlap with that of the third replicate, while the evidence against the second replicate is weaker.
The second and third replicates exhibit more similar batch effects. Importantly, the batch effect allows us to control for experimental variability and isolate the effects of rotenone on neuronal damage more clearly.
}

The intra-class correlation coefficient $\text{ICC} = {\rho^2}/{(\sigma^2 + \rho^2)} \in (0,1)$ measures the proportion of the total variance in the model that is made up of the within-group variance. Recall that we set the within-group variance to be $\sigma^2=1$ to enforce identifiability in our model.
In Figure~\ref{fig:mcmc_repl_icc}, we see that the ICC is concentrated around \na{0.65 (95\% posterior credible interval (0.60, 0.70)),}
which indicates that almost two-thirds of the variance is explained by the heterogeneity due to the worm effect.
This is particularly insightful, as it suggests that there is a strong association of damage experienced by neurons located in the same worm, whereas most existing methods don't account for this dependence.
Note that the presence of such a positive correlation among neurons on a given worm implies that we need a bigger sample size to achieve a desired power \na{than that indicated by typical calculations ignoring a worm effect.}

\begin{figure}[ht!]
\centering
    \includegraphics[width=0.49\textwidth]{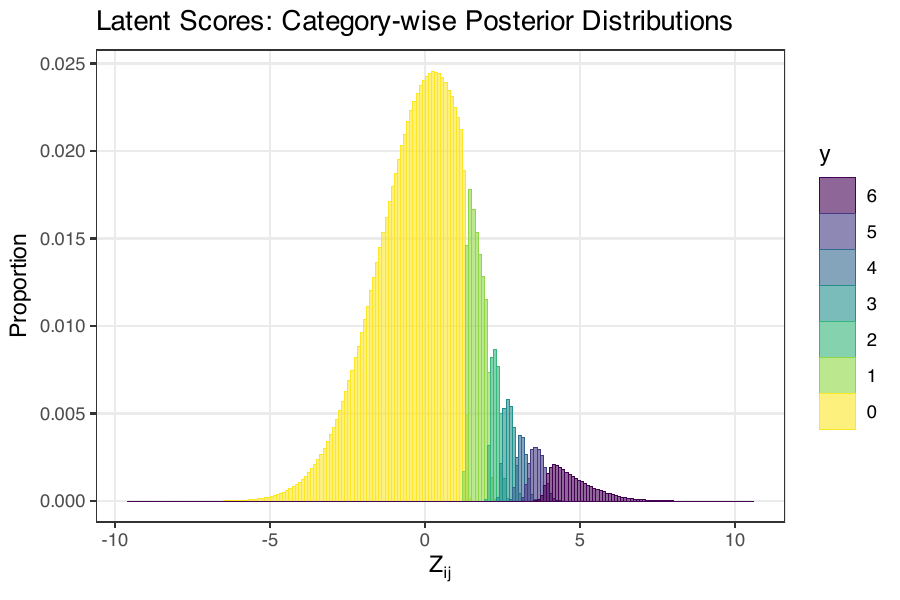}
    \put(-230,148){\makebox(0,0){\textbf{(A)}}}
    \includegraphics[width=0.49\textwidth]{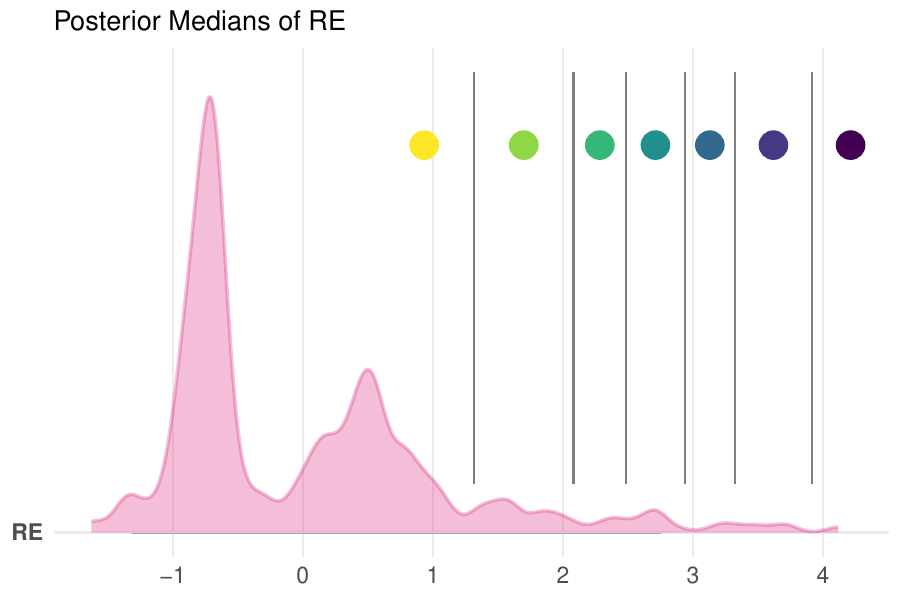}
    \put(-235,148){\makebox(0,0){\textbf{(B)}}}
    \vspace{-10pt}
    \caption{\na{(A) Posterior distribution of the latent variables $Z_{ij}$ for the $N=\text{2,099}$ individual neurons, pooled together and colored by the corresponding damage score. (B) Histogram of the posterior medians of the random effects across the $n=559$ worms. 
    The vertical black lines mark the effective cutoffs separating latent scores associated with different observed scores, with colored dots further identifying the corresponding response-regions in the latent space. 
    Though not explicitly modeled by the rank likelihood, these cutoffs are readily recoverable a posteriori, as detailed in \ref{app:performance}.}}
\label{fig:mcmc_Z_hist_RE}
\end{figure}

\subsection{Latent Variable and Marginal Worm Effects}

\na{Figure~\ref{fig:mcmc_Z_hist_RE} displays posterior summaries for the latent damage scores $Z_{ij}$ across all $N = \text{2,099}$ neurons and the $n = 559$ worm-level random effects $\eta_i$.
In the left panel, we plot the posterior distribution of the latent variables $Z_{ij}$, colored by the corresponding damage score.}
We can clearly appreciate the power of the rank likelihood approach for inference.
Even though we do not explicitly model the cutpoints \na{$\delta_\ell$}
in our model, samples respect the constraints that each latent variable will not intersect with the samples from a different damage score than its own \na{-- strictly within each sampling iteration, with boundaries that may appear slightly blurred when averaging across iterations.} 

\na{
In the right panel, we plot a histogram of the posterior medians of the worm-specific random effects.
The distribution is highly asymmetric, with roughly three components: a dominant mode at low latent scores, a smaller secondary mode at intermediate latent scores, and a long right tail at higher latent scores.
To provide further insight into these subgroups, we juxtapose the thresholds separating latent scores associated with different observed damage scores, implicitly inferred by the rank likelihood.
These are estimated as the posterior means of the corresponding category boundaries.
The first peak suggests that many worms are robust to damage, as even the largest treatment effect quantified in Figure~\ref{ref:treatment} would not be sufficient to shift the average latent scores of these worms into the damage-$1$ region.
The second mode corresponds to intermediate worms, still typically showing no damage but close enough to the boundary that high exposure could push them into higher observed categories.}
The right tail suggests that there are a few worms that are particularly susceptible, and so they experience a strong effect on the margin (i.e., they get damaged a lot more than the other worms).

\section{Discussion and Extensions}\label{sec:discussion}

\na{In this paper, we analyzed a {\it C. elegans} neurotoxicity assay to study how rotenone exposure induces neuronal damage and, in particular, whether maternal exposure sensitizes offspring to damage at lower dose levels. 
We proposed a Bayesian ordinal regression model that combines the rank likelihood framework with a novel coding of treatment effects that generalizes to any number of ordinal dose levels and captures within-worm dependence through worm-level random effects. The resulting model admits an efficient Gibbs sampler, and our simulation study demonstrates that it achieves higher statistical power than both structurally faithful and simpler frequentist and Bayesian competitors -- often relying on ad-hoc tests whose assumptions are violated by the data and fail to capture all of the critical inferential targets our model does. 
}

\na{Applied to the {\it C. elegans} assay, our model provides evidence that maternal rotenone exposure does not induce neurodegeneration directly in offspring, but sensitizes them to subsequent toxic insult: the offspring generation exhibits substantially amplified neuronal damage upon later-life rotenone exposure, with this sensitization effect detectable even under mild parental developmental treatment.
Beyond highlighting a previously overlooked strong within-worm dependence in damage scores, the worm-level random effects additionally uncover substantial heterogeneity in neuronal susceptibility, ranging from two groups with marked and mild resistance to severe damage to a small subset of particularly vulnerable worms.
}

\na{
Our modeling framework can be readily extended to other popular neurotoxicity scoring systems that combine quantitative and qualitative damage information.
Examples include non-functional neuron counts \citep{berkowitz2008application, tucci2011modeling}, break-per-neuron counts \citep{gonzalez2014exposure}, damage proportions \citep{hartman2019genetic}, or continuous fluorescence area \citep{luo2019age}. These naturally lend themselves to an ordinal interpretation, and can seamlessly be integrated into our Bayesian ordinal regression. 
In parallel, our framework naturally extends to the analysis of different {\it C. elegans} neurotoxicity assays, such as the multi-neuron-type DNT paradigm of \citet{huayta2025assessment}, targeting dopaminergic, glutamatergic, and cholinergic neurodegeneration induced by lead, cadmium, and benzo(a)pyrene.
}

\section*{Acknowledgements}

This project has received funding from the United States National Institutes of Health (R01Al167850, R35ES035049, and P42ES010356).
The authors thank David B. Dunson and Christina M. Bergemann for their helpful feedback and comments.



\putbib
\end{bibunit}

\begin{bibunit}


\newpage

~\\

\begin{center}
{\sffamily\bfseries\Large{Supplementary Materials for \\[3pt]
``Order-Restricted Bayesian Ordinal Regression for \\[3pt]
the Modeling of Neuron Degeneration in \textit{C. elegans}''}}
\end{center}

\appendix
\renewcommand{\thesection}{Appendix \Alph{section}}
\renewcommand{\thesubsection}{\Alph{section}.\arabic{subsection}}
\renewcommand{\theequation}{\Alph{section}.\arabic{equation}}
\renewcommand{\thefigure}{\Alph{section}.\arabic{figure}}
\setcounter{section}{0}
\setcounter{subsection}{0}
\setcounter{equation}{0}
\setcounter{figure}{0}

\section{Gibbs Sampling Schemes}\label{sec_gibbs}

\na{
Posterior inference for the proposed isotonic rank-based regression model can be performed via a Gibbs sampling scheme, appealing for its computational efficiency.
In this section, we provide the associated implementation details, starting with a high-level description, update schemes, and pseudocodes.
The derivations for the corresponding full conditionals are deferred to \ref{sec:derivations}, while parameter-expansion techniques used to augment selected updates are presented in \ref{sec:px}.
}

\na{
Gibbs sampling can proceed by leveraging two alternative representations of the model -- in principle equivalent, but potentially leading to different Marko Chain Monte Carlo (\textsc{mcmc}) mixing in specific settings: a regular sampler based on the standard formulation of the likelihood as in equation~\eqref{Z_model_v0}, and a collapsed Gibbs sampler \citep{Park_vanDyk_2009_Collapsed}, which exploits the marginalization of the random effects $\eta_i$. 
In fact, integrating out $\eta_i$ in equation~\eqref{Z_model_v0} leads to
\begin{equation}\label{Z_model_marg}
\begin{aligned}
	(Z_{i\,1}, \dots, Z_{i\, n_i})^\top
    &\sim \, \mathcal{N}_{n_i} \Big(\, \psi_i \, \1_{n_i} \, , \, \sigma^2 \, \mathbb{I}_{n_i} + \rho^2 \, \1_{n_i}\1_{n_i}^\top \,\Big) \;.
\end{aligned}
\end{equation}
The resulting marginalized likelihood thus explicitly encodes within-worm dependence through the off-diagonal terms of the covariance.
This mirrors the marginal-likelihood approach used in frequentist mixed-effects models, and serves an inferential purpose also in a Bayesian setting.
In fact, \textsc{mcmc} mixing under the rank likelihood can sometimes be hindered by the latent nature of the model.
The setting of the {\it C. elegans} assay is particularly challenging, since each worm contributes only a small number of neurons (typically $4$), leading to weak identification of the random effects $\eta_i$.
By contrast, equation~\eqref{Z_model_marg} can be exploited within a collapsed Gibbs sampling scheme that removes $\eta_i$ from the conditioning chain, breaking the strong posterior dependence that otherwise hinders \textsc{mcmc} mixing -- as illustrated in Figure~\ref{fig:acf_ess_px_marg}
from \ref{app:performance}.}

\subsection{Vectorized Formulation and Sampling Schemes}

\na{
To simplify the exposition of both regular and collapsed Gibbs sampling schemes, let us now introduce a vectorized notation as follows.
Recall that $\bY=\operatorname{vec}\big(\{ \{ Y_{ij} \}_{j=1}^{n_i} \}_{i=1}^n\big)$,  $\bZ=\operatorname{vec}\big(\{ \{ Z_{ij} \}_{j=1}^{n_i} \}_{i=1}^n\big)$, and $\mathcal{R}(\bY) := \{\bZ\in \Re^{\sum_{i=1}^n n_i} \mid Y_{ij} < Y_{i'j'} \Rightarrow Z_{ij} < Z_{i'j'} \}$.
Let $P=T\cdot(G+1)\cdot(C+1)$ and $M=\max_{1\leq i \leq n} n_i$. 
We introduce design matrices $\bW^{(0)} \in \{0,1\}^{N \times R}$, $\bW \in \{0,1\}^{N \times (R-1)}$, $\bQ \in \{0,1\}^{N \times n}$, and $\bX \in \{0,1\}^{N \times P}$ defined row-wise.
To lighten the notation, we use the double-index $(i,j)$ to refer to the $\big(j + \sum_{i' < i} n_{i'}\big)$-th row of each matrix, corresponding to neuron $j$ of worm $i$.
With this convention, we have
\begin{equation*}
\begin{aligned}
\bW^{(0)}_{(ij),r} &= \mathbbm{1}_{\big(\mathrm{rep}(i)=r \big)} \\
\bQ_{(ij),i'} &= \mathbbm{1}_{\big(i=i'\big)}
\end{aligned}
\qquad\quad
\begin{aligned}
\bW &= \bW^{(0)}_{\bullet,-R} - \bW^{(0)}_{\bullet,R} \otimes \mathbf{1}_{R-1} \\
\bX_{(ij),(tgc)} &= \mathbbm{1}_{\big(\mathrm{treat}(i) \geq t \big)} 
\mathbbm{1}_{\big(\mathrm{gen}(i) = g \big)}
\mathbbm{1}_{\big(\mathrm{rech}(i) = c \big)} 
\end{aligned}
\end{equation*}
The use of $\bW$ in place of $\bW^{(0)}$ readily enforces the zero-sum constraint on the replicate intercepts by effectively fixing $\mu_R = - \sum_{r=1}^{R-1}\mu_r$.
Let us also introduce the vectors $\bmu = (\mu_1,\dots,\mu_{R-1})^\top$, $\balpha=\operatorname{vec}\big(\{\{\{ \alpha_{tgc} \}_{t=1}^T \}_{g=0}^G \}_{c=0}^C \big)$, and $\boeta = (\eta_1,\dots,\eta_n)^\top$.
}

\na{
Equipped with the vectorized notation, the rank-likelihood can thus be rewritten as
\begin{equation}\label{full_lik}
    p(\bZ \mid \bY, \bmu, \balpha, \boeta, \rho^2 ) \propto \phi_N\big(\, \bZ - (\bW \bmu + \bX\balpha + \bQ \boeta ) \, ,\, \sigma^2 \, \mathbb{I}_N \, \big) \cdot \mathbbm{1}_{\big(\bZ \in \mathcal{R}(\bY)\big)} \;, 
\end{equation}
whereas marginalizing out the random effects $\boeta$ leads to
\begin{equation}\label{full_lik_marg}
    p(\bZ \mid \bY, \bmu, \balpha, \rho^2 ) \propto \phi_N\big(\, \bZ - (\bW \bmu + \bX\balpha ) \, ,\, \sigma^2 \, \mathbb{I}_N + \rho^2 \bQ \bQ^\top\, \big) \cdot \mathbbm{1}_{\big(\bZ \in \mathcal{R}(\bY) \big)} \; .
\end{equation}
Note that $\bQ \bQ^\top$ is a block-diagonal matrices with blocks $\{ \1^\top \1_{n_i}^\top\}_{i=1}^n$. Conversely, $ \bQ^\top\bQ = \text{diag}\big(\{ n_i \}_{i=1}^n \big)$.}

\setcounter{savedfigure}{\value{figure}}
\renewcommand{\figurename}{Table}
\addtocounter{table}{0}
\setcounter{figure}{\value{table}}
\begin{figure}[ht!]
\centering
\begin{tabular}{l|l|l|}
  & \quad Regular Gibbs Sampler \;~ & \quad Collapsed Gibbs Sampler \;~ \\
 \hline
1. \; Latent Scores \quad & \quad  $p(\bZ \mid \bY, \bmu, \balpha, \boeta, \textcolor{lightgray}{\rho^2} )$ \quad\quad
&  \quad $p(\bZ \mid \bY, \bmu, \balpha, \rho^2 )$ \\
2. \; Coefficients \quad & \quad  $p(\balpha \mid \textcolor{lightgray}{\bY}, \bZ, \bmu, \boeta, \textcolor{lightgray}{\rho^2} )$ \quad\quad
&  \quad $p(\balpha \mid \textcolor{lightgray}{\bY},\bZ, \bmu, \rho^2 )$ \\
3. \; Intercepts \quad & \quad  $p(\bmu \mid \textcolor{lightgray}{\bY},\bZ , \balpha, \boeta, \textcolor{lightgray}{\rho^2} )$ \quad\quad 
&  \quad $p(\bmu \mid \textcolor{lightgray}{\bY},\bZ, \balpha, \rho^2 )$ \\
4. \; RE variance \quad & \quad  $p(\rho^2 \mid \textcolor{lightgray}{\bY},\textcolor{lightgray}{\bZ, \bmu, \balpha,}\, \boeta )$ \quad\quad
& \quad $p(\rho^2 \mid \textcolor{lightgray}{\bY}, \bZ, \bmu, \balpha )$ \\
5. \; Random Effects \quad & \quad $p( \boeta \mid \textcolor{lightgray}{\bY}, \bZ, \bmu, \balpha, \rho^2 ) $\quad\quad
&  \quad $p( \boeta \mid \textcolor{lightgray}{\bY}, \bZ, \bmu, \balpha, \rho^2 )$ \\
\end{tabular}
\caption{\na{Regular and collapsed Gibbs sampling schemes for the proposed model. The light-gray coloring highlights variables that naturally get dropped from the conditioning sets, due to conditional independence in the corresponding model representation.}}
\label{tab_gibbs_schemes}
\end{figure}
\renewcommand{\figurename}{Fig.}
\setcounter{figure}{\value{savedfigure}}

\na{
Samples from the resulting posterior distributions can be generated by targeting the sequential updates in either scheme of Table~\ref{tab_gibbs_schemes}, whose steps are unpacked in Algorithm~\ref{alg:regular} and \ref{alg:collapsed} below.
Note that samples of the random effects $\boeta$ can be recovered even under the collapsed sampler, by drawing from their standard full conditional distribution in step~5.
We obtain competitive computational performances by exploiting conditional independence to parallelize the update of the latent scores in step~1.
This strategy leads to competitive runtimes, almost 10-fold smaller than a \texttt{stan}-based Bayesian competitor -- as documented in Table~\ref{tab:test_desc} from \ref{app:sim_study_details}.
Despite substantial improvements brought by the random effect marginalization, \textsc{mcmc} mixing can still be suboptimal -- as illustrated in Figure~\ref{fig:acf_ess_px_marg} from \ref{app:performance}.
We therefore augment the updates of $\boeta$, 
$\balpha$, 
and $\rho^2$ via parameter expansion (\textsc{px}) techniques \citep{Murray2013}, as detailed in \ref{sec:px}, which proved sufficient to achieve good mixing across all model parameters.
}

\subsection{Regular Gibbs Sampler: Pseudocode}

\na{
Algorithm~\ref{alg:regular}
provides the pseudocode to draw a sample from the posterior distribution of the model parameters using the regular Gibbs sampler.
For ease of notation, we define the score-based index sets $\mathcal{G}(\ell) = \big\{ (i,j) \mid Y_{ij} = \ell \big\}$, $\forall \, \ell \in \{0,\dots,L\}$, and the worm-based ones $\mathcal{J}(i) = \big\{ (i,j) \mid \bQ_{(ij),i}=1 \big\}$, $\forall \, i \in \{1,\dots,n\}$.
We thus use the notation $\bZ_{\mathcal{G}(\ell)}=\operatorname{vec}\big( \{ Z_{ij} \}_{(i,j) \in \mathcal{G}(\ell)} \big)$ for sub-vectors,
and introduce the convention $\max \bZ_{\mathcal{G}(-1)} = -\infty$ and $\min \bZ_{\mathcal{G}(L+1)} = \infty$.
Following \cite{neelon2004bayesian}, let us also introduce a dedicated notation for the zero-inflated truncated normal distribution $\alpha \sim \mathcal{ZI}\text{--}\mathcal{N}^+ \big({\pi}, {\lambda}, {\nu}^2 \big)$, as in the prior from equation~\ref{eq_priors}.}

\vspace{15pt}

\na{
\hrule \vspace{3pt}
\refstepcounter{algorithm}
Algorithm~\thealgorithm\phantomsection\label{alg:regular}: Regular Gibbs Sampler -- Single \textsc{mcmc} Iteration \vspace{3pt} 
\hrule 
\begin{enumerate}
\item[1.] For $\ell=0, \dots, L$: \vspace{-1pt}
\begin{itemize}
    \item[] Compute\; $l_h = \max \bZ_{\mathcal{G}(\ell-1)}$ and $u_h = \min \bZ_{\mathcal{G}(\ell+1)}$\\[2pt]
    For $(i,j) \in \mathcal{G}(\ell)$: \; Sample\; $\bZ_{ij} \sim \mathcal{TN}\big( (\bW \bmu + \bX\balpha + \bQ \boeta)_{ij} \,, \, \sigma^2 \,; \, (l_h,u_h) \big)$
\end{itemize}
\item[2.] For $t=1, \dots, T$, $\; g=0, \dots, G$, $\; c=0, \dots, C$ : \vspace{-1pt}
\begin{itemize}
\item[] Sample\; $\alpha_{tgc} \sim \mathcal{ZI}\text{--}\mathcal{N}^+ \big( \widetilde{\pi}_{tgc}, \widetilde{\lambda}_{tgc}, \widetilde{\nu}_{tgc}^2 \big)$ \;with\; 
\item[] \hspace{20pt} $\widetilde{\nu}_{tgc}^2 = \big({\nu}^{-2} + \sigma^{-2} \bX_{\bullet, (tgc)}^\top\bX_{\bullet, (tgc)} \big)^{-1}$
\item[] \hspace{20pt} $\widetilde{\lambda}_{tgc} =  \widetilde{\nu}_{tgc}^2 \big({\nu}^{-2}\lambda + \sigma^{-2} \bX_{\bullet, (tgc)}^\top (\bZ - \bW \bmu - \bQ \boeta - \bX_{\bullet, -(tgc)}\balpha_{-(tgc)}) \big)$
\item[] \hspace{20pt} $\widetilde{\pi}_{tgc} = \displaystyle{\frac{
\pi \, \phi(\lambda,\nu^2)\,/\,\Phi(\lambda/\nu)
}{
\pi \, \phi(\lambda,\nu^2)\,/\,\Phi(\lambda/\nu)
+ (1-\pi) \,
\phi\big(\widetilde{\lambda}_{tgc},\widetilde{\nu}_{tgc}^2\big)\,/\,\Phi\big(\widetilde{\lambda}_{tgc}/\widetilde{\nu}_{tgc}\big)
}}$
\end{itemize}
\item[3.] Sample\; $ \bmu \sim \mathcal{N}_{R-1} \big( \, \sigma^{-2} \, \mathbf{\Xi} \,\bW^\top(\bZ - \bX\balpha - \bQ\boeta) , \mathbf{\Xi} \big)$ \;with\; 
\vspace{2pt}
\item[] \hspace{20pt} $\mathbf{\Xi} = ( \varphi^{-2}\bI_{R-1} + \sigma^{-2} \bW^\top \bW)^{-1}$ \vspace{5pt}
\item[4.] Sample\; $\rho^{2} \sim \mathcal{I}nv\mathcal{G}a \big(a + \frac{n}{2}, b + \frac{1}{2}\boeta^\top \boeta\big)$ \vspace{2pt}
\item[5.] For $i = 1, \dots, n$: \; Sample\; $\eta_i \sim \mathcal{N}\big( \rho_i^{2} \, \bQ_{\bullet,i}^\top (\bZ-\bX\balpha - \bW \bmu ) , \rho_i^{2}  \big)$ \;with\; 
\vspace{2pt}
\item[] \hspace{20pt} $\rho_i^{2} = (\rho^{-2} + n_i \, \sigma^{-2})^{-1} $
\end{enumerate}
\hrule 
}

\vspace{20pt}

\na{
In step 1 of Algorithm~\ref{alg:regular}, the shared truncation extrema within each iteration of the loop over response values arise from conditional independence after conditioning on the latent scores associated with all remaining response groups. 
Consequently, the resulting blockwise updates admit a parallel implementation via vectorized sampling from univariate truncated normal distributions.
This is particularly important because the update of the latent score $\bZ$ is the computational bottleneck in large-$n$-small-$p$ scenarios, as in the \textit{C. elegans} assay.\\
A similar vectorization underlies the rank-likelihood implementation of the \texttt{R} package \texttt{mtlm} \citep{hoff2026mtlm}, which does not support random effects, parameter constraints, collapsed sampling, or parameter expansion.
While \texttt{mtlm} relies on global ordering constraints, explicitly defining the neighborhood index sets $\mathcal{J}(\ell)$ becomes essential in the more elaborate collapsed Gibbs sampler, where the full conditional distributions involve multivariate truncated normals.
Further details are provided in \ref{sec:derivations}.
}

\subsection{Collapsed Gibbs Sampler: Pseudocode}

\na{Algorithms~\ref{alg:collapsed} provides the pseudocode to draw a sample from the posterior distribution of the model parameters using the collapsed Gibbs sampler.} 

\vspace{15pt}
\na{
\hrule \vspace{3pt}
\refstepcounter{algorithm}
Algorithm~\thealgorithm\phantomsection\label{alg:collapsed}: Collapsed Gibbs Sampler -- Single \textsc{mcmc} Iteration
 \vspace{3pt}
\hrule 
\begin{enumerate}
\item[1.] For $\ell=0, \dots, L$: \vspace{-1pt}
\begin{itemize}
    \item[] Compute\; $l_h = \max \bZ_{\mathcal{G}(\ell-1)}$ and $u_h = \min \bZ_{\mathcal{G}(\ell+1)}$\\[2pt]
    For $i = 1,\dots,n$: \; Define \; $\mathcal{L}(i,\ell) = \mathcal{J}(i) \cap \mathcal{G}(\ell)$
    \; and \; $n_{i\ell}= \vert \mathcal{L}(i,\ell) \vert$
    \item[] \hspace{20pt} Sample\; $\bZ_{\mathcal{L}(i,\ell)} \sim \mathcal{TN}_{n_{i\ell}}\big( (\bW \bmu + \bX\balpha + \bQ \boeta)_{\mathcal{L}(i,\ell) } +  \1_{n_{i\ell}} \xi_{i\ell}  \,, \, \boldsymbol{\Sigma}_{i\ell} \,; \, (l_h,u_h)^{n_{i\ell}} \big)$
    \item[] \hspace{20pt} with\; $\boldsymbol{\Sigma}_{i\ell} = \sigma^{2} \mathbb{I}_{n_{i\ell}} + \displaystyle{ \frac{\sigma^2 \,\rho^2}{\sigma^2+\rho^2 (n_i-n_{i\ell})}} \1_{n_{i\ell}}\1_{n_{i\ell}}^\top$ \, and 
    \item[] \hspace{20pt} $\xi_{i\ell}=\displaystyle{\frac{\rho^2}{\sigma^2+\rho^2  (n_i-n_{i\ell})}} \1_{n_i-n_{i\ell}}^\top \big( \bZ_{\mathcal{J}(i) \setminus \mathcal{L}(i,\ell)} - (\bW \bmu + \bX\balpha + \bQ \boeta)_{\mathcal{J}(i) \setminus \mathcal{L}(i,\ell)} \big)$
\end{itemize}
\item[2.] For $t=1, \dots, T$, $\; g=0, \dots, G$, $\; c=0, \dots, C$ : \vspace{-1pt}
\begin{itemize}
\item[] Sample\; $\alpha_{tgc} \sim \mathcal{ZI}\text{--}\mathcal{N}^+ \big( \widetilde{\pi}_{tgc}, \widetilde{\lambda}_{tgc}, \widetilde{\nu}_{tgc}^2 \big)$ \;with\; 
\item[] \hspace{20pt} $\widetilde{\nu}_{tgc}^2 = \big({\nu}^{-2} + \bX_{\bullet, (tgc)}^\top (\sigma^{2} \mathbb{I}_N + \rho^2 \bQ\bQ^\top )^{-1} \bX_{\bullet, (tgc)} \big)^{-1}$ 
\item[] \hspace{20pt} $\widetilde{\lambda}_{tgc} =  \widetilde{\nu}_{tgc}^2 \big({\nu}^{-2}\lambda + \bX_{\bullet, (tgc)}^\top (\sigma^{2} \mathbb{I}_N + \rho^2 \bQ\bQ^\top )^{-1} (\bZ - \bW \bmu - \bX_{\bullet, -(tgc)}\balpha_{-(tgc)}) \big)$
\item[] \hspace{20pt} $\widetilde{\pi}_{tgc} = \displaystyle{\frac{
\pi \, \phi(\lambda,\nu^2)\,/\,\Phi(\lambda/\nu)
}{
\pi \, \phi(\lambda,\nu^2)\,/\,\Phi(\lambda/\nu)
+ (1-\pi) \,
\phi\big(\widetilde{\lambda}_{tgc},\widetilde{\nu}_{tgc}^2\big)\,/\,\Phi\big(\widetilde{\lambda}_{tgc}/\widetilde{\nu}_{tgc}\big)
}}$
\end{itemize}
\item[3.] Sample\; $ \bmu \sim \mathcal{N}_{R-1} \big( \,  \mathbf{\Xi} \,\bW^\top (\sigma^{2} \mathbb{I}_N + \rho^2 \bQ\bQ^\top )^{-1} (\bZ - \bX\balpha) , \mathbf{\Xi} \big)$ \;with\; 
\vspace{2pt}
\item[] \hspace{20pt} $\mathbf{\Xi} = \big( \varphi^{-2}\bI_{R-1} + \bW^\top (\sigma^{2} \mathbb{I}_N + \rho^2 \bQ\bQ^\top )^{-1} \bW \big)^{-1}$
\item[4.] Sample\; $\rho^{2} \sim p(\rho^2 \mid \bZ, \bmu, \balpha ) \propto (\rho^2)^{-a-1} \, \exp\big( -b/\rho^2 - f(\rho^2) \big) \, \displaystyle{\prod_{i=1}^n} (\sigma^2+\rho^2  n_i)^{-1/2} $
\vspace{-1pt}
\item[] \hspace{20pt} $f(\rho^2) = \displaystyle{\frac{1}{2}} (\bZ-\bW\bmu - \bX \balpha)^\top \bQ \,
\text{diag}\bigg( \bigg\{ \displaystyle{\frac{1}{n_i}} \cdot \displaystyle{\frac{1}{\sigma^2+\rho^2 n_i}} \bigg\}_{i=1}^n \bigg) \bQ^\top  (\bZ-\bW\bmu - \bX \balpha)$ \vspace{2pt}
\item[5.] For $i = 1, \dots, n$: \; Sample\; $\eta_i \sim \mathcal{N}\big( \rho_i^{2} \, \bQ_{\bullet,i}^\top (\bZ-\bX\balpha - \bW \bmu ) , \rho_i^{2}  \big)$ \;with\; 
\vspace{2pt}
\item[] \hspace{20pt} $\rho_i^{2} = (\rho^{-2} + n_i \, \sigma^{-2})^{-1} $
\end{enumerate}
\hrule 
}

\vspace{20pt}

\na{
The update of the latent scores $\bZ$ remains parallelizable across worms within each response-level index set, but is no longer separable across neurons belonging to the same worm. 
In fact, random effect marginalization induces multivariate truncated normal full conditionals that encode within-worm dependence.
State-of-the-art implementations for multivariate truncated normal sampling, such as the \texttt{R} packages \texttt{tmvtnorm} or \texttt{TruncatedNormal}, do not support vectorized draws, even under shared covariance and common truncation regions.
Nevertheless, partial vectorization can still be recovered by embedding such an update within an inner coordinate-wise Gibbs scheme, cycling over the components of each multivariate truncated normal draw.
To parallelize this, we group worms according to the number of neurons falling into each response stratum $\mathcal{G}(\ell)$.
In practice, we rewrite the update by iterating explicitly over the dimensions $m=1,\dots,M$ of the truncated normals -- rather than the worm index $i=1,\dots,n$ -- and vectorize the coordinate-wise Gibbs sampling across the sets of worms $\mathcal{M}(m,\ell) = \big\{ i \in \{1,\dots,n\} \mid  \vert \mathcal{L}(i,\ell) \vert = m \big\}$.
See \ref{sec:derivations} for further details. 
}

\na{
The full conditional distribution $p(\rho^2 \!\mid\! \bZ, \bmu, \balpha ) $ takes the form of a generalized inverse-gamma-type density. 
To the best of our knowledge, it does not belong to a standard conjugate family, and it does not admit direct sampling.
While Metropolis-within-Gibbs or slice sampling could be employed, the one-dimensional nature of the problem makes inverse transform sampling particularly convenient. To this end, numerical CDF inversion can be achieved by simply evaluating the unnormalized density over a pre-specified grid. 
Conversely, note that computing the matrix inversion $(\sigma^{2} \mathbb{I}_N + \rho^2 \bQ\bQ^\top )^{-1}$ would require a $\mathcal{O}(N^3)$ cost. 
To avoid this bottleneck in large-$N$ settings, as in the {\it C. elegans} assay, we recommend leveraging the Woodbury matrix inversion formula \citep{golub1996matrix}
\begin{equation*}
    (\sigma^{2} \mathbb{I}_N + \rho^2 \bQ\bQ^\top )^{-1} =
    \sigma^{-2} \mathbb{I}_N - \sigma^{-2} \bQ \,
\text{diag}\bigg( \bigg\{ \displaystyle{\frac{\rho^2}{\sigma^2+\rho^2 n_i}} \bigg\}_{i=1}^n \bigg) \bQ^\top
\end{equation*}
}

\section{Derivations of Full Conditionals Distributions}\label{sec:derivations}

\na{
In this section, we provide the derivations of the full conditional distributions appearing in Algorithms~\ref{alg:regular} and \ref{alg:collapsed}.
}

\subsection{Latent scores}
 
\na{
Both equations \eqref{full_lik} and \eqref{full_lik_marg} could be readily seen as corresponding to the kernels of $N$-dimensional multivariate truncated normal distributions.
Joint update of the full vector $\bZ$ is thus limited by its large dimension and by a composite truncation region, inducing non-trivial dependencies across its coordinates. 
Conversely, updating $\bZ$ becomes easy via blocking it by response value and sampling sequentially each subset $\bZ_{\mathcal{G}(\ell)}$ -- that is, all the $Z_{ij}$ with $Y_{ij}=\ell$ -- given all other blocks $\bZ_{-\mathcal{G}(\ell)}$.
Following \citet{hoff2009first}, we thus rewrite the feasible region $\mathcal{R}(\bY)$ as 
\begin{equation*}
\begin{aligned}
\mathcal{R}(\bY) &= 
\bigcap_{i=1}^n \bigcap_{j=1}^{n_i}\Big[\Big(\bigcap_{y_{ij} < y_{i'j'}} \{Z_{ij} < Z_{i'j'} \} \Big) \cap \Big(\bigcap_{y_{ij}> y_{i'j'}} \{Z_{ij}> Z_{i'j'} \} \Big) \Big] \\
&= \bigcap_{i=1}^n \bigcap_{j=1}^{n_i}\big\{Z_{ij} \in \big(\max_{Y_{ij} > Y_{i'j'}} Z_{i'j'}, \min_{Y_{ij} < Y_{i'j'}} Z_{i'j'} \big) \big\} = \bigcap_{\ell=0}^{L} \bigcap_{(i,j) \in \mathcal{G}(\ell)} \big\{Z_{ij} \in (l_\ell, u_\ell) \big\} \;.
\end{aligned}
\end{equation*}
where the group-wise extrema $l_\ell = \max \bZ_{\mathcal{G}(\ell-1)}$ and $ u_\ell = \min \bZ_{\mathcal{G}(\ell+1)} $ mask the dependence on the remaining latent scores $\bZ_{-\mathcal{G}(\ell)}$.
Thus, conditioned on $\bZ_{-\mathcal{G}(\ell)}$, all elements of the subvector $\bZ_{\mathcal{G}(\ell)}$ share the same truncation interval $(l_\ell, u_\ell)$.
To leverage this, we first make explicit the factorization in the rank likelihood from equation~\eqref{full_lik}, and then highlight all and only the occurrences of elements of $\bZ_{\mathcal{G}(\ell)}$ 
\begin{equation*}\label{full_lik_fact}
\begin{aligned}
    p(\bZ \mid \bY, \bmu, \balpha, \boeta, \rho^2 ) &\propto
    \mathbbm{1}_{\big(\bZ \in \mathcal{R}(\bY)\big)} \cdot \prod_{i=1}^n \prod_{j=1}^{n_i}
    \phi\big(\, Z_{ij} - R_{ij} \, ,\, \sigma^2 \, \big) \\
    &\propto \prod_{(i,j) \in \mathcal{G}(\ell)} \phi\big(\, Z_{ij} - R_{ij} \, ,\, \sigma^2 \, \big) \cdot \mathbbm{1}_{\big( Z_{ij} \in (l_\ell, u_\ell) \big)}
    \;, 
\end{aligned}
\end{equation*}
where we introduced the linear predictor $\bR = \bW \bmu + \bX\balpha + \bQ \boeta$.
It is then easy to recognize the product of kernels of one-dimensional truncated normal distributions, so that
\begin{equation*}
    \big( \bZ_{\mathcal{G}(\ell)} \mid \bY, \bZ_{-\mathcal{G}(\ell)}, \bmu, \balpha, \boeta, \rho^2 \big) \sim \bigotimes_{(i,j) \in \mathcal{G}(\ell)}
    \mathcal{TN}\big( R_{ij} \,, \, \sigma^2 \,; \, (l_\ell,u_\ell) \big)
    \;,
\end{equation*}
where $\bigotimes$ denotes the joint distribution of independent components indexed by $(i,j) \in \mathcal{G}(\ell)$.
}

\subsubsection{Collapsed Sampler}

\na{
The collapsed sampler allows a similar reasoning. First, we make explicit the factorization in the marginalized version of the rank likelihood from equation~\eqref{full_lik_marg} as
\begin{equation*}
\begin{aligned}
    p(\bZ &\mid \bY, \bmu, \balpha, \rho^2 ) \propto \mathbbm{1}_{\big(\bZ \in \mathcal{R}(\bY) \big)} \cdot 
    \prod_{i=1}^n \phi_{n_i} \big(\, \bZ_{\mathcal{J}(i)} - \bR_{\mathcal{J}(i)} \, ,\, \sigma^2 \, \mathbb{I}_{n_i} + \rho^2 \1_{n_i} \1_{n_i}^\top\, \big)
    \; ,
\end{aligned}
\end{equation*}
where we now redefined the linear predictor as $\bR = \bW \bmu + \bX\balpha $.
As before, we want to highlight all occurrences of  $\bZ_{\mathcal{G}(\ell)}$ given $\bZ_{-\mathcal{G}(\ell)}$.
To this end, we partition each $\bZ_{\mathcal{J}(i)}$ by response value, giving the sub-block $\bZ_{\mathcal{L}(i,\ell)}$ -- of size $n_{i\ell}$ -- and its complement $\bZ_{\mathcal{J}(i) \setminus \mathcal{L}(i,\ell)}$ -- of size $n_i - n_{i\ell}$.
Recall that we defined $\mathcal{L}(i,\ell) = \mathcal{J}(i) \cap \mathcal{G}(\ell)$ in Algorithms~\ref{alg:collapsed}.
Then, we rewrite the unconstrained multivariate normal kernel for each $\bZ_{\mathcal{J}(i)} $ to highlight the conditional distribution $\bZ_{\mathcal{L}(i,\ell)} \mid \bZ_{\mathcal{J}(i) \setminus \mathcal{L}(i,\ell)}$.
Leveraging the Sherman–Morrison identity, the precision matrix of the full $i$-th block reads
\begin{equation*}
\bLambda_i = \big(\sigma^2 \mathbb{I}_{n_i} + \rho^2 \1_{n_i}\1_{n_i}^\top \big)^{-1} = \sigma^{-2} \bigg( \mathbb{I}_{n_i} -
\frac{\rho^2}{\sigma^2 + \rho^2 n_i} \, \1_{n_i} \1_{n_i}^\top \bigg) \;.
\end{equation*}
The covariance for $\bZ_{\mathcal{L}(i,\ell)} \mid \bZ_{\mathcal{J}(i) \setminus \mathcal{L}(i,\ell)}$ is then obtained as
\begin{equation*}
\begin{aligned}
\boldsymbol{\Sigma}_{i\ell} &
= \big( (\bLambda_i)_{\mathcal{L}(i,\ell),\, \mathcal{L}(i,\ell)} \big)^{-1}
= \sigma^{2} \bigg( \mathbb{I}_{n_{i\ell}} -
\frac{\rho^2}{\sigma^2 + \rho^2 n_i} \, \1_{n_{i\ell}} \1_{n_{i\ell}}^\top \bigg)^{-1} \\
&= \sigma^{2} \bigg( \mathbb{I}_{n_{i\ell}} - \Big( n_{i\ell} -
\frac{\rho^2}{\sigma^2 + \rho^2 n_i}  \Big)\, \1_{n_{i\ell}} \1_{n_{i\ell}}^\top \bigg)
= \sigma^{2} \bigg( \mathbb{I}_{n_{i\ell}} +
\frac{\rho^2}{\sigma^2 + \rho^2 (n_i-n_{i\ell})} \, \1_{n_{i\ell}} \1_{n_{i\ell}}^\top \bigg) \;.
\end{aligned}
\end{equation*}
Conversely, the conditional mean for $\bZ_{\mathcal{L}(i,\ell)} \mid \bZ_{\mathcal{J}(i) \setminus \mathcal{L}(i,\ell)}$ requires 
\begin{equation*}
\begin{aligned}
&= \boldsymbol{\Sigma}_{i\ell} \, \big((\bLambda_i)_{\mathcal{L}(i,\ell) ,\, \mathcal{J}(i) \setminus \mathcal{L}(i,\ell)} \big) \\
&=\sigma^{2} \bigg( \mathbb{I}_{n_{i\ell}} +
\frac{\rho^2}{\sigma^2 + \rho^2 (n_i-n_{i\ell})} \, \1_{n_{i\ell}} \1_{n_{i\ell}}^\top \bigg) \sigma^{-2} \bigg(
-\frac{\rho^2}{\sigma^2 + \rho^2 n_{i}} \, \1_{n_{i\ell}} \1_{n_{i}-n_{i\ell}}^\top \bigg) \\
&= -\frac{\rho^2}{\sigma^2 + \rho^2 n_{i}} \, \bigg( 1+ 
\frac{\rho^2 n_{i\ell}}{\sigma^2 + \rho^2 (n_i-n_{i\ell})} \bigg) \1_{n_{i\ell}} \1_{n_{i}-n_{i\ell}}^\top 
= - \frac{\rho^2 }{\sigma^2 + \rho^2 (n_i-n_{i\ell})}  \1_{n_{i\ell}} \1_{n_{i}-n_{i\ell}}^\top \;.
\end{aligned}
\end{equation*}
Defining $\xi_{i\ell}= \displaystyle{ \frac{\rho^2 }{\sigma^2 + \rho^2 (n_i-n_{i\ell})}} \1_{n_{i}-n_{i\ell}}^\top \big(\bZ_{\mathcal{J}(i) \setminus \mathcal{L}(i,\ell)} - \bR_{\mathcal{J}(i) \setminus \mathcal{L}(i,\ell)} \big)$ as in Algorithms~\ref{alg:collapsed},\\[5pt]
the conditional mean for $\bZ_{\mathcal{L}(i,\ell)} \mid \bZ_{\mathcal{J}(i) \setminus \mathcal{L}(i,\ell)}$ thus reads
\begin{equation*}
\bR_{\mathcal{L}(i,\ell)} - \boldsymbol{\Sigma}_{i\ell} \, \big((\bLambda_i)_{\mathcal{L}(i,\ell) ,\, \mathcal{J}(i) \setminus \mathcal{L}(i,\ell)} \big)\, \big(\bZ_{\mathcal{J}(i) \setminus \mathcal{L}(i,\ell)} - \bR_{\mathcal{J}(i) \setminus \mathcal{L}(i,\ell)} \big) 
= \bR_{\mathcal{L}(i,\ell)} + \1_{n_{i\ell}} \xi_{i\ell}
\; .
\end{equation*}
Putting all this together, the marginalized likelihood finally becomes
\begin{equation*}\label{full_lik_marg_fact}
\begin{aligned}
    p(\bZ &\mid \bY, \bmu, \balpha, \rho^2 ) \propto \prod_{i=1}^n 
    \phi_{n_{i\ell}} \big(\, \bZ_{\mathcal{L}(i,\ell)} - \bR_{\mathcal{L}(i,\ell)} - \1_{n_{i\ell}} \xi_{i\ell} \, ,\, \boldsymbol{\Sigma}_{i\ell} \, \big)
    \cdot \mathbbm{1}_{\big( \bZ_{\mathcal{L}(i,\ell)} \in (l_\ell, u_\ell)^{n_{i\ell}} \big)}
    \; ,
\end{aligned}
\end{equation*}
where we can recognize the kernel of multivariate truncated normal distributions
\begin{equation*}
    \big( \bZ_{\mathcal{G}(\ell)} \mid \bY, \bZ_{-\mathcal{G}(\ell)}, \bmu, \balpha, \boeta, \rho^2 \big) \sim \bigotimes_{i=1}^n \,
    \mathcal{TN}_{n_{i\ell}}\big( \bR_{\mathcal{L}(i,\ell)} + \1_{n_{i\ell}} \xi_{i\ell} \,, \, \boldsymbol{\Sigma}_{i\ell} \,; \, (l_\ell,u_\ell)^{n_{i\ell}} \big)
    \;.
\end{equation*}
}

\subsubsection{Code Vectorization for the Collapsed Sampler}

\na{
To implement parallel updates, we further reshape the above full conditional by highlighting sub-blocks of truncated normal with shared dimension, covariance, and constraints 
\begin{equation*}
    \big( \bZ_{\mathcal{G}(\ell)} \mid \bY, \bZ_{-\mathcal{G}(\ell)}, \bmu, \balpha, \boeta, \rho^2 \big) \sim \bigotimes_{m=1}^M \Bigg( \bigotimes_{i  \in \mathcal{M}(m,\ell)}
    \mathcal{TN}_{m}\big( \bR_{\mathcal{L}(i,\ell)} + \1_{n_{i\ell}} \xi_{i\ell} \,, \, \boldsymbol{\Sigma}_{m} \,; \, (l_\ell,u_\ell)^{m} \big) \Bigg)
    \;.
\end{equation*}
where $\boldsymbol{\Sigma}_{m} = \sigma^{2} \bigg( \mathbb{I}_{m} +
\displaystyle{\frac{\rho^2}{\sigma^2 + \rho^2 (n_i-m)}} \, \1_{m} \1_{m}^\top \bigg)$.
No state-of-the-art \texttt{R} package allows vectorized sampling from multivariate truncated normal distributions. However, vectorized updates can still be achieved via custom routines for coordinate-wise Gibbs sampling from each $\mathcal{TN}_{m}$ distribution. 
Concretely, this requires explicit for loops only over an outer index $m \in \{ 1,\dots, M\}$ and an inner one $m' \in \{1,\dots,m\}$, while performing matrix operations over the index set $\mathcal{M}(m,\ell)$.
This gives crucial computational advantages in scenarios with a large number of random effect groups $n$, like the {\it C. elegans} assay.
}

\subsection{Treatment Coefficients}

\na{
We begin by recalling a well-known result to handle mixture prior distributions, such as the zero-inflated truncated normal prior proposed by \cite{neelon2004bayesian}.}
\begin{lem}\label{lemma:mixtures}
    Suppose $\btheta\in\bTheta\in\bbR^d$ is a parameter of interest in some parameter space $\bTheta$. Let $\mathfrak{L}(\btheta\mid\by)$ be the likelihood, and place a mixture prior on $\btheta$ given by $p(\btheta) = \sum_{k=0}^K \pi_k p_k(\btheta)$, where there are $K$ mixture components and $\sum_{k=0}^K \pi_k = 1$. Then the posterior is given by \\[-10pt]
    $$p(\btheta\mid\by) = \sum_{k=0}^K \pi_k(\btheta\mid\by) p_k(\btheta\mid\by)$$
    ~\\[-20pt]
    with\\[-10pt]
    $$
    \pi_k(\btheta) = \frac{\pi_k \int_{\bTheta} \mathfrak{L}(\btheta\mid\by)p_k(\btheta)\,d\btheta}{\sum_{k=0}^K \pi_k \int_{\bTheta} \mathfrak{L}(\btheta\mid\by)p_k(\btheta)\,d\btheta}
    \qquad \qquad 
    p_k(\btheta\mid\by) = \frac{ \mathfrak{L}(\btheta\mid\by)p_k(\btheta)}{\int_{\bTheta} \mathfrak{L}(\btheta\mid\by)p_k(\btheta)\,d\btheta}
    \;.$$
\end{lem}

\na{
Equipped with these results, we give a unified derivation for the full conditional of each $\alpha_{tgc}$ within both regular and collapsed samplers.
To this end, let us first define the respective base residuals and the covariance matrix as
\begin{equation*}
\begin{aligned}
    &\text{Regular Gibbs Sampler:} \qquad& \bR^{(o)}& =\bZ - \bW \bmu - \bQ \boeta \qquad& \bOmega &= \sigma^{2} \mathbb{I}_N  \\
    &\text{Collapsed Gibbs Sampler:} \qquad& \bR^{(o)} &=\bZ - \bW \bmu \qquad & \bOmega &= \sigma^{2} \mathbb{I}_N + \rho^2 \bQ\bQ^\top 
    \; ,
\end{aligned}    
\end{equation*}
and the overall residual $\bR^{(tgc)} = \bR^{(o)} - \bX_{\bullet, -(tgc)}\balpha_{-(tgc)} $.
Accordingly, the effective likelihood $\mathfrak{L}(\btheta\mid\by)$ in Lemma~\ref{lemma:mixtures} reads
$\displaystyle{ \phi_N\big(\, \bR^{(tgc)} - \bX_{\bullet,(tgc)}\alpha_{tgc} \, ,\, \bOmega \, \big) }$.
}

\na{
The prior $\alpha_{tgc} \sim \mathcal{ZI}\text{--}\mathcal{N}^+ \big(\pi, \lambda, \nu^2\big)$ is the two-component mixture
$\pi \cdot p_0(\alpha_{tgc}) + (1-\pi) \cdot p_1(\alpha_{tgc})$, with spike $p_0(\alpha_{tgc}) = \delta_0(\alpha_{tgc})$ and slab $p_1(\alpha_{tgc}) = \mathbbm{1}_{(\alpha_{tgc}>0)} \, \phi(\alpha_{tgc} - \lambda,\nu^2) / \Phi(\lambda/\nu)$.
By Lemma~\ref{lemma:mixtures}, the posterior full conditional of  $\alpha_{tgc}$ inherits the same mixture structure with updated component posteriors $\widetilde{p}_0$, $\widetilde{p}_1$, and updated weights $\widetilde{\pi}$, $1-\widetilde{\pi}$.
Under the spike, the posterior remains a point mass at zero $\widetilde{p}_0(\alpha_{tgc} \mid \,\relbar) = \delta_0(\alpha_{tgc})$, 
where the hyphen is shorthand for conditioning on all remaining variables.
Under the slab, the combination of the truncated normal prior with the effective likelihood gives
\begin{equation*}
\begin{aligned}
\widetilde{p}_1(\alpha_{tgc} \mid \,\relbar) &\propto \mathbbm{1}_{(\alpha_{tgc}>0)} \, \phi_N\big(\, \bR^{(tgc)} - \bX_{\bullet,(tgc)}\alpha_{tgc} \, ,\, \bOmega \, \big) \cdot \phi(\alpha_{tgc} - \lambda,\nu^2) \\
&\propto \mathbbm{1}_{(\alpha_{tgc}>0)} \,  \exp\bigg(-\frac{1}{2} \alpha_{tgc}^2 \big( \nu^{-2} + \bX_{\bullet,(tgc)}^\top \bOmega^{-1} \bX_{\bullet,(tgc)} \big) \bigg) \cdot \\
& \qquad \exp\bigg( \alpha_{tgc} \big( \nu^{-2}\lambda + \bX_{\bullet,(tgc)}^\top \bOmega^{-1} \bR^{(tgc)} \big) \bigg) \;.
\end{aligned}
\end{equation*}
Completing the square yields a positive truncated normal $ \mathcal{TN}\big(\widetilde{\lambda}_{tgc},\widetilde{\nu}_{tgc}^2,(0,\infty)\big)$ with
\begin{equation*}
\begin{aligned}
    \widetilde{\nu}_{tgc}^2 &= \big( \nu^{-2} + \bX_{\bullet,(tgc)}^\top \bOmega^{-1} \bX_{\bullet,(tgc)} \big)^{-1}
    \qquad\quad
    \widetilde{\lambda}_{tgc} = \widetilde{\nu}_{tgc}^2 \, \big( \nu^{-2}\lambda + \bX_{\bullet,(tgc)}^\top \bOmega^{-1} \bR^{(tgc)} \big) \;.
\end{aligned}
\end{equation*}
Conversely, the posterior spike weight reads
\begin{equation*}
    \widetilde{\pi}_{tgc} = \frac{\pi \, \varpi_0}{\pi \, \varpi_0 + (1-\pi) \, \varpi_1} \quad \qquad \varpi_k = \int \phi_N\big( \bR^{(tgc)} - \bX_{\bullet,(tgc)}\alpha_{tgc}, \bOmega \big) \, p_k(\alpha_{tgc}\mid \,\relbar) \, d\alpha_{tgc} \;.
\end{equation*}
The spike marginal $\varpi_0$ is just the likelihood evaluated at $\alpha_{tgc}=0$, namely 
$ \phi_N\big(\bR^{(tgc)},\bOmega\big)$.
Conversely, completing the square in $\alpha_{tgc}$ -- as done above for the slab posterior -- and recognizing the resulting positive truncated normal kernel gives the closed form expression 
\begin{equation*}
\begin{aligned}
    \varpi_1 &= \phi_N\big( \bR^{(tgc)},\bOmega\big) \cdot
    \frac{ \phi(\widetilde{\lambda}_{tgc},\widetilde{\nu}_{tgc}^2) \, / \, \Phi(\widetilde{\lambda}_{tgc}/\widetilde{\nu}_{tgc}) }
         { \phi(\lambda,\nu^2) \, / \, \Phi(\lambda/\nu) } \;,
\end{aligned}
\end{equation*}
so that the likelihood pre-factor cancels in the ratio $\varpi_0/\varpi_1$ and we obtain
\begin{equation*}
\widetilde{\pi}_{tgc} = \frac{\pi \, \phi(\lambda,\nu^2)/\Phi(\lambda/\nu)}{\pi \, \phi(\lambda,\nu^2)/\Phi(\lambda/\nu) + (1-\pi) \, \phi(\widetilde{\lambda}_{tgc},\widetilde{\nu}_{tgc}^2)/\Phi(\widetilde{\lambda}_{tgc}/\widetilde{\nu}_{tgc})} \;.
\end{equation*}
The explicit forms of $\widetilde{\nu}_{tgc}^2$ and $\widetilde{\lambda}_{tgc}$ as in Algorithms~\ref{alg:regular} and \ref{alg:collapsed} are simply obtained by replacing the corresponding definitions for $\bOmega$ and $\bR^{(tgc)}$.
}

\subsection{Replicate Intercepts}

\na{
Analogously, we provide a unified derivation for the full conditional of $\bmu$ within both regular and collapsed samplers.
We redefine the auxiliary variables
\begin{equation*}
\begin{aligned}
    &\text{Regular Gibbs Sampler:} \qquad& \bR& =\bZ - \bX\balpha - \bQ \boeta \qquad& \bOmega &= \sigma^{2} \mathbb{I}_N  \\
    &\text{Collapsed Gibbs Sampler:} \qquad& \bR &=\bZ - \bX\balpha \qquad & \bOmega &= \sigma^{2} \mathbb{I}_N + \rho^2 \bQ\bQ^\top 
    \; .
\end{aligned}    
\end{equation*}
Under both representations, the effective likelihood of the latent scores is proportional to $\phi_N\big(\, \bR - \bW \bmu \, ,\, \bOmega \, \big)$.
Combining it with the Gaussian prior $\bmu \sim \mathcal{N}_{R-1}(\mathbf{0}, \varphi^2 \mathbb{I}_{R-1})$ we have
\begin{equation*}
\begin{aligned}
    p(\bmu \mid \,\relbar) &\propto \exp\bigg( -\tfrac{1}{2} \big( \bR - \bW \bmu \big)^\top \bOmega^{-1} \big( \bR - \bW \bmu \big) \bigg) \cdot \exp\bigg( -\tfrac{1}{2\varphi^2} \bmu^\top \bmu \bigg) \\
    &\propto \exp\bigg( -\tfrac{1}{2} \bmu^\top \big( \varphi^{-2} \mathbb{I}_{R-1} + \bW^\top \bOmega^{-1} \bW \big) \bmu + \bmu^\top \bW^\top \bOmega^{-1} \bR \bigg) \;.
\end{aligned}
\end{equation*}
Completing the square yields the Gaussian full conditional
\begin{equation*}
\bmu \mid \,\relbar \, \sim \, \mathcal{N}_{R-1}\big( \, \boldsymbol{\Xi} \, \bW^\top \bOmega^{-1} \bR \, , \, \boldsymbol{\Xi} \, \big) \qquad \text{with} \qquad \boldsymbol{\Xi} = \big( \varphi^{-2} \mathbb{I}_{R-1} + \bW^\top \bOmega^{-1} \bW \big)^{-1} \;.
\end{equation*}
The explicit forms in Algorithms~\ref{alg:regular} and \ref{alg:collapsed} are recovered by replacing the corresponding definitions for $\bOmega$ and $\bR$.
}

\subsection{Random Effect Variance}\label{sec_cond_dist_RE_var}

\na{
We now derive the full conditional of $\rho^2$ given all remaining variables. Unlike for $\balpha$ and $\bmu$, this derivation differs between the regular and the collapsed Gibbs samplers.
The inverse-gamma prior $\rho^2 \sim \text{InvGa}(a,b)$ is fully conjugate in the regular case, whereas conjugacy is lost upon marginalization of $\boeta$.
We begin with the regular Gibbs sampler case, where, conditioned on $\boeta$, $\rho^2$ is independent from everything else.
Thus
\begin{equation*}
\begin{aligned}
    p(\rho^{2} \mid \,\relbar) &\propto p(\rho^{2}) \cdot \prod_{i=1}^n p(\eta_i \mid \rho^{2}) \propto (\rho^{2})^{-a-1} \exp(-b / \rho^{2} ) \cdot \prod_{i=1}^n (\rho^{2})^{-1/2} \exp\big(\!-\!\tfrac{1}{2}  \eta_i^2 / \rho^{2} \big) \\
    &\propto (\rho^{-2})^{-a - n/2 - 1} \exp\big( \!- ( b + \!\tfrac{1}{2} \boeta^\top \boeta ) / \rho^{2} \big) \;,
\end{aligned}
\end{equation*}
which is the kernel of $\mathcal{I}nv\mathcal{G}a \big( a + n/2 \, , \, b + \tfrac{1}{2} \boeta^\top \boeta \big)$, matching Step~4 of Algorithm~\ref{alg:regular}.
}

\subsubsection{Collapsed Gibbs Sampler}

\na{
In the collapsed sampler, $\rho^2$ enters the model directly through the marginalized covariance $\bOmega = \sigma^2 \mathbb{I}_N + \rho^2 \bQ \bQ^\top$ of the latent scores $\bZ$ after the random effects $\boeta$ have been integrated out. 
Defining the residual $\bR = \bZ - \bW \bmu - \bX \balpha$, the full conditional reads
\begin{equation*}
    p(\rho^2 \mid \,\relbar) \propto \big(\rho^2\big)^{-a-1} \exp\big( -b/\rho^2 \big) \cdot \det\big(\bOmega\big)^{-1/2} \exp\bigg( -\frac{1}{2} \bR^\top \bOmega^{-1} \bR \bigg) \;.
\end{equation*}
We now exploit the block-diagonal structure of $\bQ\bQ^\top$, with blocks $\1_{n_i}\1_{n_i}^\top$ for $i=1,\dots,n$, to simplify both the determinant and the quadratic form as functions of $\rho^2$.
For the determinant, applying the matrix determinant lemma block-by-block gives
\begin{equation*}
    \det\big(\bOmega\big) = \det\Big( \mathbb{I}_N + \rho^2 \bQ \bQ^\top \Big) = \prod_{i=1}^n \det\big( \sigma^2 \mathbb{I}_{n_i} + \rho^2 \1_{n_i}\1_{n_i}^\top \big) = \prod_{i=1}^n \sigma^{2(n_i-1)} \big( \sigma^2 + \rho^2 n_i \big) \;,
\end{equation*}
so that 
\begin{equation*}\label{eq_h_rho2}
\det\big(\bOmega\big)^{-1/2} \propto h(\rho^2) = \prod_{i=1}^n \big( \sigma^2 + \rho^2 n_i \big)^{-1/2}
\end{equation*}
as a function of $\rho^2$.
For the quadratic form, the Woodbury identity reported in \ref{sec_gibbs} yields
\begin{equation*}
    \bR^\top \bOmega^{-1} \bR = \sigma^{-2} \bR^\top \bR - \sigma^{-2} \bR^\top \bQ \, \text{diag}\bigg( \bigg\{ \frac{\rho^2}{\sigma^2 + \rho^2 n_i} \bigg\}_{i=1}^n \bigg) \bQ^\top \bR \;.
\end{equation*}
The first term does not depend on $\rho^2$ and can be absorbed in the proportionality constant. 
In the second term, the product of $\sigma^{-2}$ and the diagonal entries can be rewritten as
\begin{equation*}
    \sigma^{-2} \frac{\rho^2}{\sigma^2 + \rho^2 n_i} = \frac{1}{n_i} \bigg( \frac{1}{\sigma^2} - \frac{1}{\sigma^2 + \rho^2 n_i} \bigg) \;,
\end{equation*}
so that, dropping again the $\rho^2$-free part, the overall quadratic form collapses to 
\begin{equation*}\label{eq_f_rho2}
    f(\rho^2) = \frac{1}{2} \bR^\top \bQ \, \text{diag}\bigg( \bigg\{ \frac{1}{n_i} \cdot \frac{1}{\sigma^2 + \rho^2 n_i} \bigg\}_{i=1}^n \bigg) \bQ^\top \bR \;.
\end{equation*}
Combining all the steps, the full conditional reads
\begin{equation*}
    p(\rho^2 \mid \,\relbar) \propto \big(\rho^2\big)^{-a-1} h(\rho^2) \, \exp\big( -b/\rho^2 - f(\rho^2) \big) \, 
    \;,
\end{equation*}
matching Step~4 of Algorithms~\ref{alg:collapsed}. To the best of our knowledge, this density does not belong to a standard conjugate family.
As discussed in \ref{sec_gibbs}, the one-dimensional nature of the problem makes inverse-transform sampling via numerical CDF inversion on a pre-specified grid a particularly convenient strategy.
}

\subsubsection{Vectorized evaluation of $f(\rho^2)$}

\na{
To efficiently evaluate $f(\rho^2)$ on the grid, it is convenient to rewrite it by grouping worms with the same number of neurons. 
Recall that $\bQ^\top \bR \in \Re^n$ has $i$-th entry equal to the within-worm sum $\1_{n_i}^\top \bR_{\mathcal{J}(i)} = \sum_{j \in \mathcal{J}(i)} R_j$.
Reshuffling the sum to highlight an index $m = 1,\dots,M$ and the index sets $\mathcal{F}(m) = \big\{ i \in \{1,\dots,n\} \mid n_i = m \big\}$ we have
\begin{equation*}
    f(\rho^2) = \frac{1}{2} \sum_{m=1}^M \frac{1}{m} \cdot \frac{1}{\sigma^2 + \rho^2 m} \sum_{i \in \mathcal{F}(m)} \big( \1_{n_i}^\top \bR_{\mathcal{J}(i)} \big)^2 \;.
\end{equation*}
The inner sum over $\mathcal{F}(m)$ does not depend on $\rho^2$ and can be pre-computed once before traversing the grid, leaving only $M$ scalar additions per grid point. The same reshuffling applies to the $\log$-determinant prefactor, since 
$$\prod_{i=1}^n (\sigma^2 + \rho^2 n_i)^{-1/2} = \prod_{m=1}^M (\sigma^2 + \rho^2 m)^{-|\mathcal{F}(m)|/2} \,.$$
}

\subsection{Random Effects}\label{sec_cond_dist_RE}

\na{
We now derive the full conditional of $\boeta$ given all other variables. Recall that this update is used in both versions of the sampler: it is part of the cyclical scheme in the regular case, and is performed ex post in the collapsed one to recover samples of $\boeta$ from the targeted joint posterior.
Define the residual $\bR = \bZ - \bW \bmu - \bX \balpha $, with $\bR \sim \mathcal{N}_N\big(\bQ \boeta ,\sigma^2 \bI_N \big)$.
Combined with the prior $\boeta \sim \mathcal{N}_n(0, \rho^2 \bI_n)$, this gives
\begin{equation*}
\begin{aligned}
    p(\boeta \mid \,\relbar) &\propto \exp\bigg( -\tfrac{1}{2\sigma^2} \big\| \bR - \bQ \boeta \big\|^2 \bigg) \cdot \exp\bigg( -\tfrac{1}{2\rho^2} \big\|\boeta \big\|^2 \bigg) \\
    &\propto \exp\bigg( -\tfrac{1}{2} \boeta^\top \big( \rho^{-2} \bI_n + \sigma^{-2} \bQ^\top \bQ \big) \boeta + \, \sigma^{-2}  \boeta^\top \bQ^\top \bR \bigg) \\ 
    &\propto \prod_{i=1}^n \exp\bigg( -\tfrac{1}{2} \eta_i^2 \big( \rho^{-2} + n_i \sigma^{-2} \big) + \eta_i \, \sigma^{-2} \1_{n_i}^\top \bR_{\mathcal{J}(i)} \bigg) \;,
\end{aligned}
\end{equation*}
where we highlighted the worm-specific residual $\bR_{\mathcal{J}(i)} = \bZ_{\mathcal{J}(i)} - (\bW \bmu + \bX \balpha)_{\mathcal{J}(i)} $ and leveraged the diagonal structure of $\bQ^\top \bQ $, reflecting conditional independence.
Completing the square and noting that $\1_{n_i}^\top \bR_{\mathcal{J}(i)} = \bQ_{\bullet,i}^\top (\bZ - \bW \bmu - \bX \balpha)$, we obtain
\begin{equation*}
\eta_i \mid \,\relbar \, \sim \, \mathcal{N}\big( \, \rho_i^2 \, \sigma^{-2} \, \bQ_{\bullet,i}^\top (\bZ - \bX \balpha - \bW \bmu) \, , \, \rho_i^2 \, \big) \qquad \text{with} \qquad \rho_i^2 = \big( \rho^{-2} + n_i \sigma^{-2} \big)^{-1} \;.
\end{equation*}
}

\section{Parameter Expansion}\label{sec:px}

\na{
Mixing of the Gibbs chain can be slow when the model contains strongly coupled variables, especially when these live in a latent space.
Parameter expansion (\textsc{px}) is a general strategy to alleviate this issue by augmenting the model with redundant, non-identifiable variables that loosen the posterior dependence among inferential parameters \citep{LiuWu1999, MengVanDyk1999,Murray2013}.
Most often, this amounts to reparameterizing the model by introducing a scale or transformation parameter $\tau$ equipped with a working prior $p(\tau)$, 
and running Gibbs updates on the expanded parameter space.
The auxiliary variable $\tau$ is not a quantity of interest in itself, but merely a tool to improve mixing, while valid inference is preserved for the original identifiable parameters.
We employ \textsc{px} for the updates of $\boeta$, 
$\balpha$, and $\rho^2$. The \textsc{px} update for $\boeta$ closely follows \citet{Murray2013}, while those of $\balpha$ and $\rho^2$ require methodological innovations.
}

\subsection{Parameter Expansion for Random Effects}

\na{
Following \cite{Murray2013}, we apply \textsc{px} to the random effect update via an unidentified rescaling of the (residual) latent scores. Specifically, we introduce a scalar working parameter $\tau \in \Re_+$ and augment the model with $\widetilde{\bR} \sim \mathcal{N}_N\big( \tau \, \bQ \boeta \, , \, \tau^2 \sigma^2 \, \mathbb{I}_N \big)$. 
Such an augmented model is observationally equivalent to the original one since $\widetilde{\bR} / \tau$ is equal in distribution to the original residual $\bR$.
In setting a conjugate \textsc{px} prior $\tau^{2} \sim \mathcal{I}nv\mathcal{G}a (n_0/2, n_0/2)$, the authors claim maximal benefit under the diffuse improper prior $p(\tau^2) \propto \tau^{-2}$, arising from the limit $n_0 \to 0^+$.
Appendix B of \cite{Murray2013} proves that the corresponding \textsc{px} Gibbs sampler step $p(\boeta,\tau \!\mid \,\relbar)$ still yields samples from the desired stationary distribution $p(\boeta \mid \,\relbar)$ of
the identifiable parameters.
}

\na{Operationally, the \textsc{px} update is decoupled as $p(\boeta,\tau^2 \mid \,\relbar) = p(\boeta \mid \tau^2, \,\relbar) \, p(\tau^2 \mid \,\relbar)$, where the random effects have been marginalized out in the full conditional for $\tau$. 
Note that $\tau$ is purely instrumental and only appears in the expanded update of $\boeta$.
It is not carried over from one iteration of the Gibbs sampler to the next, and induces no effective rescaling of the residual in the updates of any other parameter.
As such, $\widetilde{\bR}$ and $\bR$ effectively encode the same information within every augmented sweep of $\boeta$.
We thus work directly with
the augmented likelihood
${p(\bR \mid \boeta, \tau^2, \relbar )} = \phi_N\big(\bR - \tau\bQ\boeta, \tau^2 \sigma^2 \, \mathbb{I}_N \big)$, treating $\tau^2$ as an auxiliary unidentified scaling of the residual variance.
We first draw from $p(\tau^2 \! \mid \,\relbar)$, where marginalizing $\boeta$ out of the augmented likelihood gives
\begin{equation*}
    p({\bR} \mid \tau^2, \relbar ) = \phi_N\big( \bR \, , \, \tau^2 \bOmega \big) \qquad \text{with} \qquad \bOmega = ( \sigma^2 \mathbb{I}_N + \rho^2 \bQ \bQ^\top )  \;.
\end{equation*}
Combining with the working prior $p(\tau^2) \propto \tau^{-2}$ yields
\begin{equation*}
\begin{aligned}
    p(\tau^2 \mid \,\relbar) 
    &\propto \big(\tau^2\big)^{-N/2 - 1} \exp\bigg(\! -\frac{1}{2\tau^2} \, {\bR}^\top \bOmega^{-1} {\bR} \bigg) \;,
\end{aligned}
\end{equation*}
which is the kernel of $\mathcal{I}nv\mathcal{G}a\big(N/2, {\bR}^\top \bOmega^{-1} {\bR}/2\big)$. 
The exponent is efficiently computed by leveraging the Woodbury identity from \ref{sec_gibbs}, rewriting ${\bR}^\top \bOmega^{-1} {\bR}$ as
\begin{equation*}
    {\bR}^\top (\sigma^2 \mathbb{I}_N + \rho^2 \bQ\bQ^\top)^{-1} {\bR} = \sigma^{-2} {\bR}^\top {\bR} - \sigma^{-2} \, {\bR}^\top \bQ \, \text{diag}\bigg( \bigg\{ \frac{\rho^2}{\sigma^2 + \rho^2 n_i} \bigg\}_{i=1}^n \bigg) \bQ^\top {\bR} \;.
\end{equation*}
Conditional on 
$\tau$, the update of the random effects $\boeta \sim p(\boeta \!\mid\! \tau, \,\relbar)$ follows a regular form, analogous to the one derived in Section~\ref{sec_cond_dist_RE} with unchanged variance and a rescaled mean.
Worm-by-worm conditional independence is preserved, and
\begin{equation*}
\eta_i \mid \,\relbar \, \sim \, \mathcal{N}\big( \, \tau^{-1} \, \rho_i^2 \, \sigma^{-2} \, \bQ_{\bullet,i}^\top (\bZ - \bX \balpha - \bW \bmu) \, , \, \rho_i^2 \, \big) \qquad \text{with} \qquad \rho_i^2 = \big( \rho^{-2} + n_i \sigma^{-2} \big)^{-1} \;.
\end{equation*}
}

\subsection{Parameter Expansion for Random Effect Variance}

\na{
The within-worm variance $\rho^2$ in the collapsed sampler still suffers from non-decaying autocorrelations, as illustrated in Figure~\ref{fig:acf_ess_px_marg}.
However, applying a similar rationale as before is more nuanced in this case.
We focus here only on the collapsed sampler, with random effects $\boeta$ already integrated out.
We leverage the same parameter expansion scheme used for $\boeta$, using the augmented residual likelihood
$
p({\bR} \mid \tau^2, \rho^2, \relbar ) = \phi_N\big( \bR \, , \, \tau^2 \bOmega \big)
$, where $\bOmega = \sigma^2 \mathbb{I}_N + \rho^2 \bQ\bQ^\top$. 
We aim again to decouple the \textsc{px} update as
$
p(\rho^2,\tau^2 \mid \,\relbar) = {p(\rho^2 \mid \tau^2, \,\relbar) }\, {p(\tau^2 \mid \,\relbar)}
$,
where $\rho^2$ has been marginalized out in the full conditional for $\tau^2$. 
}

\na{
The conditional $p(\rho^2 \mid \tau^2, \,\relbar)$ is obtained by directly plugging the augmented likelihood into the collapsed update of Appendix~\ref{sec_cond_dist_RE_var}. Combining with the original inverse-gamma prior $\rho^2 \sim \mathcal{I}nv\mathcal{G}a(a, b)$ and the diffuse improper prior $p(\tau^2) \propto \tau^{-2}$ yields 
\begin{equation*}
\begin{aligned}    
    p(\rho^2, \tau^2 \mid \relbar) &\propto (\tau^2)^{-1} \big(\rho^2\big)^{-a-1} \exp\!\big( -b/\rho^2 \big) \cdot \det\big(\tau^2 \bOmega\big)^{-1/2} \exp\!\bigg( -\tfrac{1}{2} \tau^{-2} \bR^\top \bOmega^{-1} \bR \bigg) \\ 
    &\propto (\tau^2)^{-N/2-1} \, \exp\big(- \widetilde{b}\, \tau^{-2} \big) \big(\rho^2\big)^{-a-1} h(\rho^2) \, \exp\big( -b/\rho^2 - \tau^{-2} f(\rho^2) \big) \;,
\end{aligned}
\end{equation*}
with $f(\rho^2)$ and $h(\rho^2)$ defined as in equations~\eqref{eq_f_rho2} and~\eqref{eq_h_rho2}, respectively, while
\begin{equation*}
\widetilde{b}= \frac{1}{2} \sigma^{-2} \bR^\top \Big( \bI_N - \bQ \, \text{diag}\big( \{ {1}/{n_i} \}_{i=1}^n \big) \bQ^\top \Big) \bR  = \tfrac{1}{2} \sigma^{-2} \bR^\top \Big( \bI_N - \bQ (\bQ^\top \bQ)^{-1}\bQ^\top \Big) \bR \;.
\end{equation*}
Here we used $\text{diag}\big( \{ {1}/{n_i} \}_{i=1}^n \big) = (\bQ^\top \bQ)^{-1}$.
The full conditional $p(\rho^2 \mid \tau^2, \,\relbar) $ is then fully analogous to the non-\textsc{px} case
\begin{equation*}
    p(\rho^2 \mid \tau^2, \,\relbar) \propto \big(\rho^2\big)^{-a-1} h(\rho^2) \, \exp\big( -b/\rho^2 - \tau^{-2} f(\rho^2) \big) \, 
    \;,
\end{equation*}
up to the rescaling of the quadratic form by $\tau^{-2}$.
The same algebraic manipulations of Appendix~\ref{sec_cond_dist_RE_var} apply (i.e. block-diagonal determinant \& Woodbury identity), and the inverse-CDF grid sampling routine of Appendix~\ref{sec_cond_dist_RE_var} extends with minor modifications.
}

\na{
The collapsed full conditional $p( \tau^2 \mid \,\relbar)$ of the \textsc{px} variable reads
\begin{equation*} 
    p(\tau^2 \mid \, \relbar) = \int_0^\infty p(\rho^2, \tau^2 \mid \, \relbar) \, d\rho^2 \propto (\tau^2)^{-N/2-1} \, \exp\big(- \widetilde{b}\, \tau^{-2} + \log \mathfrak{I}(\tau^2) \big) \;,
\end{equation*}
where $\mathfrak{I}(\tau^2) = \int_0^\infty e^{-S(\tau^2, \rho^2) } \,d\rho^2$
and $$ S(\tau^2, \rho^2) = (a+1) \log (\rho^2) - \log h(\rho^2) + b/\rho^2 + \tau^{-2} f(\rho^2)  \,.$$
This poses a twofold challenge: the normalizing constant $\mathfrak{I}(\tau^2)$ is not available in closed form, and direct sampling from $p(\tau^2 \mid \, \relbar)$ is not feasible.
In principle, numerical methods could be employed both to approximate the integral and to sample $\tau^2$ exactly -- for instance, via an inner griddy-Gibbs procedure.
However, since $\tau^2$ is introduced solely as an auxiliary device to improve mixing on the inferential parameters, we prefer to avoid the additional computational overhead and algorithmic complexity that such a treatment would entail.
}

\na{
To avoid this extra burden, we proposed to combine the following approximations. 
First, we Taylor-expand the term $\log \mathfrak{I}(\tau^2)\approx \log \mathfrak{I}(1) + \mathfrak{E}(1) \big(\tau^{-2}-1\big)$ around $\tau^{-2}=1$ in the exponent, where
$$ \mathfrak{E}(\tau^2) = \frac{\partial \log \mathfrak{I}(\tau^2)}{\partial (\tau^{-2})} = -\frac{1}{\mathfrak{I}(\tau^2)} \int_0^\infty f(\rho^2) \, e^{-S(\tau^2, \rho^2) } \,d\rho^2$$
The expansion point is motivated by the rationale of \textsc{px} itself, under which $\tau^2$ is expected to fluctuate around 1 -- i.e. centered on the un-expanded model.
Crucially, this restores conditional conjugacy in the resulting approximate full conditional for $\tau^{2}$,
\begin{equation*} 
    \widehat{p}\,(\tau^2 \mid \, \relbar) \propto (\tau^2)^{-N/2-1} \, \exp\big(- (\widetilde{b} - \mathfrak{E}(1) ) \, \tau^{-2} \big) \;,
\end{equation*}
which is the kernel of an inverse-gamma distribution $\mathcal{I}nv\mathcal{G}a \big(N/2, {\widetilde{b} - \mathfrak{E}(1)} \big)$.
}

\na{
Second, we employ a Laplace approximation \citep{TierneyKadane1986} to compute the integrals in $\mathfrak{I}(\tau^2)$ and $\mathfrak{E}(\tau^2)$, yielding $\mathfrak{E}(1) \approx -f(\rho_{\text{saddle}}^2(1))$, where $\rho_{\text{saddle}}^2(1)$ is the solution of $\partial S(1, \rho^2)/\partial \rho^2 = 0$.
To simplify this saddle-point equation, we employ a third and final approximation, replacing
\begin{equation*}
\begin{aligned}
    h(\rho^2) &\approx \widehat{h}(\rho^2) = \big( \sigma^2 + \rho^2 N/n \big)^{-n/2} \\
    f(\rho^2) &\approx \widehat{f}(\rho^2) = \frac{1}{2} \, \frac{1}{\sigma^2 + \rho^2 N/n} \, \big(\bR^\top \bQ \, \text{diag}\big( \{ {1}/{n_i} \}_{i=1}^n \big) \bQ^\top \bR \big) \\
    &\qquad\qquad\hspace{-4pt} = \frac{1}{2} \, \frac{1}{\sigma^2 + \rho^2 N/n} \, \big(\bR^\top \bQ (\bQ^\top \bQ)^{-1}\bQ^\top \bR \big) \;. 
\end{aligned}
\end{equation*} 
This amounts to approximating the worm-specific neuron counts $n_i$ with their average $N/n$, in all and only occurrences of $\sigma^2 + \rho^2 n_i$. Crucially, the resulting approximate saddle-point equation $\widehat{S}'(1, \rho^2) = 0$ is cubic in $\rho^2$ and can be solved with built-in routines such as \texttt{cubic} from the \texttt{R} package \texttt{RConics}.
}

\na{
The \textsc{px} scheme proposed above is structurally analogous to the parameter-expanded Gibbs sampler of \citet{Murray2013}, but differs in that we approximated the auxiliary update of $\tau^2$.
Despite leaving the conditional $p(\rho^2 \mid \tau^2,\relbar)$ exact, target preservation as in the exact \textsc{px-da} construction is no longer guaranteed.
The resulting sampler rather falls within the framework of approximate-kernel \textsc{mcmc} \citep{johndrow2017optimal}.
Accordingly, the bias of its stationary distribution is controlled by the total-variation distance $\big\|\widehat{p}\,(\tau^2 \mid \,\relbar) - p(\tau^2 \mid \,\relbar)\big\|_{\textsc{tv}}$ between the approximate and true full conditionals of $\tau^2$.
Since \textsc{px} is itself only effective in the regime where $\tau^2$ concentrates around 1, both approximations are evaluated in their region of highest accuracy: the Taylor expansion has error $\mathcal{O} \big((\tau^{-2}-1)^2\big)$ and the Laplace plug-in has error $\mathcal{O}(1/S''(1,\rho_\star^2(1)))$ controlled by the saddle concentration.
Additionally, since $\tau^2$ is purely auxiliary, the practically relevant quantity determining the quality of our approximation is the discrepancy between the induced marginal posteriors for $\rho^2$, rather than the total variation on $\tau^2$.
Empirically, we found the two marginals for the random effect variance to be in close agreement.
The induced bias on inference for $\rho^2$ is therefore expected to be negligible in any regime where \textsc{px} delivers meaningful mixing gains.}

\na{We note that the order in which the Taylor expansion and the Laplace approximation are applied could be inverted.
Applying Laplace directly to $\mathfrak{I}(\tau^2)$ would yield the same leading-order coefficient $\mathfrak{E}(1) \approx -f(\rho_\star^2(1))$, plus a curvature correction of order $1/S''(\rho_\star^2(1))$ involving higher derivatives of $S$ and $f$ at the saddle.
The two sequences therefore agree at leading order in the saddle concentration. The subleading discrepancy is of the same magnitude as the Laplace approximation error already present in both orderings.
We find the Taylor-first version empirically adequate and adopt it for its simpler form, requiring only the saddle value of $f$.}

\subsection{Parameter Expansion for Treatment Coefficients}

\na{
A similar issue affects the mixing of the treatment coefficients $\alpha_{tgc}$.
Here, we apply \textsc{px} only to the slab component $\widetilde{p}_1(\alpha_{tgc} \mid \,\relbar)$ of the zero-inflated truncated-normal full conditional within both regular and collapsed samplers -- whereas the spike component $\widetilde{p}_0(\alpha_{tgc} \mid \,\relbar) = \delta_0(\alpha_{tgc})$ would be unaffected.
We introduce again a scalar working parameter $\tau \in \Re_+$ within the observationally equivalent residual likelihood 
${p({\bR^{(tgc)}} \mid \alpha_{tgc}, \tau^2, \,\relbar)} = \phi_N\big( \bR^{(tgc)} - \tau \, \bX_{\bullet,(tgc)} \alpha_{tgc} \, , \, \tau^2 \bOmega \big)$, and endow it with a diffuse improper prior $p(\tau^2) \propto \tau^{-2}$.
Note that here $\bR^{(tgc)}$ and $\bOmega$ are defined differently for collapsed and marginal samplers, as done in \ref{sec:derivations}.
As before, we aim to decouple the \textsc{px} update as
$\widetilde{p}_1(\alpha_{tgc}, \tau^2 \mid \,\relbar) = \widetilde{p}_1(\alpha_{tgc} \mid \tau^2, \,\relbar) \, \widetilde{p}_1(\tau^2 \mid \,\relbar)$, where the joint full conditional reads
\begin{equation*}
\begin{aligned}
\widetilde{p}_1( \alpha_{tgc}, \tau^2 \mid \,\relbar) &\propto \tau^{-2} \cdot  \frac{\phi(\alpha_{tgc} - \lambda, \nu^2)}{\Phi(\lambda/ \nu)} \, \mathbbm{1}_{(\alpha_{tgc} > 0)} \cdot \det(\tau^2 \bOmega)^{-1/2} \\
&\qquad \exp\!\Big( -\tfrac{1}{2} (\tau^{-1}\bR^{(tgc)} - \bX_{\bullet,(tgc)}\alpha_{tgc})^\top \bOmega^{-1} (\tau^{-1}\bR^{(tgc)} - \bX_{\bullet,(tgc)}\alpha_{tgc}) \Big) \\
\end{aligned}
\end{equation*}
Introducing
\begin{equation*}
\begin{aligned}
    \widetilde{\nu}_{tgc}^2 &= \big( \nu^{-2} + \bX_{\bullet,(tgc)}^\top \bOmega^{-1} \bX_{\bullet,(tgc)} \big)^{-1}
    \qquad\quad
    \widetilde{\lambda}_{tgc}(\tau) = \widetilde{\nu}_{tgc}^2 \big( \nu^{-2}\lambda + \tau^{-1} \bX_{\bullet,(tgc)}^\top \bOmega^{-1} \bR^{(tgc)} \big) 
\end{aligned}
\end{equation*}
to complete the square in $\alpha_{tgc}$ in the exponent, we have
\begin{equation*}
\begin{aligned}
\widetilde{p}_1( \alpha_{tgc}, \tau^2 \mid \,\relbar) &\propto  (\tau^2)^{-N/2-1} \cdot  
\exp\!\bigg( \tau^{-1} \widetilde{\nu}_{tgc}^2 \, \big(\bX_{\bullet,(tgc)}^\top \bOmega^{-1} \bR^{(tgc)}\big) \, \lambda/\nu^2 \bigg) \cdot
\Phi\big(\widetilde{\lambda}_{tgc}(\tau) / \widetilde{\nu}_{tgc}\big) \, \cdot 
\\
&\qquad
\exp\!\bigg(-\tfrac{1}{2} \tau^{-2} \Big(\bR^{(tgc)\top} \bOmega^{-1} \bR^{(tgc)} - \widetilde{\nu}_{tgc}^2 \big( \bX_{\bullet,(tgc)}^\top \bOmega^{-1} \bR^{(tgc)} \big)^2 \Big) \bigg) \cdot
\\
&\qquad \bigg( \frac{\phi\big(\alpha_{tgc} - \widetilde{\lambda}_{tgc}(\tau) , \widetilde{\nu}_{tgc}^2\big)}{\Phi\big(\widetilde{\lambda}_{tgc}(\tau) / \widetilde{\nu}_{tgc}\big)} \, \mathbbm{1}_{(\alpha_{tgc} > 0)} \bigg)
\end{aligned}
\end{equation*}
The last line is recognized as the kernel of a positive truncated normal, giving the conditional full conditional 
${\widetilde{p}_1(\alpha_{tgc} \mid \tau^2, \,\relbar)} = \mathcal{TN}\big(\widetilde{\lambda}_{tgc}(\tau), \widetilde{\nu}_{tgc}^2, (0,\infty)\big)$.
This is analogous to the non-augmented case, with the same variance, but with a scaling factor $\tau^{-1}$ within the location parameter.
}

\na{
Conversely, the auxiliary update $\widetilde{p}_1(\tau^2 \mid \,\relbar) = \int_0^\infty \widetilde{p}_1(\alpha_{tgc}, \tau^2 \mid \,\relbar) \, d\alpha_{tgc} $ reads
\begin{equation*}
\begin{aligned}
\widetilde{p}_1( \tau^2 \mid \,\relbar)
& \propto  (\tau^2)^{-N/2-1} \exp\!\Big(-\tfrac{1}{2}\widetilde{a} \, \tau^{-2} + (\widetilde{c}\cdot\widetilde{d})\, \tau^{-1} + \log \Phi \big(\widetilde{c} \,\tau^{-1}  + \widetilde{d} \,\big)\,\Big)
\end{aligned}
\end{equation*}
where, for readability, we wrote $\widetilde{\lambda}_{tgc}(\tau) / \widetilde{\nu}_{tgc} = \widetilde{c} \,\tau^{-1} + \widetilde{d}$ and introduced
\begin{equation*}
\begin{aligned}
    \widetilde{c} &= \widetilde{\nu}_{tgc} \, \bX_{\bullet,(tgc)}^\top \bOmega^{-1} \bR^{(tgc)}
    \qquad\qquad\qquad
    \widetilde{d} = \widetilde{\nu}_{tgc} \, \lambda/\nu^2 \\[3pt]
    \widetilde{a} &= 
    \bR^{(tgc)\top} \bOmega^{-1} \bR^{(tgc)} - \widetilde{c}^2 =
    \bR^{(tgc)\top} \bOmega^{-1} \bR^{(tgc)} - \widetilde{\nu}_{tgc}^2 \big( \bX_{\bullet,(tgc)}^\top \bOmega^{-1} \bR^{(tgc)} \big)^2 \\
    & \qquad\qquad\qquad\qquad\qquad\hspace{10pt}
    = \bR^{(tgc)\top} \big( \bOmega + \nu \, \bX_{\bullet,(tgc)}^\top\bX_{\bullet,(tgc)} \big)^{-1} \bR^{(tgc)}
    \;.
    \end{aligned}
\end{equation*}
As in the \textsc{px} for $\rho^2$, this density does not allow sampling via a standard routine. 
We thus proceed as before, applying a first-order Taylor expansion at $\tau^{-2} = 1$ (i.e. around the unexpandend model) to restore conditional conjugacy. 
Highlighting their dependence on $\tau^{-2}$, this entails the last two terms in the exponent.
Overall, this gives the approximate full conditional
\begin{equation*}
\begin{aligned}
\widehat{p}_1( \tau^2 \mid \,\relbar)
& \propto  (\tau^2)^{-N/2-1} \exp\!\Big(-\tfrac{1}{2}\widetilde{a} \, \tau^{-2} \Big) \cdot \\
&\qquad \exp\!\bigg( \tau^{-2} \cdot \bigg[ \frac{\partial}{\partial (\tau^{-2})} \Big( (\tau^{-2})^{1/2}\, (\widetilde{c}\cdot\widetilde{d}) + \log \Phi \big( (\tau^{-2})^{1/2} \, \widetilde{c} + \widetilde{d} \,\big) \Big) \bigg]_{\tau^{-2}=1} \,\bigg)
\end{aligned}
\end{equation*}
Simple algebra then leads to the kernel of an inverse-gamma distribution 
$$\mathcal{I}nv\mathcal{G}a \bigg(
\tfrac{1}{2}N, \tfrac{1}{2} \widetilde{a} - \tfrac{1}{2} \widetilde{c} \, \Big( \widetilde{d}+\phi\big( \widetilde{c} + \widetilde{d}\,\big)/\Phi\big(\widetilde{c} + \widetilde{d}\,\big) \Big) \bigg) \,. $$
}

\na{
In contrast to the \textsc{px} scheme for $\rho^2$, only the Taylor expansion is needed here, with no integral approximation required.
This is because the normalizing constant $\Phi(\widetilde\lambda/\widetilde\nu)$ of $\widetilde{p}_1( \alpha_{tgc} \mid \tau^2, \,\relbar)$ is available in closed form.
The same approximate-\textsc{mcmc} analysis \citep{johndrow2017optimal} carries over, with Taylor error $\mathcal{O}\big((\tau^{-2}-1)^2\big)$ in the concentration regime $\tau^2 \approx 1$ where \textsc{px} is expected to be effective.
As discussed in the \textsc{px} scheme for $\rho^2$, the practically relevant quantity determining the quality of the proposed approximation is the discrepancy between the induced marginal posteriors for $\alpha_{tgc}$, which we found to be in close agreement empirically
}

\section{Details on Power Calculations}\label{app:sim_study_details}

In this appendix, we provide additional details of the simulation study we described in Section~\ref{sec:simulations},  including typical runtimes and a description of how power was computed for each considered method.

\na{
We begin by providing full details for the data generation mechanism.
We simulate synthetic data from a latent-Gaussian formulation, corresponding to the \textsc{da} representations of the cumulative probit model.
Specifically, for each neuron $j$ on worm $i$ we draw a latent score
$Z_{ij} \sim \mathcal{N}(\psi_i + \eta_i, \, \sigma^2)$,
with $\eta_i \overset{iid}{\sim} \mathcal{N}(0, \rho^2)$ and $\psi_i$ as in equation~\eqref{Z_model_v0}, and map it to an ordinal damage score $Y_{ij} \in \{0, \dots, L\}$ through a fixed set of cutpoints $\boldsymbol{\delta} = (\delta_0, \dots, \delta_{L-1})$.
The hyperparameters $\boldsymbol{\delta}$, $\boldsymbol{\mu}$, $\sigma^2$, and $\rho^2$ are first calibrated by fitting an unconstrained cumulative probit model with worm-level random effects to the target {\it C.\ elegans} data.
This yields the values 
$\boldsymbol{\delta} = (2.07,\, 2.86,\, 3.28,\, 3.74,\, 4.13,\, 4.74)$, 
$\boldsymbol{\mu} = (0.92,\, 0.56,\, 0.39)$, 
$\sigma^2 = 1$, and $\rho^2 = 2.05$.
The number of neurons per worm is set to four, matching the median experimental design, and the number of worms in each replicate-treatment-generation-rechallenge stratum is drawn from the empirical distribution observed in the original {\it C.\ elegans} assay.
Consistent with the focus introduced above, we set all treatment coefficients to zero except $\alpha_{\text{low}\text{-}P0\text{-}R}$ and $\alpha_{\text{high}\text{-}P0\text{-}R}$, which we vary one at a time over the grid $\{0,\, 0.25,\, 0.5,\, 0.75,\, 1,\, 1.25,\, 1.5,\, 1.75,\, 2,\, 2.25,\, 2.5\}$, holding the other one fixed at zero.
For each of these configurations, we generate 1,000 independent random datasets.
}

\setcounter{savedfigure}{\value{figure}}
\renewcommand{\figurename}{Table}
\addtocounter{table}{1}
\setcounter{figure}{\value{table}}
\begin{figure}[H]
\centering
\begin{tabular}{l|l|r}
    \toprule
    \multicolumn{1}{c|}{Method} & \multicolumn{1}{c|}{$H_0$ Rejection Threshold} & Runtime [sec] \\
    \midrule
    \textsc{bayes-rank}      & $0.086$ & 114.154 \\ 
    \textsc{bayes-cum-logit} & $0.774$ & 993.953 \\ 
    \textsc{iso-cum-logit}   & $0.026$ & 12.893 \\
    \textsc{trend-cum-logit} & $0.025$ & 7.130  \\
    $Multinomial$   & $0.0475$          & 22.035 \\
    $ANOVA$         & $4\times 10^{-4}$ & 0.005 \\
    $Lin.\ Reg.$    & $1\times 10^{-3}$ & 0.003 \\
    $t$-$test$      & $7\times 10^{-4}$ & 0.003 \\
    $\chi^2$-$test$ & $8\times 10^{-3}$ & 0.001 \\
    \bottomrule
\end{tabular}
\caption{\na{Rejection thresholds for each competing method, calibrated via simulation to achieve a nominal Type-I error of 5\%.
Power curves are then obtained by varying the effect size and recording the rejection rate of each method against its own calibrated threshold.
The rightmost column reports typical runtimes in seconds of each considered methodology on the synthetic data considered in our simulation study.}}
\label{tab:test_desc}
\end{figure}
\renewcommand{\figurename}{Fig.}
\setcounter{figure}{\value{savedfigure}}

\na{
In the subsections below, we report the mathematical details for each considered model.
Recall that, to place all models on equal footing, we calibrate the rejection threshold of each method by simulating data under the effective null configuration $\alpha_{\text{low}\text{-}P0\text{-}R} = \alpha_{\text{high}\text{-}P0\text{-}R} = 0$, to yield an empirical Type-I error of approximately $5\%$.
The resulting calibrated thresholds are reported in Table~\ref{tab:test_desc}.
}
\na{
Therein, we also report runtimes in seconds on a typical synthetic dataset from the simulation study for each considered method.
The code was run on a 13-inch MacBook Pro (2018) with a 2.7 GHz quad-core Intel Core i7 processor and 16Gb RAM (macOS 14.6.1).}


\na{
The higher runtime of \textit{Multinomial} relative to \textsc{trend-cum-logit} is consistent with the former requiring three separate model fits rather than one.
For a direct comparison between the two Bayesian competitors, Table~\ref{tab:test_desc} reports the runtime of \textsc{bayes-cum-logit} under the same number of iterations as \textsc{bayes-rank} (15,000 total, 5,000 burn-in), revealing that \textsc{bayes-rank} is almost 10 times faster.
Note that, since both models achieve comparably good mixing, comparisons of effective samples per second are driven primarily by raw wall time.
This computational overhead explains our choice to run fewer iterations for \textsc{bayes-cum-logit} in the full power-curve assessment (6,000 total, 1,000 burn-in), where each routine needs to be executed $\text{1,000} \times 2 \times 10 = \text{20,000}$ times.
Finally, regarding warm-up of \textsc{bayes-rank}, we recommend a two-phase strategy.
Running burn-in under the regular sampler allows reaching the relevant posterior region at a cheaper computational cost -- roughly at half cost per-iteration of the collapsed version, the latter requiring multivariate truncated normal updates and more involved code vectorization.
Switching to the collapsed sampler for the core iterations then ensures reliable posterior inference, leveraging its superior mixing in the small N/n ratio regime characteristic of the {\it C.\ elegans} assay.
}

\subsection{\textsc{bayes-rank}: Bayesian Isotonic Rank Likelihood}

\na{
We fit the proposed model to each simulated dataset via the Gibbs sampler of Algorithms~\ref{alg:collapsed}.
As derived in Section~\ref{sec:simulations}, the single-stratum hypothesis test
$H_0: \beta_{Tgc} = 0$ versus $H_1: \beta_{Tgc} > 0$
reduces to a test on $\beta_{\text{high}\text{-}P0\text{-}R} = \alpha_{\text{low}\text{-}P0\text{-}R} + \alpha_{\text{high}\text{-}P0\text{-}R}$.
We adopt as test statistic the posterior probability of the null,
\begin{equation*}
    \widehat{S}_{\textsc{bayes-rank}} \,=\, \mathbb{P}\big[\, \alpha_{\text{low}\text{-}P0\text{-}R} + \alpha_{\text{high}\text{-}P0\text{-}R} = 0 \,\mid\, \bY \,\big]
    \,\approx\, \frac{1}{Q}\sum_{q=1}^Q \mathbbm{1}_{\big(\alpha_{\text{low}\text{-}P0\text{-}R}^{(q)} + \alpha_{\text{high}\text{-}P0\text{-}R}^{(q)} \, = \, 0\big)} \;,
\end{equation*}
where $q = 1,\dots,Q$ indexes the Gibbs iterates after burn-in and the superscript $^{(q)}$ denotes the $q$-th retained draw.
The null is rejected whenever $\widehat{S}_{\textsc{bayes-rank}} < \tau_{\textsc{bayes-rank}}$, with calibrated threshold $\tau_{\textsc{bayes-rank}} = 0.07$ as reported in Table~\ref{tab:test_desc}.
}


\subsection{\textsc{bayes-cum-logit}: Bayesian Ordinal Logit via \texttt{brms} and \texttt{mo()}}

\na{
We fit a Bayesian isotonic cumulative-logit mixed model leveraging the wrapper \texttt{mo()} from the \texttt{R} package \texttt{brms} \citep{burkner2020mo}.
As described in Section~\ref{sec:simulations}, the \texttt{mo($\,$)} wrapper reparametrizes the cumulative dose effects as $\beta_{tgc} \,=\, T \, \widetilde{\beta}_{gc} \, \textstyle{\sum_{s=1}^t} \zeta_{sgc}$ ,
with overall magnitude $\widetilde{\beta}_{gc} \geq 0$ and normalized increments $\boldsymbol{\zeta}_{\bullet gc} = (\zeta_{1gc}, \dots, \zeta_{Tgc}) \sim \mathcal{D}ir(k)$ on the simplex $\sum_t \zeta_{tgc} = 1$, separately for each rechallenge--generation stratum.
The linear predictor $\psi_i + \eta_i$ thus has
\begin{equation*}
\begin{aligned}
\psi_i &= \sum_{r=1}^R \mu_r \, \mathbbm{1}_{\big(\text{rep}(i)= r \big)} + \sum_{t=1}^T \sum_{g=0}^G \sum_{c=0}^C
\, T \, \widetilde{\beta}_{gc} \, \zeta_{tgc} \,
\mathbbm{1}_{\big(\text{treat}(i) \geq t \big)}
\mathbbm{1}_{\big(\text{gen}(i) = g \big)}
\mathbbm{1}_{\big(\text{rech}(i) = c \big)} \;,
\end{aligned}
\end{equation*}
Restricting attention to the rechallenged parent stratum, the hypothesis test reduces to $H_0: \widetilde{\beta}_{P0\text{-}R} = 0$ versus $H_1: \widetilde{\beta}_{P0\text{-}R} > 0$.
We adopt as test statistic the posterior mean
\begin{equation*}
    \widehat{S}_{\texttt{brms+mo}} \,=\, \mathbb{E}\big[\widetilde{\beta}_{P0\text{-}R} \mid \bY\big] \,\approx\, \frac{1}{Q}\sum_{q=1}^Q \widetilde{\beta}_{P0\text{-}R}^{(q)} \;,
\end{equation*}
where $\widetilde{\beta}_{P0\text{-}R}^{(q)}$ denotes the $q$-th post-burn-in draw.
The null is rejected whenever $\widehat{S}_{\texttt{brms+mo}} > \tau_{\texttt{brms+mo}}$, with calibrated threshold $\tau_{\texttt{brms+mo}}$ reported in Table~\ref{tab:test_desc}.
As mentioned in Section~\ref{sec:simulations}, mixing was found to be brittle and highly sensitive to prior choices; satisfactory performance was obtained under
$\mu_r \sim \mathcal{N}(0,3)$, $\boldsymbol{\zeta}_{\bullet gc} \sim \mathcal{D}ir(\mathbf{1})$, $\widetilde{\beta}_{gc} \sim \mathcal{G}a(2,1)$, and $\rho \sim \mathrm{t}_3^+(0,2)$, where $\mathrm{t}_3^+(0,2)$ denotes a half-Student-$t$ with $3$ degrees of freedom, location $0$, and scale $2$.
}

\subsection{\textsc{iso-cum-logit}: Constrained Ordinal Logit \& Likelihood Ratio Test}

\na{
We fit a frequentist isotonic cumulative-logit mixed model that enforces $\alpha_{tgc}\geq 0$ and test
$H_0: \boldsymbol{\alpha} = \boldsymbol{0}$ against $H_1: \boldsymbol{\alpha} \geq \boldsymbol{0}$ via the likelihood-ratio statistic
\begin{equation*}
    \mathcal{T} \,=\, 2 \big[ \mathfrak{L}(\hat{\boldsymbol{\gamma}}_1) - \mathfrak{L}(\hat{\boldsymbol{\gamma}}_0) \big] \;,
\end{equation*}
where $\mathfrak{L}(\boldsymbol{\gamma}) = \log \int p(\bY \mid \boldsymbol{\gamma}, \boldsymbol{\eta}) \, p(\boldsymbol{\eta} \mid \rho^2) \, d\boldsymbol{\eta}$ denotes the marginal log-likelihood with the random effects $\boldsymbol{\eta}$ integrated out, and $\hat{\boldsymbol{\gamma}}_0$ and $\hat{\boldsymbol{\gamma}}_1$ are the maximum-likelihood estimators of $\boldsymbol{\gamma} = (\boldsymbol{\alpha}^\top, \boldsymbol{\mu}^\top, \rho^2)^\top$ under $H_0$ and $H_1$, respectively.
We obtain the constrained and unconstrained maximizers as described in Section~\ref{sec:simulations}, by extracting $\mathfrak{L}(\boldsymbol{\gamma})$ from \texttt{ordinal::clmm} and optimizing it under the relevant constraints via \texttt{nlminb}.
}

\na{
We restrict attention to the rechallenged parent stratum, so that only the subvector $\boldsymbol{\alpha}_{P0\text{-}R} = (\alpha_{\text{low}\text{-}P0\text{-}R}, \alpha_{\text{high}\text{-}P0\text{-}R})^\top$ is constrained, with $B = 2$ constrained components.
Under $H_0$, the likelihood-ratio statistic is asymptotically distributed as a chi-bar-squared \citep{agresti2002analysis,davidov2011constrained,davis2011constrained}
\begin{equation*}
    \mathcal{T} \,\sim\, \bar{\chi}^2 \,=\, \textstyle{\sum_{d=0}^B} \, \rho_d(B, \mathbf{V}) \, \chi^2_d \;.
\end{equation*}
The mixing weights admit a closed-form representation  
\begin{equation*}
    \rho_d(B, \mathbf{V}) \,=\, \sum_{\zeta \subset \{1,\dots,B\}, \, |\zeta|=d} \mathbb{P}\big[ \bZ^{\star}_\zeta > \boldsymbol{0} \big] \, \mathbb{P}\big[ \bZ_\zeta > \boldsymbol{0} \mid \bZ^{\star}_\zeta = \boldsymbol{0} \big] 
\end{equation*}
involving multivariate Gaussian orthant probabilities, where $\bZ \sim \mathcal{N}_B(\boldsymbol{0}, \mathbf{V})$ partitions into the sub-vector $\bZ_\zeta = (Z_s)_{s \in \zeta}$ on the active set $\zeta$ and $\bZ^{\star}_\zeta = (Z_s)_{s \notin \zeta}$ on its complement, and $\mathbf{V}$ is the asymptotic covariance matrix of the constrained components $\boldsymbol{\alpha}_{P0\text{-}R}$ under $H_0$ \citep{davis2011constrained}.
In practice, $\mathbf{V}$ is unknown and replaced by a plug-in estimate $\widehat{\mathbf{V}}$ obtained as the $\boldsymbol{\alpha}_{P0\text{-}R}$-block of the inverse Hessian of $-\mathfrak{L}(\boldsymbol{\gamma})$ at $\hat{\boldsymbol{\gamma}}_0$, returned by \texttt{nlminb} after constrained optimization.
}

\na{
The weights $\rho_d(B, \widehat{\mathbf{V}})$ are then computed via the \texttt{R} package \texttt{ic.infer}.
We note that direct evaluation of $\rho_d(B, \widehat{\mathbf{V}})$ relies on numerical orthant probabilities of $B$-dimensional Gaussians, which is computationally feasible only for small $B$.
Bootstrap-based approximations are recommended for larger constrained parameter spaces.
In our single-stratum setting $B = 2$, so the weights are computable in closed form.
}

\na{
The test statistic is
$
    \widehat{S}_{\textsc{iso-cum-logit}} \,=\, 1 - \bar{F}\big( \mathcal{T} ; \, R, \widehat{\mathbf{V}} \big) \;,
$
where $\bar{F}(\cdot; R, \mathbf{V})$ denotes the cumulative distribution function of $\bar{\chi}^2$, evaluated via the weights above.
The null is rejected whenever $\widehat{S}_{\textsc{iso-cum-logit}} < \tau_{\textsc{iso-cum-logit}}$, with $\tau_{\textsc{iso-cum-logit}}$ calibrated as reported in Table~\ref{tab:test_desc}.
}
\na{
Recall that, to the best of our knowledge, no off-the-shelf package jointly accommodates all three substantive features required by our setting (ordinal Logit, random effects, parameter constraints), as detailed in Table~\ref{tab:packages_coverage}.
This motivated the custom optimization routine described above and in Section~\ref{sec:simulations}.
}

\setcounter{savedfigure}{\value{figure}}
\renewcommand{\figurename}{Table}
\addtocounter{table}{1}
\setcounter{figure}{\value{table}}
\begin{figure}[H]
\centering
\begin{tabular}{l|cccc}
    \toprule
    \texttt{R} Package & GLM & Ordinal Logit & Random Effects & Constraints \\
    \midrule
    \texttt{ordinal} & \cmark & \cmark & \cmark & \xmark \\
    \texttt{restriktor}   & \cmark & \xmark & \xmark & \cmark \\
    \texttt{ic.infer}     & \xmark & \xmark & \xmark & \cmark \\
    \texttt{CLME}         & \xmark & \xmark & \cmark & \cmark \\
    \texttt{varTestnlme}  & \cmark & \xmark & \cmark & \cmark \\
    \texttt{glmmTMB}      & \cmark & \xmark & \cmark & \cmark \\
    \texttt{lavaan}       & \cmark & \pmark & \pmark & \cmark \\
    \bottomrule
\end{tabular}
\caption{\na{Coverage of available \texttt{R} packages for fitting frequentist cumulative-logit mixed models with order restrictions on the regression coefficients. The four columns indicate whether each package supports, respectively: generalized linear models (GLM), the cumulative-logit link for ordinal responses (Ordinal Logit), random effects (Random Effects), and order restrictions on regression coefficients (Constraints). The symbol \cmark{} denotes a supported feature, \xmark{} a not supported one, and \pmark{} a conditionally supported one.
Specifically, the \texttt{lavaan} package supports both ordinal logit and random effects separately, but not jointly.}}
\label{tab:packages_coverage}
\end{figure}
\renewcommand{\figurename}{Fig.}
\setcounter{figure}{\value{savedfigure}}

\subsection{\textsc{trend-cum-logit}: Unrestricted Ordinal Logit \& Contrast-Based Test}

\na{
We fit an unconstrained cumulative-logit mixed model with $\beta_{tgc} \in \Re$ via \texttt{ordinal::clmm} and test for a positive trend through contrasts on the unconstrained maximum-likelihood estimates $\widehat{\beta}_{tgc}^{(u)}$.
Restricting attention to the rechallenged parent stratum, for any prespecified weight vector $\boldsymbol{\varpi} = (\varpi_1, \varpi_2)^\top$, we define the contrast
\begin{equation*}
    \psi_{P0\text{-}R}(\boldsymbol{\varpi}) \,=\, \varpi_1\, \beta_{\text{low}\text{-}P0\text{-}R} + \varpi_2\, \beta_{\text{high}\text{-}P0\text{-}R} \;,
\end{equation*}
and test $H_0: \psi_{P0\text{-}R}(\boldsymbol{\varpi}) \leq 0$ against $H_1: \psi_{P0\text{-}R}(\boldsymbol{\varpi}) > 0$ via the one-sided Wald statistic
\begin{equation*}
    \mathcal{T}(\boldsymbol{\varpi}) \,=\, \frac{ \widehat{\psi}_{P0\text{-}R}(\boldsymbol{\varpi}) }{ \widehat{\mathrm{SE}}\big(\widehat{\psi}_{P0\text{-}R}(\boldsymbol{\varpi})\big) } \,\sim\, \mathcal{N}(0,1) \quad \text{under } H_0 \;,
\end{equation*}
where the standard error is obtained from the asymptotic covariance of $\widehat{\boldsymbol{\beta}}^{(u)}$, namely $\widehat{\mathrm{SE}}^2(\widehat{\psi}) = \boldsymbol{\varpi}^\top \widehat{\mathrm{Cov}}(\widehat{\boldsymbol{\beta}}_{P0\text{-}R}^{(u)}) \, \boldsymbol{\varpi}$.
As for \textsc{iso-cum-logit}, $\widehat{\mathrm{Cov}}(\widehat{\boldsymbol{\beta}}_{P0\text{-}R}^{(u)})$ is obtained as the corresponding block of the inverse Hessian of $-\mathfrak{L}(\boldsymbol{\gamma})$ at the unconstrained MLE returned by \texttt{nlminb}.
}

\na{
A delicate choice in this approach is the weight vector $\boldsymbol{\varpi}$. Different contrasts are sensitive to different dose-response shapes, and no single choice dominates uniformly across plausible monotone alternatives \citep{farrar2022comparison}. Common choices in the toxicology literature include linear weights $\boldsymbol{\varpi} \propto (1, 2, \dots, T) - \sum_{t=1}^T t/T$ proportional to dose ranks, Helmert-type weights comparing each dose to the average of lower doses, and Stewart--Ruberg weights designed for sensitivity to convex or concave patterns.
}

\na{
Contrary to a diffused practice, we do not constrain the weights $\boldsymbol{\varpi}$ to sum to zero, and exclude the control group in the contrast -- since we always enforce null coefficients in the model. 
To mitigate sensitivity to the choice of $\boldsymbol{\varpi}$, we follow standard practice and adopt a multi-contrast strategy combining several complementary alternatives:
\begin{itemize}
    \item[(i)] {Uniform weights}: $\boldsymbol{\varpi}_{\text{u}} = (0.5, 0.5)$, comparing the average of treated doses against zero;
    \item[(ii)] {Low-pooled weights}: $\boldsymbol{\varpi}_{\text{lp}} = (0, 1)$, comparing the high dose against the control while pooling the low dose with the control (sensitive to threshold-type responses where only the high dose elicits damage).
\end{itemize}
For each contrast, we compute the one-sided $p$-value $p_k = 1 - \Phi(\mathcal{T}(\boldsymbol{\varpi}_k))$, $k \in \{\text{u},\text{lp}\}$, and combine them through a Bonferroni adjustment.
The resulting test statistic is
$\widehat{S}_{\textsc{trend-cum-logit}} \,=\, K \cdot \min_{k} p_k $, with $K = 2$ the number of contrasts considered.
The null is rejected whenever $\widehat{S}_{\textsc{trend-cum-logit}} < \tau_{\textsc{trend-cum-logit}}$, with $\tau_{\textsc{trend-cum-logit}}$ calibrated as reported in Table~\ref{tab:test_desc}.
}

\subsection{$Multinomial$: Nested Unrestricted Ordinal Logit Models}

\na{
We fit three nested unconstrained cumulative-logit mixed models on the ordinal damage scores via \texttt{clmm2} \citep{christensen2019tutorial}, that is $\bbP[Y_{ij} \leq \ell] = G\big( \delta_\ell - \psi_i - \eta_i \big)$ with link function $G(r)=e^{r}/(1+e^{r})$, worm-level random effects $\eta_i \overset{iid}{\sim} \mathcal{N}(0, \rho^2)$, and
\begin{equation*}
\psi_i = \textstyle{\sum_{r=1}^R} \mu_r \, \mathbbm{1}_{(\text{rep}(i) = r)} + \textstyle{\sum_{t=1}^T  \sum_{g=0}^G \sum_{c=0}^C} \beta_{tgc} \,
    \mathbbm{1}_{\big(\text{treat}(i) = t \big)}
    \mathbbm{1}_{\big(\text{gen}(i) = g \big)}
    \mathbbm{1}_{\big(\text{rech}(i) = c \big)} \; .
\end{equation*}
The nested model differs only in the restrictions on the three-way coefficients in the rechallenged parent stratum.
The base model set $\beta_{\text{low}\text{-}P0\text{-}R} = \beta_{\text{high}\text{-}P0\text{-}R} = 0$, (i.e. no treatment effects in the target stratum).
The intermediate one keeps $\beta_{\text{low}\text{-}P0\text{-}R}$ free, but set $\beta_{\text{high}\text{-}P0\text{-}R} = 0$ (i.e. allowing only the low-dose effect).
The full model keeps both $\beta_{\text{low}\text{-}P0\text{-}R}$ and $\beta_{\text{high}\text{-}P0\text{-}R}$ free.

Both intermediate and full models are then separately compared to the null one via a likelihood-ratio $\chi^2$ test implemented through \texttt{anova}, yielding two $p$-values ($p_1$ and $p_2$).
The test statistic is the Bonferroni-adjusted minimum
$
    \widehat{S}_{\textit{Multinomial}} \,=\, K \cdot \min_{k=1,2} p_k 
$
-- with $K = 2$.
The null is rejected whenever $\widehat{S}_{\textit{Multinomial}} < \tau_{\textit{Multinomial}}$, with $\tau_{\textit{Multinomial}}$ as in Table~\ref{tab:test_desc}.
}

\subsection{ANOVA}

\na{
We fit two nested linear regression models on the log-damage score, with treatment treated as a categorical predictor and three-way interaction effects $\beta_{tgc}$ across treatment level, generation, and rechallenge strata
\begin{equation*}
\begin{aligned}
& \log (1+Y_{ij}) \overset{iid}{\sim} \mathcal{N} \big(\psi_i \,, \sigma^2 \big)
\\[3pt]
\psi_i &= \textstyle{\sum_{r=1}^R} \mu_r \, \mathbbm{1}_{(\text{rep}(i) = r)} + \textstyle{\sum_{t=1}^T  \sum_{g=0}^G \sum_{c=0}^C} \, \beta_{tgc} \,
    \mathbbm{1}_{\big(\text{treat}(i) = t \big)}
    \mathbbm{1}_{\big(\text{gen}(i) = g \big)}
    \mathbbm{1}_{\big(\text{rech}(i) = c \big)}
\end{aligned} \;,
\end{equation*}
fit via \texttt{lm}.
The full model includes all $\beta_{tgc}$, while the reduced model sets $\beta_{\text{low}\text{-}P0\text{-}R} = \beta_{\text{high}\text{-}P0\text{-}R} = 0$, removing the treatment effects in the rechallenged parent stratum.
The two models are compared via an $F$-test implemented through \texttt{anova}, with test statistic
$
    \widehat{S}_{\textit{ANOVA}} \,=\, 1 - F_{F_{(\nu_1, \nu_2)}}\!\big(\mathcal{T}_{\textit{ANOVA}}\big) \;,
$
where $\mathcal{T}_{\textit{ANOVA}}$ is the $F$ statistic and $F_{F_{(\nu_1, \nu_2)}}$ denotes the cumulative distribution function of an $F$ distribution with $(\nu_1, \nu_2)$ degrees of freedom.
The null is rejected whenever $\widehat{S}_{\textit{ANOVA}} < \tau_{\textit{ANOVA}}$, with $\tau_{\textit{ANOVA}}$ as in Table~\ref{tab:test_desc}.
}

\newpage

\subsection{Linear Regression}

\na{We fit a linear regression model for the log-damage score $\log Y_{ij}$ on treatment exposure -- treated as a continuous predictor -- with three-way interactions across generation and rechallenge strata
\begin{equation*}
\begin{aligned}
& \log (1+Y_{ij}) \overset{iid}{\sim} \mathcal{N} \big(\psi_i \,, \sigma^2 \big)
\\[3pt]
\psi_i &= \textstyle{\sum_{r=1}^R} \mu_r \, \mathbbm{1}_{(\text{rep}(i) = r)} + \textstyle{\sum_{g=0}^G \sum_{c=0}^C} \beta_{gc} \cdot \text{dose}_i \cdot \mathbbm{1}_{(\text{gen}(i) = g)} \mathbbm{1}_{(\text{rech}(i) = c)} + \varepsilon_{ij}
\end{aligned} \;,
\end{equation*}
fit via \texttt{lm}, where with $\text{dose}_i \in \{0, 0.03, 0.5 \}$ for control, 0.03 $\mu M$, and 0.5 $\mu M$ treatment subgroups, respectively.
We extract the two-sided $p$-value for the coefficient $\beta_{P0\text{-}R}$ of interest from the \texttt{lm} output and use it as test statistic
$
    \widehat{S}_{\textit{Lin.\ Reg.}} \,=\, 2\big(1 - \Phi\!\big(\vert \mathcal{T}_{\textit{Lin.\ Reg.}} \vert\big)\big) \;,
$
where $\mathcal{T}_{\textit{Lin.\ Reg.}} = \widehat{\beta}_{P0\text{-}R} / \widehat{\mathrm{SE}}(\widehat{\beta}_{P0\text{-}R})$ is the corresponding Wald statistic.
The null is rejected whenever $\widehat{S}_{\textit{Lin.\ Reg.}} < \tau_{\textit{Lin.\ Reg.}}$, with $\tau_{\textit{Lin.\ Reg.}}$ as in Table~\ref{tab:test_desc}.
}

\subsection{$t$-Test}

\na{Within the rechallenged parent stratum}, we run 3 pairwise $t$-tests via the \texttt{R} function \texttt{pairwise.t.test}, comparing the log-damage scores across control, 0.03 $\mu M$, and 0.5 $\mu M$ treatment subgroups.
To correct for multiple testing, we perform a Bonferroni adjustment on the $K = 3$ pairwise test.
\na{The overall test statistic is then
$
    \widehat{S}_{t\text{-test}} \,=\, \min_{k = 1, \dots, K} \, K \cdot p_k \;,
$
where $p_k$ is the unadjusted $p$-value of the $k$-th pairwise comparison.
The null is rejected whenever $\widehat{S}_{t\text{-test}} < \tau_{t\text{-test}}$, with $\tau_{t\text{-test}}$ as in Table~\ref{tab:test_desc}.
}

\subsection{$\chi^2$-Test}

\na{
Restricting attention again to the rechallenged parent stratum, we test for association between treatment level and (binned) log-damage scores via a Pearson $\chi^2$ test on the resulting contingency table, implemented through \texttt{chisq.test}.
We adopt as test statistic the associated $p$-value
$
    \widehat{S}_{\chi^2\text{-test}} \,=\, 1 - F_{\chi^2_{(\nu)}}\!\big(\mathcal{T}_{\chi^2}\big) \;,
$
where $\mathcal{T}_{\chi^2}$ is the Pearson statistic and $F_{\chi^2_{(\nu)}}$ denotes the cumulative distribution function of a $\chi^2$ distribution with $\nu$ degrees of freedom.
The null is rejected whenever $\widehat{S}_{\chi^2\text{-test}} < \tau_{\chi^2\text{-test}}$, with $\tau_{\chi^2\text{-test}}$ as in Table~\ref{tab:test_desc}.
}

\section{Further Results for the {\it C. elegans} assay}\label{app:performance}

\na{This appendix complements the analysis from Section~\ref{sec:analysis}}, providing further results on the {\it C. elegans} assay and a detailed overview of the model performance.

\subsection{Goodness-of-Fit Assessment}\label{sec:eff_size}

In the plots \na{from Section~\ref{sec:analysis}}, we are able to ascertain some insights about which effects are stronger than others and which ones are meaningful in a Bayesian sense. However, all of the aforementioned analyses are conducted on a latent scale, making it difficult to interpret the observable impact. We want to translate this to an effect size on the observed scale,
\na{
by generating damage score predictions from our model.
A complicating factor is that the rank likelihood framework does not directly model $\bY$ given $\bZ$, so generating predictions requires a sensible mapping from latent to observed scale.
Two main strategies have been proposed in the literature.
The first is purely rank-based, as implemented in the \texttt{R} package \texttt{mtlm} \citep{hoff2026mtlm}: rather than estimating cutpoints, each posterior draw of the latent score for a new observation is compared against the latent draws of the training observations, and the predicted ordinal category is assigned by majority vote among the observed scores of the latent neighbors.}

\na{
The second strategy directly estimates the cutpoints $\boldsymbol{\widehat{\delta}} = (\widehat{\delta}_0, \dots, \widehat{\delta}_{L-1})$ of $\bZ$ and uses them to map the predicted latent score to an observed category.
\citet{Murray2013} estimate these cutpoints by inverting the empirical cumulative distribution function of the observed responses, setting $\widehat{\delta}_\ell = \mathrm{F}^{-1}\!\big( \widehat{\mathrm{F}}_Y(\ell) \big)$, where $\widehat{\mathrm{F}}_Y(\ell) = N^{-1} \sum_{ij} \mathbbm{1}_{(Y_{ij} \leq \ell)}$ is the empirical c.d.f.\ of the observed scores and $\mathrm{F}^{-1}$ is the quantile function of the latent distribution.
Alternatively, the $\widehat{\delta}_\ell$ can be estimated from the in-sample draws of the latent scores as
\begin{equation}\label{eq_cutpoints}
    \widehat{\delta}_\ell = \frac{1}{Q}\sum_{q=1}^Q \delta_\ell^{(q)}, \qquad
    \delta_\ell^{(q)} = \max \bZ_{\mathcal{G}(\ell)}^{(q)}, \qquad
    \forall \, \ell=0,\ldots, L-1 \, , \;\; \forall \, q=1,\ldots, Q \;,
\end{equation}
where $q=1,\ldots, Q$ indexes the Gibbs iterates after burn-in and $\bZ_{\mathcal{G}(\ell)}^{(q)}$ denotes draws in the $q$-th iteration.
Regardless of the cutpoint estimation strategy, damage score predictions for a given statistical unit are obtained by sampling all components of the linear predictor to generate latent scores
$ \widetilde{Z}^{(q)} = \widetilde{\psi}^{(q)} + \widetilde{\eta}^{(q)} + \widetilde{\epsilon}^{(q)}$,
where the tildes denote posterior predictive draws, and mapping each $\widetilde{Z}^{(q)}$ to the observed scale through the estimated cutpoints: $\widetilde{Y}^{(q)} = \ell$ if $\widehat{\delta}_{\ell-1} < \widetilde{Z}^{(q)} \leq \widehat{\delta}_\ell$, with $\widehat{\delta}_{-1} = -\infty$ and $\widehat{\delta}_L = +\infty$.
}

\begin{figure}[ht!]
\centering
    \includegraphics[width=0.49\textwidth]{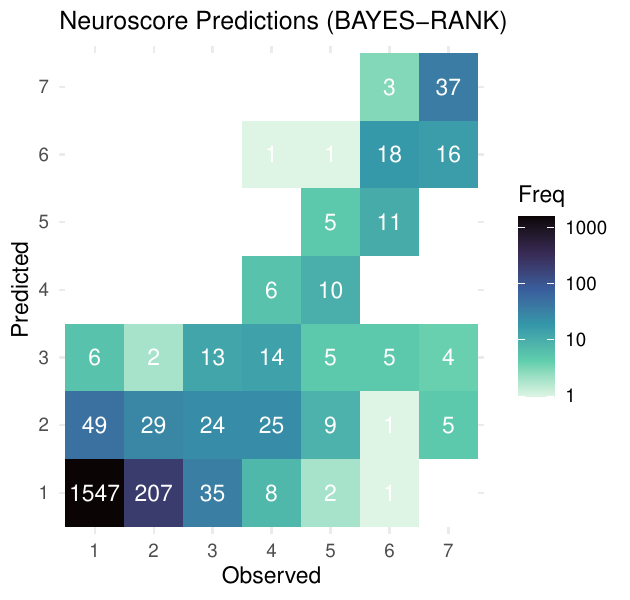}
    \put(-225,210){\makebox(0,0){\textbf{(A)}}}
    \includegraphics[width=0.49\textwidth]{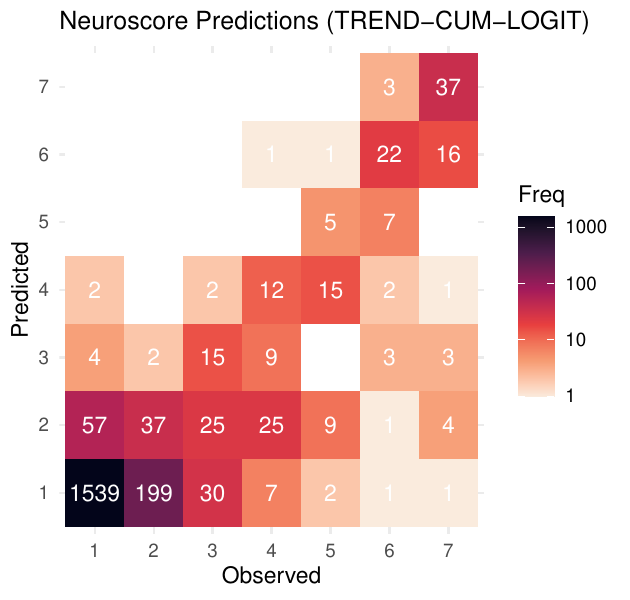}
    \put(-225,210){\makebox(0,0){\textbf{(B)}}}
    \vspace{-10pt}
    \caption{\na{Confusion matrices for in-sample damage score predictions in the {\it C. elegans} assay, supporting the goodness-of-fit of the proposed procedure.
    {(A)} Predictions from our isotonic rank-likelihood model, obtained by mapping posterior predictive draws of the latent score $\widetilde{Z}^{(q)}$ to the observed scale via the estimated cutpoints of equation~\eqref{eq_cutpoints}.
    {(B)} Predictions from an unconstrained ordinal logit model.
    Cell counts are displayed on a $\log_{10}$ color scale, with white cells corresponding to zero counts.}}
    \label{fig:predictions}
\end{figure}

\na{
Figure~\ref{fig:predictions} displays the resulting confusion matrix, with cutpoints estimated as in equation~\eqref{eq_cutpoints} and final score predictions obtained as the posterior median of  $\{ \widetilde{Y}^{(q)} \}_{q=1}^Q$.
We compare it against in-sample predictions from an unconstrained ordinal logit model with worm-specific random intercepts fit with the \texttt{R} package \texttt{ordinal} \citep{christensen2019tutorial}, as in the \textsc{trend-cum-logit} competitor discussed above.
The two confusion matrices are in close agreement, confirming the goodness-of-fit of our procedure.
}

\subsection{Model Parameters: Mixing and Summary statistics}

\na{Table \ref{tab:rep_icc} provides quantitative summaries for some of the figures in Section~\ref{sec:analysis}. It also reports}
the $\widehat{R}$ convergence diagnostic and the bulk and tail effective sample sizes (ESS) \citep{vehtari2021rank}. The $\widehat{R}$ being close to or equal to 1 indicates that almost all of the parameters converged to their stationary distributions. 
\na{The ESS shows good mixing.}

Figure \ref{fig:coeff_increment} also reports the marginal posteriors for the incremental treatment effects $\alpha_{tgc}$, complementing the information on the cumulative ones $\beta_{tgc}$ from Figure~\ref{ref:treatment}.
Figure~\ref{fig:coeff_trace} provides trace plots and autocorrelation functions (\textsc{acf}) of the model parameters, confirming generally good mixing.

\newpage

\setcounter{savedfigure}{\value{figure}}
\renewcommand{\figurename}{Table}
\addtocounter{table}{1}
\setcounter{figure}{\value{table}}
\begin{figure}[H]
\centering
\begin{tabular}{rllllrrl}
\toprule
& Prob. = 0 & Median & 2.5\% & 97.5\% & Bulk ESS & Tail ESS & $\widehat{R}$ \\
\midrule
$\alpha_{\text{low}\text{-}P0\text{-}U}$ & 0.883 & 0.1 & 0 & 0.36 & 8820 & 8807 & 1 \\ 
$\alpha_{\text{high}\text{-}P0\text{-}U}$ & 0.683 & 0.25 & 0.02 & 0.7 & 6353 & 7348 & 1 \\ 
$\beta_{\text{high}\text{-}P0\text{-}U}$ & 0.596 & 0.22 & 0.01 & 0.68 & 6309 & 7378 & 1 \\ 
\midrule
$\alpha_{\text{low}\text{-}F1\text{-}U}$ & 0.913 & 0.08 & 0 & 0.29 & 9527 & 9372 & 1 \\ 
$\alpha_{\text{high}\text{-}F1\text{-}U}$ & 0.884 & 0.1 & 0 & 0.39 & 9455 & 9298 & 1 \\ 
$\beta_{\text{high}\text{-}F1\text{-}U}$ & 0.805 & 0.1 & 0 & 0.37 & 9765 & 9757 & 1 \\ 
\midrule
$\alpha_{\text{low}\text{-}P0\text{-}R}$ & 0.526 & 0.32 & 0.02 & 0.75 & 2434 & 3838 & 1.001 \\ 
$\alpha_{\text{high}\text{-}P0\text{-}R}$ & 0.038 & 0.94 & 0.28 & 1.52 & 3213 & 4156 & 1 \\ 
$\beta_{\text{high}\text{-}P0\text{-}R}$ & 0 & 1.06 & 0.5 & 1.58 & 5083 & 8027 & 1 \\ 
\midrule
$\alpha_{\text{low}\text{-}F1\text{-}R}$ & 0.032 & 0.84 & 0.26 & 1.32 & 1344 & 2532 & 1.002 \\ 
$\alpha_{\text{high}\text{-}F1\text{-}R}$ & 0.324 & 0.59 & 0.06 & 1.32 & 1670 & 2894 & 1 \\ 
$\beta_{\text{high}\text{-}F1\text{-}R}$ & 0 & 1.2 & 0.77 & 1.68 & 2739 & 7832 & 1.001 \\ 
\midrule
Replicate 1 & -- & 0.29 & 0.09 & 0.49 & 4357 & 7653 & 1 \\ 
Replicate 2 & -- & -0.07 & -0.27 & 0.14 & 4505 & 7791 & 1 \\ 
Replicate 3 & -- & -0.22 & -0.43 & -0.02 & 5030 & 8044 & 1.001 \\ 
  \midrule
$\rho^2$ & -- & 1.87 & 1.48 & 2.27 & 684 & 2962 & 1 \\ 
ICC & -- & 0.65 & 0.6 & 0.69 & 684 & 2962 & 1 \\ 
  \bottomrule
\end{tabular}
\caption{
\na{Summary statistics for the posterior distributions of treatment coefficients, replicate effects, within-worm variance $\rho^2$, and intra-class correlation coefficient (ICC), including effective sample size diagnostics. 
Recall that 
$\beta_{\text{high}\text{-}[P0/F1]\text{-}[U/R]}=\alpha_{\text{low}\text{-}[P0/F1]\text{-}[U/R]}+\alpha_{\text{high}\text{-}[P0/F1]\text{-}[U/R]}$
is the cumulative effect at high developmental dose within the associated substratum.
``Prob.\ = 0'' reports the posterior probability that a treatment coefficient equals zero; the posterior median and credible interval refer to the truncated-normal slab component. Bulk and tail ESS, and $\widehat{R}$ statistic are reported per chain.}}
\label{tab:rep_icc}
\end{figure}
\renewcommand{\figurename}{Fig.}
\setcounter{figure}{\value{savedfigure}}

\vspace{20pt}

\begin{figure}[H]
\centering
    \includegraphics[width = \linewidth]{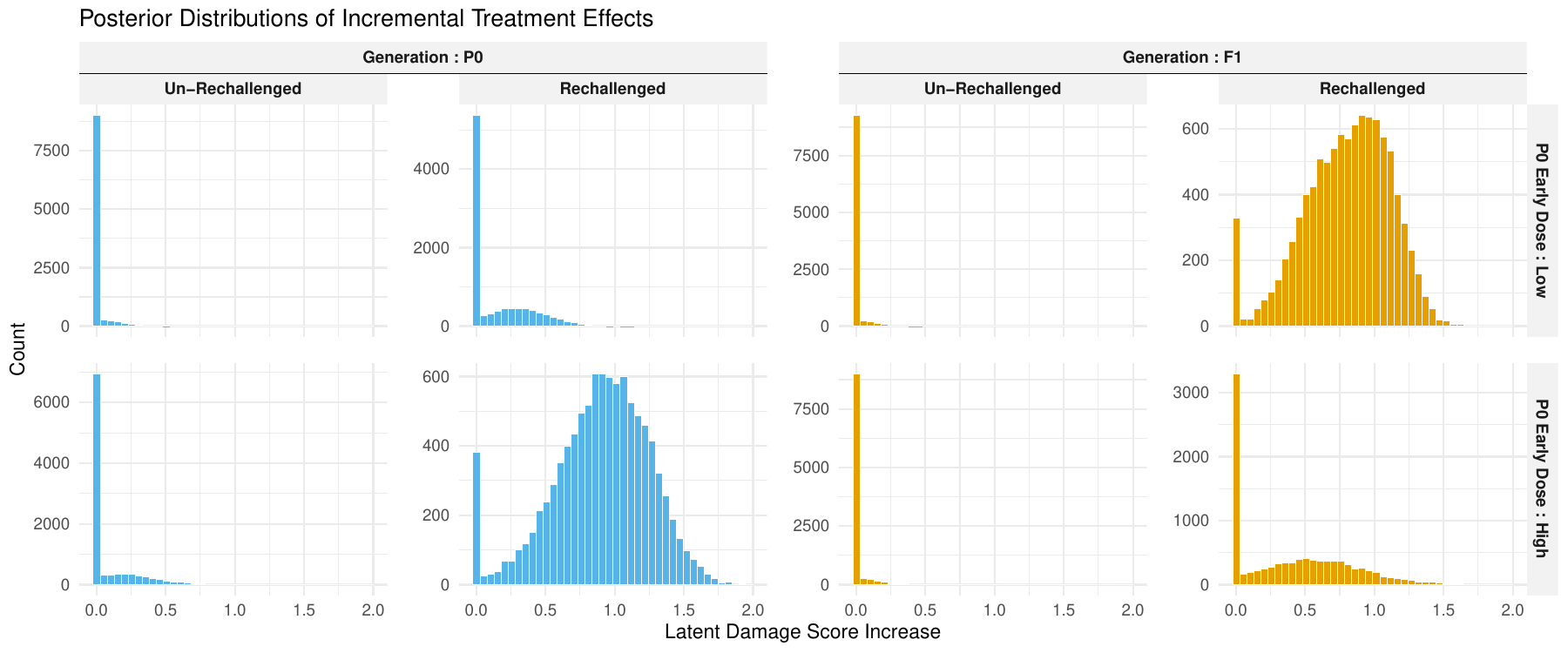}
    \vspace{-25pt}
    \caption{
    Marginal posterior distributions of incremental treatment effects $\alpha_{tgc}$ for each sub-group of worms. 
    The first row coincides with the first row of Figure~\ref{ref:treatment}.
    The second row of Figure~\ref{ref:treatment} aggregates both rows from here into a single cumulative effect.}
    \label{fig:coeff_increment}
\end{figure}

\newpage

~\vspace{10pt}

\begin{figure}[H]
\centering
    \includegraphics[width=\textwidth]{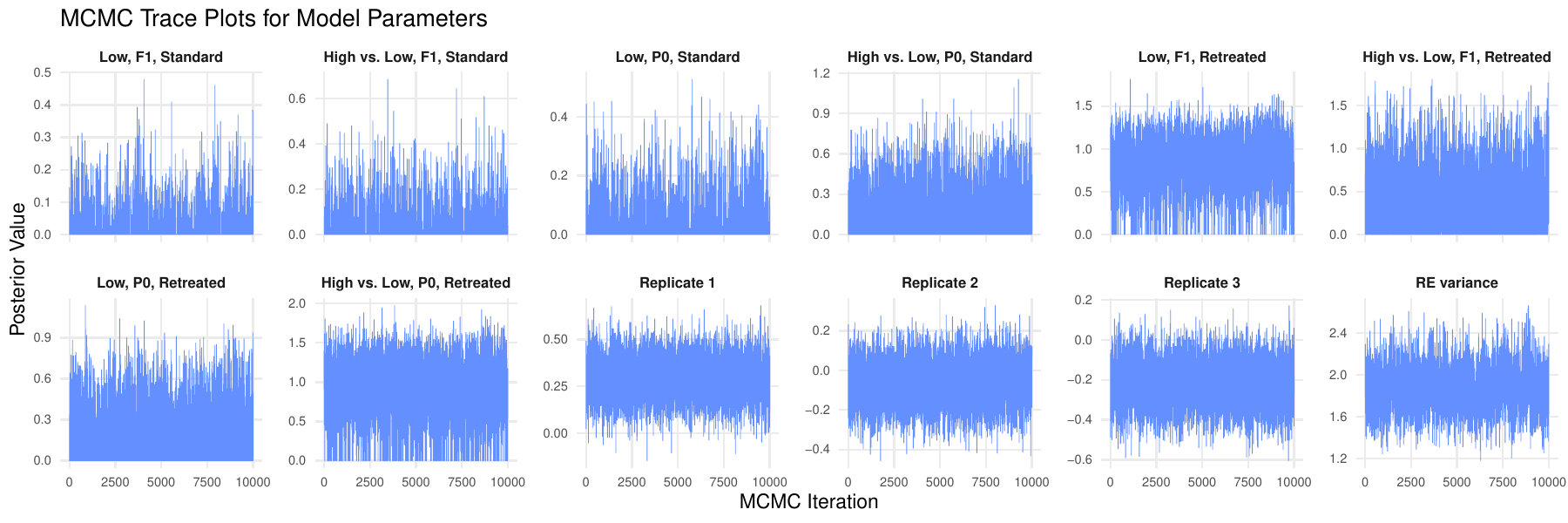}
    \put(-465,150){\makebox(0,0){\textbf{(A)}}}
    \\
    \includegraphics[width=\textwidth]{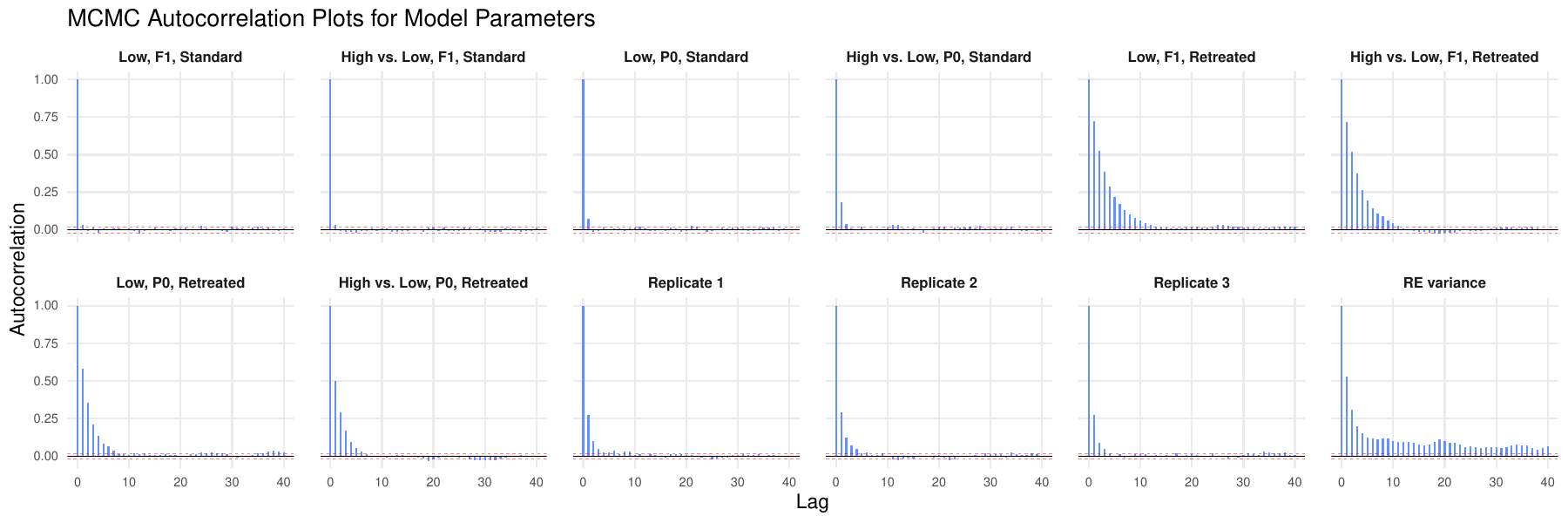}
    \put(-465,150){\makebox(0,0){\textbf{(B)}}}
    \vspace{-10pt}
    \caption{
    Trace plots (A) and autocorrelation functions (B) for treatment coefficients, replicates' batch effects, and random effects variance, displaying good \textsc{mcmc} mixing}
    \label{fig:coeff_trace}
\end{figure}

\vspace{20pt}

\subsection{Latent Scores and Random Effects}

\begin{figure}[H]
        \centering
        ~\hspace{2pt}
        \includegraphics[width=\textwidth]{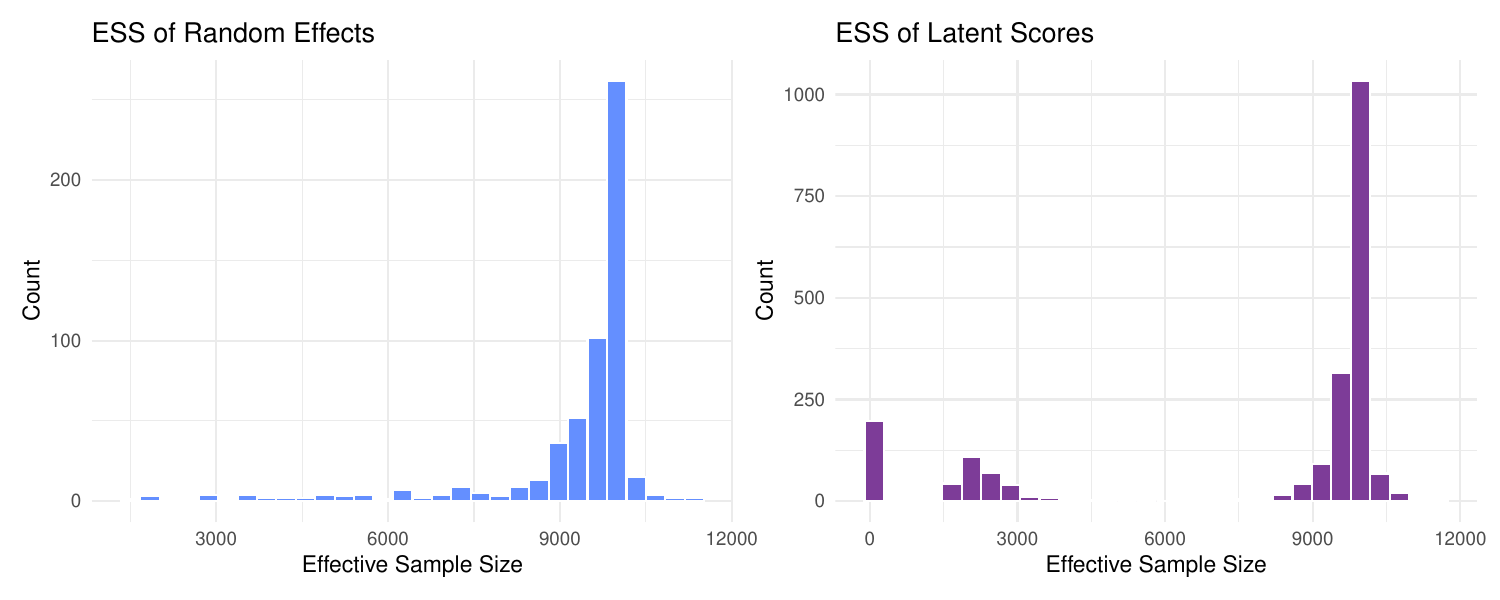}
            \put(-460,177){\makebox(0,0){\textbf{(A)}}}
            \put(-230,177){\makebox(0,0){\textbf{(B)}}}
        \vspace{-25pt}
        \caption{\na{Histograms of the ESS for the $N = \text{2,099}$ neuron-level latent scores $Z_{ij}$ (A) and for the $n = 559$ worm-level random effects $\eta_i$ (B).}}
        \label{fig:ess_Z_eta}
\end{figure}

\na{
Figure~\ref{fig:ess_Z_eta} reports histograms of the ESS for the $N = \text{2,099}$ neuron-level latent scores $Z_{ij}$ and the $n = 559$ worm-level random effects $\eta_i$.
Recall that we generated $10,000$ samples from the posterior distribution -- after discarding burn-in.
For the random effects, the ESS indicates near-independent draws.
The latent scores show a bimodal pattern: most parameters attain similarly high ESS, but a smaller subset clusters with small ESS.
The latter correspond to neurons at the boundary between damage categories, where the rank likelihood imposes tighter truncation constraints.
}

\begin{figure}[hb!]
    \includegraphics[width=0.49\textwidth]{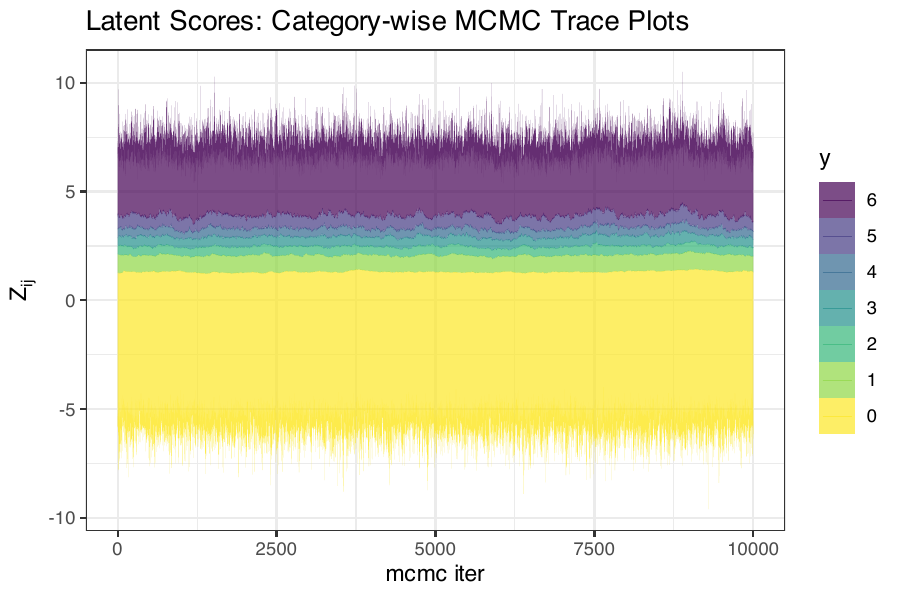} 
    \put(-225,148){\makebox(0,0){\textbf{(A)}}}
    \includegraphics[width=0.49\textwidth]{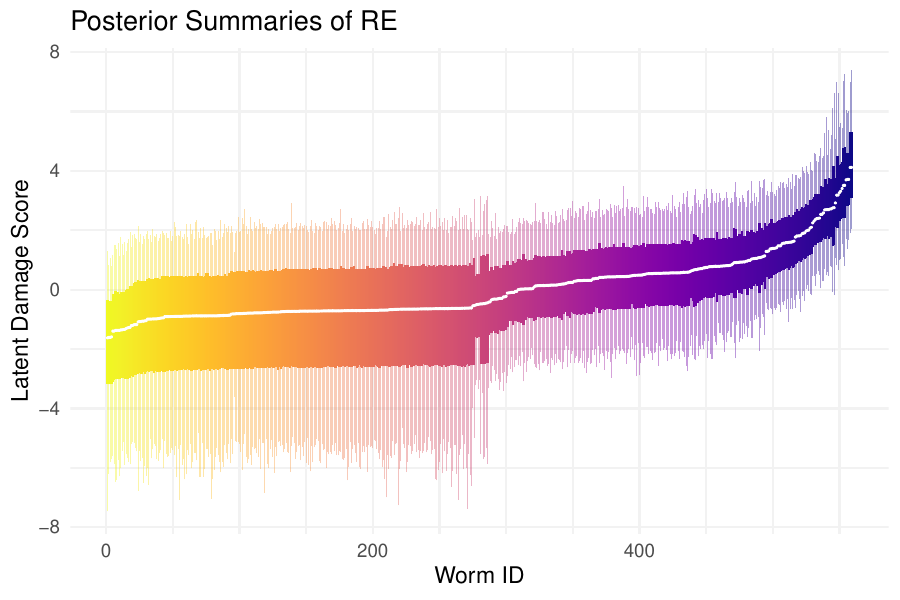}
    \put(-230,148){\makebox(0,0){\textbf{(B)}}}
    \vspace{-10pt}
    \caption{\na{(A) Traceplots of latent scores, colored by corresponding damage scores.
    (B) Marginal posterior distribution for each worm (n = 559 worms). White dots, fully-colored bands, and shaded ones represent posterior medians, $90\%$ credible intervals, and full range, respectively. The worms are colored and ordered along the x-axis based on the median of the marginal.}}
    \label{fig:traces_Z_eta}
\end{figure}

\na{
Figure~\ref{fig:traces_Z_eta} provides further diagnostic detail: traceplots of the latent scores (left) show stable, well-mixed chains within each damage-score category, with no visible drift or trend; the marginal posterior distributions across worms (right) reveal the heterogeneous random effects already discussed, ordered from least to most damage-prone, with $90\%$ credible intervals that are visibly tighter for the bulk of worms and wider for the susceptible tail on the right.}

\newpage

~\vspace{10pt}

\vspace{5pt}

\vspace{5pt}

\subsection{Collapsed Gibbs \& Parameter Expansion vs Mixing}

\begin{figure}[H]
    \includegraphics[width=\textwidth]{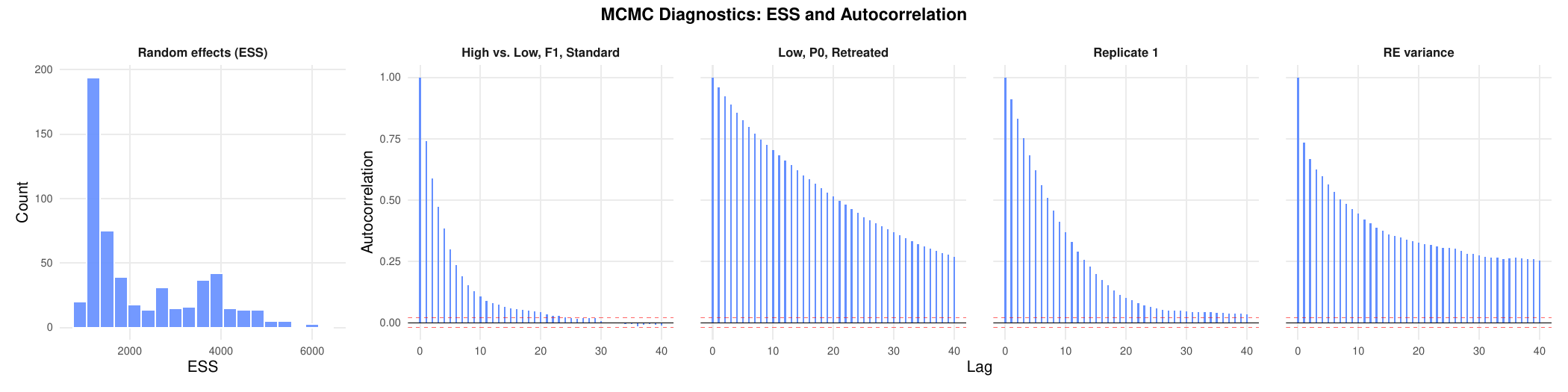} 
    \put(-465,120){\makebox(0,0)[lt]{\textbf{(A)} \, Regular Sampler}}\\[10pt]
    \includegraphics[width=\textwidth]{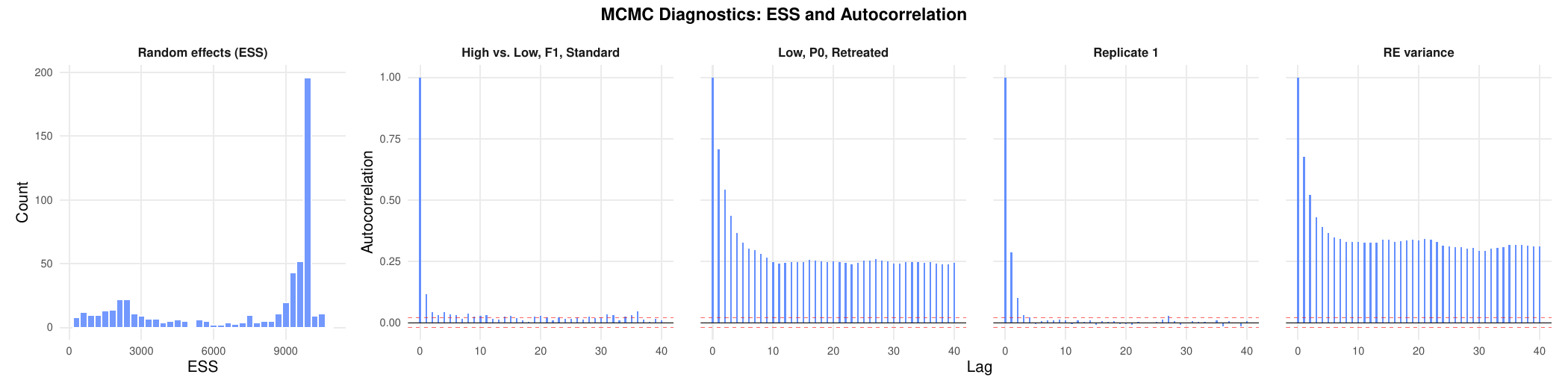} 
    \put(-465,120){\makebox(0,0)[lt]{\textbf{(B)} \, Collapsed Sampler}}\\[10pt]
    \includegraphics[width=\textwidth]{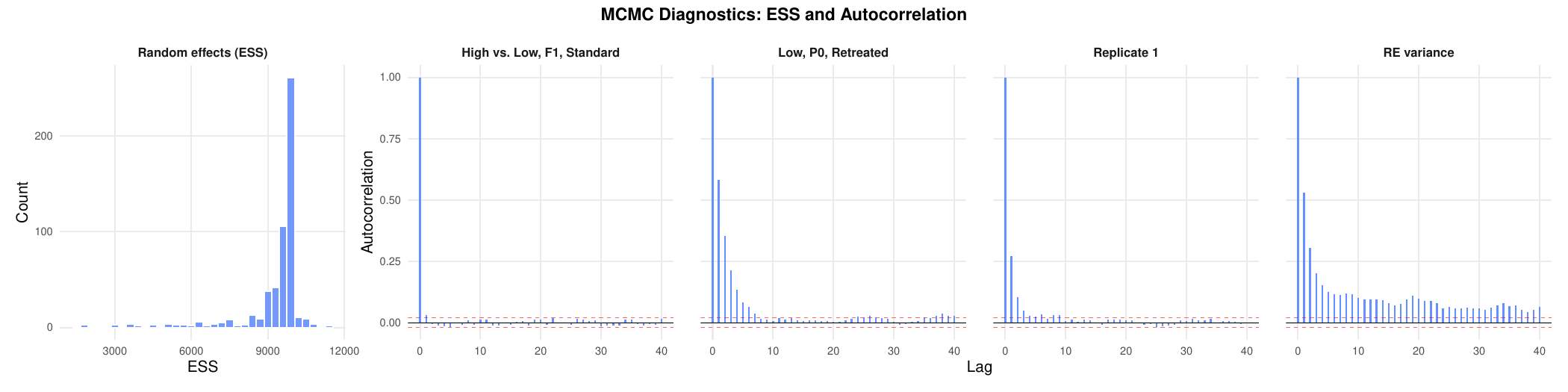} 
    \put(-465,120){\makebox(0,0)[lt]{\textbf{(C)} \, Collapsed Sampler + \textsc{px}}}
    \caption{\na{Comparing mixing across different versions of the Gibbs sampling schemes: (A) regular Gibbs sampler, no parameter expansion (B) collapsed Gibbs sampler, no parameter expansion (C) collapsed Gibbs sampler, with parameter expansion. In the first plot of each panel, we report the histogram of the ESS across random effects. In the rest, we report ACF for the random effect variance, representative treatment coefficients, and replicate effects. }}
    \label{fig:acf_ess_px_marg}
\end{figure}

\na{Figure~\ref{fig:acf_ess_px_marg} illustrates the mixing diagnostics across the three Gibbs sampling variants.
In particular, Figure~\ref{fig:acf_ess_px_marg} pairs the full distribution of the random effects' ESS (first column) with the ACF for the random effect variance and representative fixed effects (last four columns).
Under the standard Gibbs sampler (Panel A), the random effects mix poorly, exhibiting a heavily right-skewed ESS distribution concentrated at low values, while the treatment and replicate ACFs show slow, problematic decay.
The collapsed version without parameter expansion (Panel B) shifts the bulk of the random effects toward higher ESS values and improves the short-term decay across all ACF panels, but it leaves a bulk of low ESS in $\boeta$ and non-dying autocorrelation even at large lags.
Poor chain mixing is only resolved by combining the collapsed sampler with parameter expansion (Panel C).}

\newpage

\putbib
\end{bibunit}

\end{document}